\pdfoutput=1
\documentclass[journal]{IEEEtran}

\usepackage[T1]{fontenc}
\usepackage{cite}
\usepackage{graphicx}
\usepackage{acronym}
\usepackage{amsmath}
\usepackage{amssymb}
\usepackage{bm}
\usepackage{siunitx}
\usepackage{booktabs}
\usepackage{tabularx}
\usepackage{multirow}
\usepackage{array}
\usepackage{makecell}   %
\usepackage[table,xcdraw]{xcolor}
\usepackage{cite}
\usepackage{threeparttable}
\usepackage{comment}
\usepackage{caption}

\newcolumntype{L}[1]{>{\raggedright\arraybackslash}m{#1}}
\newcolumntype{C}[1]{>{\centering\arraybackslash}m{#1}}

\acrodef{0-D}{zero-dimensional}
\acrodef{1-D}{one-dimensional}
\acrodef{2-D}{two-dimensional}
\acrodef{3-D}{three-dimensional}
\acrodef{ABM}{agent-based modeling}
\acrodef{BEM}{boundary element method}
\acrodef{CFD}{computational fluid dynamics}
\acrodef{FEM}{finite element method}
\acrodef{FVM}{finite volume method}
\acrodef{FDM}{finite difference method}
\acrodef{PBS}{particle-based simulation}
\acrodef{LBM}{lattice Boltzmann method}
\acrodef{LGA}{lattice gas automata}
\acrodef{MC}{molecular communication}
\acrodef{PDE}{partial differential equation}
\acrodef{SMC}{synthetic molecular communication}
\acrodef{MIGHT}{Mixture of Inverse Gaussians for Hemodynamic Transport}
\acrodef{NMC}{natural molecular communication}
\acrodef{SSA}{stochastic simulation algorithm}
\acrodef{SPH}{smoothed particle hydrodynamics}
\acrodef{RBC}{red blood cell}
\acrodefplural{RBC}[RBCs]{red blood cells}
\acrodef{SM}{signaling molecule}
\acrodefplural{SM}[SMs]{signaling molecules}
\acrodef{Tx}{transmitter}
\acrodef{Rx}{receiver}
\acrodef{IoBNT}{Internet of Bio-Nano Things}
\acrodef{VOC}{volatile organic compound}
\acrodefplural{PDE}[PDEs]{partial differential equations}
\acrodef{Tx}{transmitter}
\acrodef{Rx}{receiver}
\acrodef{VN}[VN]{vessel network}
\acrodefplural{VN}[VNs]{vessel networks}

\acrodef{CVS}{cardiovascular system}
\acrodef{ISI}{inter-symbol interference}
\acrodef{SPION}{superparamagnetic iron-oxide nanoparticle}
\acrodefplural{SPION}[SPIONs]{superparamagnetic iron-oxide nanoparticles}
\acrodef{GFPD}{green fluorescent protein Dreiklang}
\acrodef{BD}{Brownian dynamics}
\acrodef{BDF}{backward differentiation}
\acrodef{CA}{cellular automaton}
\acrodef{CIR}{channel impulse response}
\acrodef{CLE}{chemical Langevin equation}
\acrodef{CPM}{cellular Potts model}
\acrodef{DEM}{discrete element method}
\acrodef{DPD}{dissipative particle dynamics}
\acrodef{GFRD}{Green's function reaction dynamics}
\acrodef{HI}{hydrodynamic interactions}
\acrodef{IBM}{immersed boundary method}
\acrodef{MCS}{Monte Carlo step}
\acrodef{MD}{molecular dynamics}
\acrodef{MPCD}{multi-particle collision dynamics}
\acrodef{ODE}{ordinary differential equation}
\acrodef{PINN}[PINN]{physics-informed neural network}
\acrodefplural{PINN}[PINNs]{physics-informed neural networks}
\acrodef{RDME}{reaction--diffusion master equation}
\acrodef{RRE}{reaction rate equation}
\acrodef{SD}{Stokesian dynamics}
\acrodef{SRD}{stochastic rotation dynamics}
\acrodefplural{ODE}[ODEs]{ordinary differential equations}
\acrodefplural{RRE}[RREs]{reaction rate equations}
\acrodefplural{CA}[CAs]{cellular automata}
\acrodefplural{MCS}[MCS]{Monte Carlo steps}
\acrodef{ABP}{active Brownian particles}
\acrodef{CME}{chemical master equation}
\acrodef{FBA}{flux balance analysis}
\acrodef{FSP}{finite state projection}

\acrodef{NN}{neural network}
\acrodef{ND}{nanodevice}

\acrodef{ANN}{artificial neural network}
\acrodef{FSC}{finite state controller}
\acrodef{GNN}{graph neural network}
\acrodef{MLP}{multi-layer perceptron}
\acrodef{PINN}{physics-informed \ac{NN}}
\acrodef{RNN}{recurrent neural network}
\acrodef{GAN}{generative adversarial network}
\acrodef{GP}{Gaussian process}
\acrodefplural{GP}{Gaussian processes}
\acrodef{DL}{deep learning}
\acrodef{LTI}{linear time-invariant}
\acrodef{RL}{reinforcement learning}
\acrodef{MSE}{mean squared error}

\acrodef{MOS}{metal-oxide semiconductor}
\acrodef{PID}{photo-ionization detector}
\acrodef{ML}{machine learning}

\acrodef{SDE}{stochastic differential equation}
\acrodef{RDME}{reaction-diffusion master equation}
\acrodef{PDF}{probability density function}

\acrodef{OOK}{ON-OFF keying}
\acrodef{BER}{bit error rate}
\acrodef{CAM}{chorioallantoic membrane}
\acrodef{ICG}{indocyanine green}
\acrodef{ROI}{region of interest}

\acrodef{MIMO}{multiple-input multiple-output}
\acrodef{AI}{artificial intelligence}

\definecolor{DarkGreen}{RGB}{0,150,0}
\definecolor{DarkBlue}{RGB}{0,0,200}
\definecolor{LightBlue}{RGB}{150,150,200}
\definecolor{Purple}{RGB}{128, 0, 128}
\definecolor{Orange}{RGB}{255, 128, 0}

\usepackage{pifont}

\newcommand{\na}{--}

\usepackage{stfloats}

\usepackage{booktabs,tabularx,array}

  \definecolor{N1}{HTML}{E1E9EE}\definecolor{N5}{HTML}{273237}   %
  \definecolor{G1}{HTML}{F0EFC4}\definecolor{G5}{HTML}{03411C}   %
  \definecolor{G3}{HTML}{8BB020}                                 %
  \definecolor{O1}{HTML}{F7EEBA}\definecolor{O5}{HTML}{571013}   %
  \definecolor{T1}{HTML}{CBF5FB}\definecolor{T5}{HTML}{023734}   %
  \definecolor{P1}{HTML}{FED8EF}\definecolor{P5}{HTML}{3B1844}   %
  \definecolor{B1}{HTML}{D1E2FE}\definecolor{B5}{HTML}{132D5C}   %

  \newcommand{\tablepalette}[1]{%
    \colorlet{TabHead}{#15}\colorlet{TabFirst}{#11}\colorlet{TabZebra}{#11!45!white}%
    \ifcsname tabtweak#1\endcsname\csname tabtweak#1\endcsname\fi}
  
  \tablepalette{N}                               %
  \newcolumntype{F}{>{\cellcolor{TabFirst}}l}    %
  \newcommand{\tabhead}[1]{\textcolor{white}{\bfseries #1}}
  \newcommand{\tablestyleC}[1][N]{%
    \tablepalette{#1}%
    \setlength{\aboverulesep}{0pt}\setlength{\belowrulesep}{0pt}%
    \renewcommand{\arraystretch}{1.2}%
    \rowcolors{2}{white}{TabZebra}}

\definecolor{O3}{HTML}{F17814}                                 %

 \definecolor{T3}{HTML}{19B6AE}                                 %

\usepackage{amsmath}
\allowdisplaybreaks

\begin{document}

\title{Modeling of Molecular Communication Channels: A Tutorial on the Fundamentals and the Interplay Between Theory and Experiment \vspace*{-0.3cm}}

\author{\IEEEauthorblockN{Maximilian Schäfer\textsuperscript{\scriptsize 1}, Bastian Heinlein\textsuperscript{\scriptsize 1,2}, Timo Jakumeit\textsuperscript{\scriptsize 1}, Theofilos Symeonidis\textsuperscript{\scriptsize 1}, Teena tom Dieck\textsuperscript{\scriptsize 1}, Lukas Brand\textsuperscript{\scriptsize 1}, Sebastian Lotter\textsuperscript{\scriptsize 1}, Fardad Vakilipoor\textsuperscript{\scriptsize 1}, Vahid Jamali\textsuperscript{\scriptsize 2}, and Robert Schober\textsuperscript{\scriptsize 1}} %
\IEEEauthorblockA{
\textsuperscript{\scriptsize 1}\textit{Friedrich-Alexander-Universität Erlangen-Nürnberg (FAU), Erlangen, Germany}\\
\textsuperscript{\scriptsize 2}\textit{Technical University of Darmstadt, Darmstadt, Germany}
}
\vspace*{-1cm}
}

\maketitle

\begin{abstract}
\Ac{SMC} is an emerging communication paradigm in which information is encoded into the properties of molecules instead of electromagnetic waves. Inspired by natural \ac{MC} systems, \ac{SMC} is envisioned to extend wireless communication into new domains, which enables innovative applications in healthcare, environmental science, and industry. Over the past two decades, the field has grown from theoretical research toward experimental validation and application-driven research. This tutorial revisits this development with a particular focus on the interplay between theory and experiment in channel modeling and on how this relationship has reversed. Initially, theory led and experiments mainly served as validation of channel models. Today, experiments reveal physical effects that no theoretical model anticipated, unveil which simplifying assumptions are physically justified, and thereby redefine what a channel model has to capture. This has led to a paradigm shift: the objective is no longer to describe a system in as much detail as possible, but to let the intended application and the specific research question determine the appropriate level of model complexity. From this perspective, we first present a mathematical description of molecular transport and chemical reactions in \ac{SMC} channels and review analytical, simulation, data-based, and experimental models of \ac{SMC} systems, and provide guidance for choosing among them. After having established these fundamentals, we show how observations from experiments have shaped channel models and justified their simplifying assumptions
based on two representative examples, i.e., the \ac{CVS} and airborne environments, and discuss additional aspects of this interplay based on three \ac{SMC} testbeds.
Finally, we discuss open problems in theoretical and experimental \ac{SMC} modeling.
\end{abstract}


\IEEEpeerreviewmaketitle

\section{Introduction}
\label{sec:introduction}
\Acf{SMC} is a communication paradigm in which information is encoded into the properties of \acp{SM} -- their type, concentration, or release time -- rather than into electromagnetic (EM) waves \cite{nakano13,Farsad2016,Akyildiz2008}. In this section, we motivate \ac{SMC} based on its unique properties borrowed from nature, elaborate the applications of \ac{SMC} unattainable by EM-based systems, discuss the importance of developing appropriate models of \ac{SMC} systems as well as the ongoing evolution of the corresponding methods and approaches, review existing surveys and tutorials and contrast their content with this paper, and provide an overview of the contributions and organization of this paper.

\subsection{Synthetic Molecular Communications: From Natural to Engineered Communication Systems}
\label{ssec:intro:intro}
\ac{SMC} is inspired by natural communication processes, where specific \acp{SM} are transmitted, propagated, and received by biological entities to fulfill a certain task, e.g., quorum sensing via autoinducers, adjusting body functions via hormones, or attracting a mating partner via pheromones.

The primary motivation for \ac{SMC} is to enable communication in scenarios and environments that impede or preclude the use of EM-based communication systems. For instance, inside the body, EM waves suffer strong attenuation, which is particularly severe at the high carrier frequencies imposed by nanoscale devices. Outside the body, EM-based communication can be inefficient or infeasible in so-called wave-denied environments, e.g., in metal pipe networks, tunnels, or underground \cite{guo:molecular_physical_layer_wavedenied_environments}.
By using \acp{SM} as information carriers, \ac{SMC} promises to extend the reach of communication systems into these domains. More importantly, \ac{SMC} provides a direct interface for interacting with biological systems relying on natural \ac{MC} processes \cite{Farsad2016}. Since the field was established \cite{nakano2005molecular}, it has matured over roughly two decades from predominantly theoretical work toward more experimental and application-driven research \cite{lotter2023experimental,Lotter23b} -- a development that also changed the way communication-theoretical modeling and design of \ac{SMC} systems are approached.

\subsection{Envisioned Applications of SMC}
\label{ssec:intro:applications}

\begin{figure*}
    \centering
    \includegraphics[width=\linewidth]{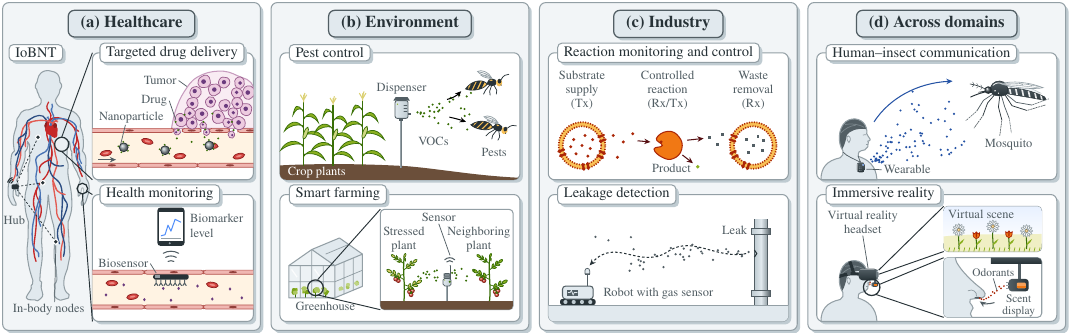}
    \caption{\textbf{Selection of envisioned applications of \ac{SMC}.}
    \textbf{(a)}~Healthcare: in-body networks of nanodevices connected to an external hub (\ac{IoBNT}), targeted drug delivery to diseased tissue, and health monitoring by implanted biosensors.
    \textbf{(b)}~Environment: pest control by dispensed \acp{VOC} and smart farming by sensing \acp{VOC} emitted by stressed plants.
    \textbf{(c)}~Industry: monitoring and control of reactions by nanodevices for substrate supply and waste removal, and localization of leakages by mobile robots with gas sensors.
    \textbf{(d)}~Across domains: human-insect communication via wearables and immersive reality via scent displays.}
    \label{fig:intro:applications}
    \vspace*{-0.3cm}
\end{figure*}

\ac{SMC} envisions various applications, which can be categorized based on their targeted field. Fig.~\ref{fig:intro:applications} illustrates selected applications of \ac{SMC}, classified into healthcare, environmental, industrial, and cross-domain applications.

\textbf{Healthcare applications} of \ac{SMC}
address medical challenges, including health monitoring \cite{felicetti2016applications}, targeted drug delivery \cite{Chude-Okonkwo_et_al:IEEECommSurvTutorials:2017,chahibi2017tdd}, in-body communication \cite{kuscu21a,Dressler15}, and the spread of infectious diseases \cite{Barros21}, see Fig.\ \ref{fig:intro:applications}(a). An overarching vision of \ac{SMC} healthcare applications is provided by the \ac{IoBNT} \cite{kuscu21a,akyildiz15}. The \ac{IoBNT} facilitates the autonomous detection and treatment of diseases inside the human body based on an in-body network of nanodevices and an external control unit. The in-body network includes engineered and natural sensors and actuators to detect biomarkers and to facilitate localized and personalized treatment \cite{Dressler15,kuscu21a}.
\textbf{Environmental applications} of \ac{SMC}
often focus on the interaction of plants and insects with their environment, see Fig.\ \ref{fig:intro:applications}(b). In particular, concepts from \ac{SMC} can be exploited to study plant-to-plant and plant-to-insect communication, in order to enable the design of novel concepts for plant engineering and plant-health monitoring for smart farming \cite{Aktas24,Ahmed22,Heil2010,Jamali23,Unluturk17plants} as well as innovative \ac{SMC}-based strategies for sustainable pest control \cite{beck2017harnessing,Vakilipoor25}.
Next to healthcare and environmental applications, \ac{SMC} is expected to also have various \textbf{industrial applications}, see Fig.\ \ref{fig:intro:applications}(c).
For example, for the design of new bioprocesses, which is preferably done at microliter scale to avoid excessive waste, concepts from \ac{MC} are promising for guiding the development of the required nanodevices and for the derivation of corresponding control algorithms \cite{tomdieck2026nanodevices,hemmerich2018microbioreactor}. Moreover, concepts from \ac{SMC} have been exploited for the design of systems for localization of leakages in pipe systems and chemical plants \cite{jakumeit2026matched, burgues2019smelling, Okcu2025smell}.
Other applications of \ac{SMC} are located at the intersection of the three established categories, such as human-insect communication or immersive reality  \cite{Aktas24,robinson:plasmodium_associated_changes_mosquitoes}, see Fig.\ \ref{fig:intro:applications}(d).

\subsection{Modeling of SMC Systems}
\label{ssec:intro:modeling}

The realization of the envisioned applications requires careful analysis, design, and optimization of the corresponding \ac{SMC} systems, which, like for any other communication system, requires suitable models for the underlying channel that are simple enough to be tractable yet accurate enough for the task the system has to carry out \cite{Jamali2019}. A suitable \ac{SMC} channel model relates the \acp{SM} released by a \ac{Tx} to the signal observed at a \ac{Rx}, thereby turning the underlying physical, chemical, and biological processes into measurable quantities that reflect the system behavior (see Section~\ref{sec:taxonomy}). Ultimately, such models allow the prediction of the performance, the optimization of the system parameters, and the exploration of the fundamental limits of \ac{SMC} systems \cite{Gohari2016,Jamali2019}, without exhaustive and costly simulations or experiments. For these reasons, channel modeling has been a central aspect of \ac{SMC} research since its inception \cite{Jamali2019,Farsad2016}.

Much of the \ac{SMC} modeling literature has been theory-led: analytical channel models were derived first, and experiments served mainly to validate them \cite{Farsad2016}. This relationship has recently been reversed. As experimental \ac{SMC} testbeds have become more common and more closely reflect the targeted application environment, experiments increasingly reveal physical effects that no analytical model anticipated, unveil which simplifying assumptions are physically justified, and thereby redefine which processes an \ac{SMC} channel model must capture \cite{lotter2023experimental,Lotter23b, Jakumeit2026sorption}. Experimental research in \ac{SMC} has thus moved from merely confirming analytical and parametric channel models to guiding their development. This conceptual reversal also implies a \emph{paradigm shift} in the goal of \ac{SMC} channel modeling: the objective is no longer to describe an idealized system, e.g., a point \ac{Tx} with impulsive release in an unbounded medium and an ideal \ac{Rx} without sensor effects, in as much detail as possible, but to let the application define a suitable model. In particular, the ultimate task of the \ac{SMC} system and the specific research question one seeks to answer determine the appropriate level of model complexity. This realization is not unique to the field of \ac{SMC}. In wireless communications, choosing among different levels of abstraction, e.g., link and network level, based on the application at hand is well established. At the physical layer, e.g., millimeter-wave channel measurements have revealed a sparse multipath structure in the angular domain, which has subsequently motivated corresponding angular channel representations and provided a basis for the choice of empirical model parameter distributions \cite{hemadeh2017millimeter}. At the network level, by contrast, these detailed propagation characteristics are often abstracted into link-level quantities such as latency, error rate, or achievable data rate \cite{poddar2023tutorial}.

One central aspect of this tutorial is this paradigm shift and the resulting interplay between experimental and theoretical channel modeling for \ac{SMC} systems. To this end, it is crucial to have a clear categorization and understanding of the different types of models, which all build on the same \emph{mathematical description} of the underlying physical processes for the transmission, propagation, and reception of \acp{SM}.
\emph{Analytical models} solve the governing equations in closed form, possibly after approximations, and facilitate insight and fast evaluation. \emph{Simulation models} solve these equations numerically and can handle more complex \ac{SMC} systems at higher computational cost. \emph{Data-based models} learn the input-output behavior from measured or simulated data and are suited if the governing equations are unknown or too costly to solve. Finally, \emph{experimental models}, i.e., experimental testbeds, realize an \ac{SMC} system physically, and aim to emulate the target application environment by reproducing selected properties of it.

\begin{figure*}[t]
\centering
\includegraphics{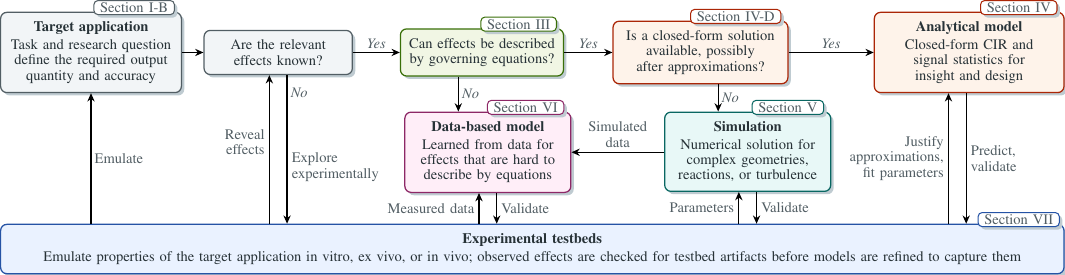}
\caption{\textbf{Choice of appropriate \ac{SMC} channel model and interplay between experimental models and analytical, simulation, and data-based models.} Starting from the target application, the decision path leads to an analytical model if the relevant effects are known, described by governing equations, and solvable in closed form; otherwise, simulation or data-based models are used.
Experimental testbeds emulate the target application and reveal unknown effects.}
\label{fig:intro:model-choice}
\vspace*{-0.3cm}
\end{figure*}

Fig.~\ref{fig:intro:model-choice} illustrates how the different types of models relate to each other and to the target application that defines which output quantity a model has to predict and to which accuracy. If the relevant physical effects are not known in advance, they have to be explored experimentally.
Otherwise, whether the governing equations are known and can be solved in analytical or closed form determines the choice among the different modeling approaches introduced above.
Experimental models are crucial for all other modeling techniques (see Fig.~\ref{fig:intro:model-choice}): in the theory-led direction, analytical, simulation, and data-based models predict and experimental models validate, whereas in the experiment-guided direction, experiments reveal physical effects, justify approximations, and provide parameters and training data.

In this tutorial, we focus on two application environments to illustrate the modeling of \ac{SMC} systems: the \textbf{\acf{CVS}} \cite{Chahibi13,Jakumeit2026MIGHT}, as it constitutes the most important domain for healthcare applications, and \textbf{airborne} environments \cite{farsad2013tabletop,heinlein:tmbmc2026}, as they are relevant for a large class of environmental and industrial applications. These environments were chosen because they are complementary in various aspects. The \ac{CVS} is an in-body, liquid environment in which \ac{SM} transport is bounded and dominated by advective flow through a complex \ac{VN}. Moreover, direct measurement of the molecular quantities of interest is challenging because of their microscopic scale and limited experimental access within the body, where safety and regulatory constraints further restrict investigation. Airborne environments, in contrast, are macroscale and gaseous, turbulent transport dominates, and low-cost hardware is readily available. Thus, corresponding environmental and industrial applications are comparatively closer to practical realization. In Sections~\ref{sec:examples-cvs} and~\ref{sec:examples-airborne}, we apply the models introduced in Sections~\ref{sec:theoretical-models}-\ref{sec:experimental-models} for the modeling of \ac{SMC} systems in the \ac{CVS} and airborne environments.

\subsection{Existing Surveys and Tutorials}
\label{ssec:intro:existingSurvTut}

There exist many surveys and tutorials on \ac{SMC} and the most relevant ones for this paper are summarized in Table~\ref{tab:intro:surveys}. An early overview of \ac{SMC} as a networking paradigm is provided in \cite{nakano12}, and a comprehensive overview of the field is given in \cite{Farsad2016}. Channel modeling is covered in detail in \cite{Jamali2019}, \ac{Tx} and \ac{Rx} architectures are reviewed in \cite{kuscu19}, and modulation techniques are surveyed in \cite{kuran21}, while information-theoretic limits are the subject of \cite{Gohari2016}. The biological and chemical foundations of \ac{SMC} and the relation between \ac{SMC} and cell biology are studied in \cite{Soldner20} and \cite{bi2021survey}, respectively. Experimental research is reviewed in \cite{lotter2023experimental,Lotter23b}. For individual application domains, such as drug delivery \cite{Chude-Okonkwo_et_al:IEEECommSurvTutorials:2017}, odor-based communication \cite{Aktas24}, and exhaled-breath analysis \cite{bhattacharjee2026breath}, as well as the role of machine learning in \ac{SMC} \cite{torresgomez2025nn}, there are also dedicated surveys.

Each of these works is organized around a specific topic or subfield of \ac{SMC}. However, none of them provides a holistic and comprehensive overview on \ac{SMC} channel modeling, particularly from the complementary theory- and application-guided perspectives that have emerged in recent years. The work most closely related to this paper is \cite{Jamali2019}, which provides a tutorial review of the mathematical foundations of \ac{SMC} channel modeling but discusses simulation, data-driven and experimental models only briefly and does not cover recent advances regarding the interplay between theoretical, experimental, and application-oriented modeling. This tutorial closes this gap by providing an overview of existing modeling techniques, investigating the interplay of experimental and theoretical modeling, and revealing how the intended application ultimately determines the required level of model complexity.

\begin{table*}[!t]
  \centering
  \caption{Selected surveys and tutorials on \ac{SMC}.}
  \vspace*{-0.2cm}
  \label{tab:intro:surveys}
  \footnotesize
  \tablestyleC
  \begin{tabularx}{\textwidth}{F l p{3.0cm} X}
  \rowcolor{N5}
  \cellcolor{N5}\tabhead{Ref.} & \tabhead{Year} & \tabhead{Topic / Focus} & \tabhead{Key Contributions \& Scope} \\
  \cite{nakano12}  & 2012 & MC networking            & Early survey; SMC paradigms (diffusion, ligand-receptor, gap junctions) and an OSI-inspired
  nanonetwork stack. \\
  \cite{Farsad2016}  & 2016 & Comprehensive survey     & Comprehensive survey; biophysics, channel modeling, modulation, coding, simulation tools, and
  testbeds. \\
  \cite{Gohari2016}  & 2016 &  Information theory        & Information-theoretic treatment; capacity for concentration- and timing-based modulation. \\
  \cite{Chude-Okonkwo_et_al:IEEECommSurvTutorials:2017}  & 2017 & Targeted drug delivery    & SMC-based drug delivery; nanocarrier design, cardiovascular
  channel models, pharmacokinetic modeling. \\
  \cite{Jamali2019} & 2019 & Channel modeling & Green's-function models, reaction-advection-diffusion, time-varying channels, and simulation. \\
  \cite{kuscu19} & 2019 & Tx/Rx architectures       & Tx/Rx design; nanomaterial-based sensors; modulation, coding, and detection. \\
  \cite{Soldner20} & 2020 & Biological building blocks & Interdisciplinary catalog of DNA, proteins, and lipids as building blocks for \ac{SMC}. \\
  \cite{kuran21} & 2021 & Modulation techniques     & CSK/MoSK/timing/spatial modulation for diffusive \ac{SMC}. \\
  \cite{bi2021survey} & 2021 & \ac{SMC} in cell biology        & Five-level communication hierarchy; quorum sensing, neuronal and DNA communication. \\
  \cite{lotter2023experimental,Lotter23b} & 2023 & Experimental \ac{SMC} & First comprehensive survey of experimental \ac{SMC}; short- and long-range testbeds by application. \\
  \cite{Aktas24} & 2025 & Odor-based \ac{SMC} & Olfaction-inspired \ac{SMC}; VOC propagation, olfactory reception, environmental monitoring. \\
  \cite{torresgomez2025nn} & 2025 & Neural networks for MC   & Data-driven channel models, NN-based detectors, and learned end-to-end \ac{SMC} systems. \\
  \cite{bhattacharjee2026breath} & 2026 & Exhaled-breath analysis   & Air-based \ac{SMC} for breath analysis; breath as Tx, sensor arrays as Rx; health monitoring. \\
  \arrayrulecolor{N5}\specialrule{1pt}{0pt}{0pt}
  \end{tabularx}
  \\[2pt]
  \parbox{\textwidth}{\raggedright\scriptsize OSI: open systems interconnection; CSK: concentration shift keying; MoSK: molecule shift keying; BER: bit error rate; ISI: inter-symbol interference; VOC: volatile organic compound; NN: neural network.}
  \vspace*{-0.5cm}
  \end{table*}

\subsection{Contributions and Organization}
\label{ssec:intro:contributions}

This tutorial covers \ac{SMC} channel modeling, encompassing the relevant analytical, simulation, data-based, and experimental aspects. The contributions of this tutorial can be summarized as follows:
\begin{itemize}
    \item We provide a fundamental tutorial overview of the mathematical description of \ac{SMC} systems and of analytical, simulation, data-based, and experimental models, as well as guidance for choosing among them.
    \item We unveil how the relationship between theory and experiment has changed in \ac{SMC} channel modeling, i.e., how experiments increasingly guide the development of analytical, simulation, and data-driven channel models rather than merely validating them.
    \item We apply the developed channel modeling concepts exemplarily to two important use cases, namely \ac{SMC} systems operating inside the \ac{CVS} and in airborne environments, covering the corresponding channel models and the experimental testbeds used to study them, and investigate additional aspects of the interplay between theory and experiment based on three \ac{SMC} testbeds.
    \item We identify challenges in theoretical and experimental modeling of \ac{SMC} systems and outline directions for future work.
\end{itemize}

The remainder of this paper is organized as follows. Section~\ref{sec:taxonomy} introduces a taxonomy for \ac{SMC} systems and a general end-to-end channel model. Section~\ref{sec:mathematical-models} presents  mathematical descriptions of the transport and environmental processes relevant in \ac{SMC} systems. Sections~\ref{sec:theoretical-models}, \ref{sec:simulation-models}, and \ref{sec:data-based-models} provide a tutorial review of analytical, simulation, and data-based models, respectively. Section~\ref{sec:experimental-models} discusses experimental models, i.e., testbeds, and their role as models of the target application environment. In Sections~\ref{sec:examples-cvs} and~\ref{sec:examples-airborne}, we apply the developed modeling concepts to two practically relevant use cases, i.e., \ac{SMC} systems operating inside the \ac{CVS} and airborne environments. Section~\ref{sec:examples-others} discusses additional aspects of the interplay between theory and experiment based on three experimental \ac{SMC} systems.
Finally, Section~\ref{sec:challenges} presents open challenges and future directions, and Section~\ref{sec:conclusion} concludes the paper.

\section{Taxonomy of SMC Systems}
\label{sec:taxonomy}
Before any model of an \ac{SMC} system can be derived, it must be specified which physical processes should be captured. In this section, we therefore first introduce a canonical end-to-end \ac{SMC} channel model comprising release, propagation, and reception processes, see Fig.~\ref{fig:taxonomy:e2echannel} for an overview. Subsequently, we discuss exemplary realizations of these processes for three relevant application environments, namely microscale liquid, macroscale liquid, and airborne \ac{SMC} systems.

\subsection{Canonical End-to-End \ac{SMC} Channel Model}
Any communication system can be canonically decomposed into a \emph{\ac{Tx}}, a \emph{channel}, and an \emph{\ac{Rx}}. The goal of channel modeling is to establish a relationship between the channel input and the channel output, whose definitions depend on the modeling scope. The components and physical processes relating the channel input to the channel output form the \emph{end-to-end channel} and may include not only the propagation process but also components of the \ac{Tx} and \ac{Rx}. For example, wireless channel models often incorporate the responses of the transmit and receive antennas (e.g., antenna gains or array responses). Channel input and output are typically chosen such that the components within the end-to-end channel can be treated as fixed. This modeling approach is particularly useful because it enables fundamental questions to be posed. For instance, for a given end-to-end channel: \emph{How should information be optimally represented at the \ac{Tx} through encoding and modulation? How should it be optimally recovered at the \ac{Rx} through demodulation and decoding?}

\begin{figure*}
    \centering
\includegraphics[width=1\linewidth]{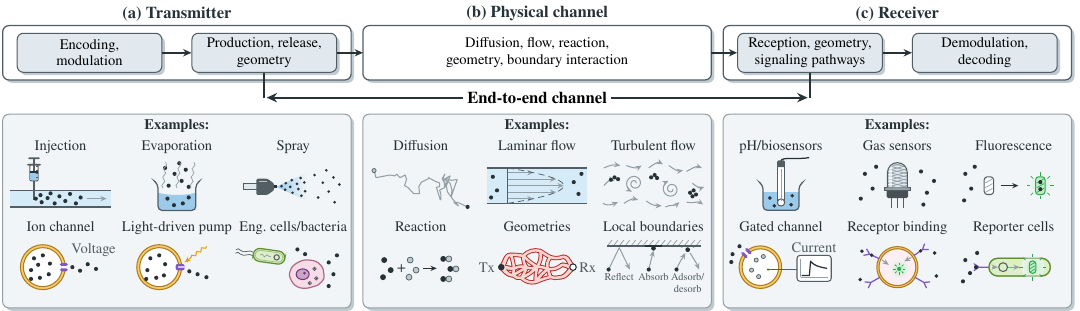}
    \caption{\textbf{End-to-end channel of an \ac{SMC} system.} The end-to-end channel comprises the release at the \ac{Tx}, the propagation in the physical channel, and the reception at the \ac{Rx}.
    \textbf{(a)} Release systems range from macroscale devices, such as syringes, evaporators, and sprayers, to synthetic microscale vesicles with ion channels or light-driven pumps, and to engineered cells.
    \textbf{(b)} Propagation is governed by diffusion, laminar or turbulent flow, and reactions, and is shaped by the geometry and by local boundaries.
    \textbf{(c)} Reception systems range from electrochemical, gas, and fluorescence sensors to synthetic microscale vesicles with ligand-gated channels, and to receptor binding and reporter cells.}
    \label{fig:taxonomy:e2echannel}
    \vspace*{-0.3cm}
\end{figure*}

This concept also applies to \ac{SMC} systems, see Fig.~\ref{fig:taxonomy:e2echannel}. Throughout this tutorial, we adopt the following definition.

\textbf{Definition:} \textit{An end-to-end \ac{SMC} channel relates a specified channel input to a specified channel output and comprises all physical processes and system components between them, which typically includes the molecular release, propagation, and reception processes.}

The modeling scope depends on the application, the objective, and the quantities that can be controlled or observed. In experimental systems, in particular, the choice of channel input and output is strongly influenced by controllability and observability. For instance, the number of ligand-bound receptors may represent the main molecular response of a microscale \ac{Rx} but may not be observable experimentally. Instead, receptor binding may activate a downstream signaling pathway that produces an observable secondary signal. Similarly, molecular release may not be directly controllable at the cellular level and may instead be triggered by an optical stimulus \cite{grebenstein19,tomdieck2026nanodevices}. If the modeling objective is to characterize the relation between this optical stimulus (i.e., channel input) and the resulting fluorescence signal (i.e., channel output), both the optical stimulation at the \ac{Tx} and the downstream signaling and fluorescence detection processes at the \ac{Rx} must be incorporated into the end-to-end channel model. In the following, we introduce several release, propagation, and reception processes in \ac{SMC} systems.

\subsection{Release Process}\label{ssec:taxonomy-release}

The release process describes how the \ac{Tx} source signal is mapped to the spatial, temporal, and molecular characteristics of the \acp{SM} entering the environment. Its role is analogous to that of the transmit antenna in wireless communication systems. Depending on the \ac{Tx} architecture, different physical mechanisms may realize this mapping (see Fig.~\ref{fig:taxonomy:e2echannel}(a)).

\textbf{Membrane-based release:} At the microscale, \acp{SM} can be released across the membrane of natural or artificial cells through passive pores, membrane channels,  active pumps, or vesicle fusion~\cite{Soldner20, lotter22, grebenstein19, tomdieck2026nanodevices}. Moreover, the opening and closing of membrane channels may be controlled by chemical, electrical, mechanical, or optical stimuli~\cite{langton2021engineering,kohata2025synthetic,Schaefer2022}. Depending on the mechanism, the release kinetics may be governed by diffusion, membrane permeability, gating dynamics, or chemical reactions.

\textbf{Injection/pumping:} In macroscale liquid environments, \acp{SM} can be directly injected into the propagation medium using syringe pumps or micropumps \cite{Wicke2021, tuccitto2017fluorescent,farsad2017novel}. The resulting release is typically characterized by the injection flow rate, duration, and \ac{SM} concentration.

\textbf{Airborne release:} In airborne \ac{SMC} systems, \acp{SM} can be released into the air passively, e.g., through evaporation from a liquid source~\cite{giannoukos2018chemical}, or actively, e.g., through spraying~\cite{farsad2013tabletop}. Hence, \acp{SM} may enter the propagation medium directly in the gas phase or initially as liquid droplets or aerosols that subsequently evaporate.

In addition to direct release by the \ac{Tx}, the release process may also occur indirectly through a secondary interaction. Examples include the activation of \acp{SM} in media modulation \cite{Brand2022MediaModulation} and the activation of mobile nanosensors by biomarkers emitted from a target, such as cancer cells~\cite{mosayebi2019cancer}.

\subsection{Propagation Process}
\label{ssec:taxonomy-propagation}

The propagation process describes the transport of \acp{SM} from their release by the \ac{Tx} to their reception by the \ac{Rx}. Analogous to wave propagation in wireless communication systems, which determines the EM field in space and time, the propagation process in \ac{SMC} systems determines the spatio-temporal distribution of \acp{SM}. Depending on the application scenario, the underlying physical transport may be dominated by diffusion, flow, or a combination of both, whereby additional chemical processes may further affect the \acp{SM}. In the following, we discuss representative propagation mechanisms commonly present in \ac{SMC} systems (see Fig.~\ref{fig:taxonomy:e2echannel}(b)).

\textbf{Diffusion-based propagation:} At microscopic scales and in fluidic environments, \acp{SM} are mainly transported by diffusion as a result of their random thermal motion \cite{Jamali2019,lotter22}. This leads to a gradual spreading of the molecular signal over space and time. Intracellular communication and communication between neighboring cells, such as synaptic communication, represent prominent examples of diffusion-based propagation.

\textbf{Advection-assisted propagation:} In larger-scale liquid environments, diffusion alone may be too slow, and \acp{SM} have to be transported by fluid advection in addition to diffusion. Examples of advection-assisted \ac{SMC} include communication in the \ac{CVS}, microfluidic channels, and pipelines \cite{vakilipoor2025cam,Jakumeit2026sorption,jakumeit2026matched,Zoofaghari19cylinder,schaefer21b,Chahibi13}. Depending on the flow profile and geometry, the interplay between advection and diffusion shapes the spatio-temporal concentration profile at the \ac{Rx}.

\textbf{Turbulent airborne propagation:} In airborne environments, \acp{SM} can be transported by air currents over long distances. In nature, airborne \ac{MC} facilitates communication among insects and plants, but it is also relevant to \ac{SMC} systems for interfacing with these biological entities, immersive virtual reality, and purely synthetic applications such as communication among robots \cite{Jamali23,Aktas24}. A key transport phenomenon in airborne systems is turbulent flow, where irregular and multiscale fluid motion leads to time-varying molecular concentration fields, often characterized by intermittent plumes and strong spatio-temporal fluctuations \cite{carde2021navigation,heinlein2026source}.

Besides the underlying transport mechanisms, the propagation process is also influenced by the geometry of the environment, local interactions with boundaries, and potential chemical reactions during propagation~\cite{scherer2025closed,Noel2014a}. Moreover, background molecules from natural or artificial sources may interfere with the received molecular signal and therefore need to be accounted for as noise in the channel model.

\subsection{Reception Process}
\label{ssec:taxonomy-reception}

The reception process describes the interaction of \acp{SM} with the \ac{Rx} and their conversion into an observable output \cite{kuscu19}. Its role is analogous to that of the receive antenna and front-end in wireless communication systems, which convert the incident EM field into an electrical signal. Depending on the \ac{Rx} architecture, different physical, chemical, or biological mechanisms may be employed for sensing and transduction. Representative reception mechanisms relevant to \ac{SMC} systems are discussed below (see Fig.~\ref{fig:taxonomy:e2echannel}(c)).

\textbf{Receptor-based sensing:} In nature, \acp{SM} are often detected through their interaction with receptors on the membrane surface or inside cells \cite{o2010essentials}. Similar ligand-receptor-based sensing mechanisms have also been investigated for engineered cells \cite{manhas2022evolution}. The binding of \acp{SM} to receptors changes their occupancy state and may either directly constitute the \ac{Rx} output or trigger downstream signaling processes that generate an observable signal, e.g., fluorescence.

\textbf{Sensing in liquids:} In macroscale liquid-based \ac{SMC} systems, sensors can detect the presence of \acp{SM} through various physical or chemical interactions without necessarily relying on receptor-binding processes. Prominent examples include electrochemical sensors that convert molecular interactions into electrical signals (e.g., pH sensors) \cite{grebenstein19}, magnetic sensors that detect magnetic nanoparticles used as information carriers (e.g., susceptometers) \cite{Wicke2021}, and optical sensors detecting fluorescence signals \cite{tuccitto2017fluorescent,vakilipoor2025cam,scherer2025closed}.

\textbf{Gas sensing:} In airborne \ac{SMC} systems, \acp{SM} can be detected using established gas-sensing technologies. Examples include \ac{MOS} sensors, whose electrical properties change upon interaction with \acp{SM}, and biosensors, which employ surfaces functionalized with biological detectors (e.g., receptors or peptides) together with suitable transduction mechanisms \cite{pelosi2018gas}. Inspired by natural olfaction, electronic noses are arrays of partially selective sensors (e.g., \ac{MOS} sensors) to recognize different molecular signatures and can serve as sophisticated \acp{Rx} for airborne \ac{SMC} systems \cite{kim2022artificial,qin2023artificial}.

In summary, most reception processes involve indirect or cascaded sensing mechanisms, where, instead of directly measuring the \ac{SM} count or concentration, the interaction of \acp{SM} with the reception site is transduced into a secondary observable signal, e.g., fluorescence, an electrical or magnetic signal. Depending on the targeted channel output and objective, these intermediate processes may need to be included in the end-to-end channel model.

This section introduced the main components of the end-to-end \ac{SMC} channel and exemplary realizations for different application scenarios. The subsequent section presents mathematical descriptions of the underlying physical and chemical phenomena, which build the basis for the different channel modeling approaches discussed Sections~\ref{sec:theoretical-models}-\ref{sec:experimental-models}.

\section{Mathematical Description of SMC Systems}
\label{sec:mathematical-models}
\begin{figure}[!t]
\centering
\includegraphics{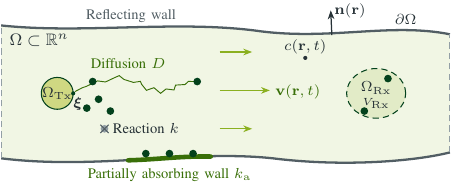}
\caption{\textbf{Spatial domain $\Omega$ of an \ac{SMC} system} with boundary $\partial\Omega$ and quantities used in Sections~\ref{sec:mathematical-models} and~\ref{sec:theoretical-models}.}
\label{fig:mathematical-domain}
\vspace*{-0.3cm}
\end{figure}

The approaches for modeling \ac{SMC} systems discussed in Sections~\ref{sec:theoretical-models}-\ref{sec:experimental-models} all build on the same mathematical descriptions of the release, propagation, and reception processes: \textit{analytical models} solve the governing equations in closed form (see Section~\ref{sec:theoretical-models}), \textit{simulation models} solve them numerically (see Section~\ref{sec:simulation-models}), \textit{data-based models} learn their solutions from data (see Section~\ref{sec:data-based-models}), and \textit{experimental models} realize them physically (see Section~\ref{sec:experimental-models}).
This section summarizes the most important governing equations of \emph{diffusion}, \emph{flow}, and \emph{chemical reactions}, as well as the interactions of \acp{SM} with the \emph{boundaries} of the environment. Throughout, we consider \acp{SM} in the spatial domain $\Omega\subset\mathbb{R}^n$, $n\in\{1,2,3\}$, of an $n$-dimensional \ac{SMC} system, with boundary $\partial\Omega$, see Fig.~\ref{fig:mathematical-domain}. A point in $\Omega$ is denoted by its $n$-dimensional position vector $\mathbf{r}$.

\textbf{Microscopic vs.\ macroscopic description:} Most phenomena in \ac{SMC} systems can be described at two levels. A \emph{microscopic} description captures the \emph{stochastic behavior} of \acp{SM}, i.e., their random positions via \acp{SDE} or, for reactions, the random numbers of \acp{SM} of each species via the \ac{CME}. A \emph{macroscopic} description captures the \emph{deterministic behavior} of the concentration $c(\mathbf{r},t)$, i.e., the average number of \acp{SM} per unit volume at position $\mathbf{r}$ and time $t$, via \acp{PDE} and \acp{ODE}. Both descriptions are connected: The \ac{PDF} $p(\mathbf{r},t)$ of the position $\mathbf{r}$ of a single \ac{SM} obeys the same \ac{PDE} as the concentration, such that $c(\mathbf{r},t)=N\,p(\mathbf{r},t)$ is the expected concentration for $N$ independently moving \acp{SM} \cite{Jamali2019,Gardiner2009}. A macroscopic description only captures this mean, not the fluctuations around it. Hence, a microscopic description is required when few \acp{SM} are involved, e.g., at the receptors of a cell or in synaptic communication \cite{lotter22}, and it yields the statistical signal models discussed in Section~\ref{ssec:theoretical-statistical}.

\subsection{Diffusion}
\label{ssec:mathematical-diffusion}
Diffusion results from Brownian motion, i.e., from the random thermal motion of \acp{SM} and their collisions with other molecules. Diffusion is present in all \ac{SMC} systems and dominates transport at small length scales~\cite{Jamali2019,Farsad2016}.

\textbf{Microscopic description:} The position $\mathbf{r}_i(t)\in\Omega$ of the $i$-th \ac{SM} obeys the \ac{SDE}
\begin{equation}
    \mathrm{d}\mathbf{r}_i(t) = \sqrt{2D}\,\mathrm{d}\mathbf{W}_i(t),
    \label{eq:math-sde-diffusion}
\end{equation}
where $\mathbf{W}_i(t)$ is an $n$-dimensional standard Wiener process and $D$ is the diffusion coefficient in \unit{\square\meter\per\second}~\cite{Gardiner2009}. A constant, scalar diffusion coefficient $D$ is valid for a homogeneous, isotropic, and dilute medium. For time step $\Delta t$, \eqref{eq:math-sde-diffusion} yields the random walk $\mathbf{r}_i(t+\Delta t) = \mathbf{r}_i(t) + \sqrt{2D\Delta t}\,\mathbf{w}_i$ with $\mathbf{w}_i\sim\mathcal{N}(\mathbf{0},\mathbf{I})$, i.e., a zero-mean Gaussian random vector with identity covariance matrix, which underlies particle-based \ac{BD} simulators (see Section~\ref{sec:simulation-models}).
For a spherical particle of radius $a$ in a liquid with dynamic viscosity $\eta$ at temperature $T$, the Stokes-Einstein relation yields $D=k_\mathrm{B}T/(6\pi\eta a)$ with Boltzmann constant $k_\mathrm{B}$. Typical values are $D\approx\qty{e-9}{\square\meter\per\second}$ for small \acp{SM} in water and \qtyrange[range-units=single]{e-10}{e-12}{\square\meter\per\second} for proteins and nanoparticles, whereas for volatile \acp{SM} in air, where the Stokes-Einstein relation does not apply, $D\approx\qty{e-5}{\square\meter\per\second}$ holds. Since the mean squared displacement grows as $2nDt$, a small \ac{SM} in water diffuses over \qty{1}{\micro\meter} in less than a millisecond but needs hours for \qty{1}{\centi\meter}.

\textbf{Macroscopic description:} Combining Fick's first law for the diffusive flux, $\mathbf{J}(\mathbf{r},t)=-D\nabla c(\mathbf{r},t)$, with gradient operator $\nabla$, and mass conservation, $\partial c(\mathbf{r},t)/\partial t = -\nabla\cdot\mathbf{J}(\mathbf{r},t)$, with divergence operator $\nabla\cdot$, yields the \emph{diffusion equation} \cite{crank:1975}
\begin{equation}
    \frac{\partial c(\mathbf{r},t)}{\partial t} = D\,\nabla^2 c(\mathbf{r},t),
    \label{eq:math-diffusion}
\end{equation}
where $\nabla^2=\nabla\cdot\nabla$ denotes the Laplace operator. In heterogeneous media, e.g., crowded cells, tissue, or soil, \eqref{eq:math-diffusion} generalizes to
\begin{equation}
    \frac{\partial c(\mathbf{r},t)}{\partial t} = \nabla\cdot\big(D(\mathbf{r})\,\nabla c(\mathbf{r},t)\big),
    \label{eq:math-diffusion-hetero}
\end{equation}
with diffusion tensor $D(\mathbf{r})$ for anisotropic media. Crowding and binding to obstacles can also cause anomalous diffusion, whose mean squared displacement grows sublinearly in time.

\subsection{Flow}
\label{ssec:mathematical-flow}
Directed transport of \acp{SM}, referred to as \emph{drift} or, if caused by bulk flow of the fluid, \emph{advection}, results either from external forces acting only on the \acp{SM} or from the motion of the fluid.

\textbf{Navier-Stokes equation:} For bulk flow of an incompressible Newtonian fluid, e.g., water, aqueous buffers, or air at low speeds, the velocity field $\mathbf{v}(\mathbf{r},t)$ and pressure $p(\mathbf{r},t)$ obey \cite{Bruus2008}
\begin{subequations}
\label{eq:math-navier-stokes}
\begin{align}
&\varrho\Big(\frac{\partial\mathbf{v}(\mathbf{r},t)}{\partial t} + \big(\mathbf{v}(\mathbf{r},t)\cdot\nabla\big)\mathbf{v}(\mathbf{r},t)\Big) \nonumber\\
&\qquad\qquad = - \nabla p(\mathbf{r},t) + \eta\,\nabla^2\mathbf{v}(\mathbf{r},t)
+ \mathbf{f}_\mathrm{b}(\mathbf{r},t), \label{eq:math-ns-momentum}\\
&\nabla\cdot\mathbf{v}(\mathbf{r},t) = 0, \label{eq:math-ns-continuity}
\end{align}
\end{subequations}
with fluid density $\varrho$, body force density $\mathbf{f}_\mathrm{b}(\mathbf{r},t)$, and $\mathbf{v}(\mathbf{r},t) =\mathbf{0}$ at solid walls (no slip). The Reynolds number $\mathrm{Re}=\varrho\bar{v}\ell/\eta$ for characteristic velocity $\bar{v}$ and length $\ell$, e.g., the mean velocity and the diameter of a tube, compares inertial with viscous forces: For small $\mathrm{Re}$, the flow is \emph{laminar}; it becomes \emph{turbulent} for large $\mathrm{Re}$, e.g., $\mathrm{Re}\gtrsim2000$ for a straight pipe \cite{Bruus2008}. Due to the non-linear term $(\mathbf{v}(\mathbf{r},t)\cdot\nabla)\mathbf{v}(\mathbf{r},t)$, closed-form solutions of \eqref{eq:math-navier-stokes} exist only in simple cases. Most \ac{SMC} studies use one of the following flow models.

\begin{figure}[!t]
    \centering
    \includegraphics[width=\linewidth]{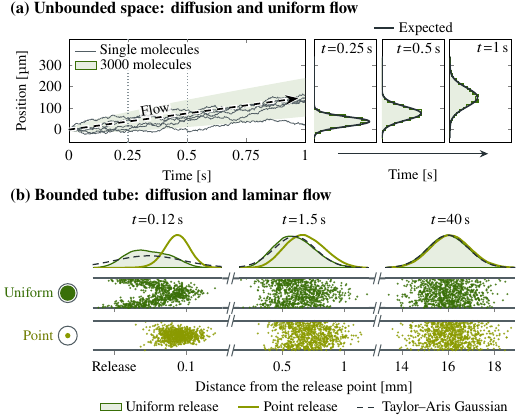}
    \vspace*{-0.2cm}
    \caption{\textbf{Diffusion- and flow-driven \ac{SM} transport.} \textbf{(a)} Unbounded space with uniform flow: \acp{SM} spread over time, biased in the direction of flow. Individual \acp{SM} take random paths.
    \textbf{(b)} Bounded tube with laminar flow: in the Taylor--Aris regime, the longitudinal distribution of \acp{SM} in a tube converges to a Gaussian, irrespective of the release distribution.
    Parameters: $D = \SI{1e-9}{\square\metre\per\second}$; (a) $\bar{v}=\SI{150}{\micro\metre\per\second}$; (b) $R=\SI{50}{\micro\metre}$, $\bar{v}=\SI{0.4}{\milli\metre\per\second}$ ($\mathrm{Pe} = 20$).}
    \label{fig:advection_diffusion_transport_intuition}
    \vspace*{-0.3cm}
\end{figure}

\emph{Uniform flow:} A constant velocity $\mathbf{v}$ trivially solves \eqref{eq:math-navier-stokes} in unbounded domains and is the most widely used flow model in \ac{SMC} \cite{Jamali2019}, see Fig.~\ref{fig:advection_diffusion_transport_intuition}(a).

\emph{Poiseuille flow:} In a straight tube of radius $R$, the non-linear term vanishes, and the resulting steady laminar flow has the parabolic profile $\mathbf{v}(\rho)=2\bar{v}(1-\rho^2/R^2)\,\mathbf{e}_z$ with radial coordinate $\rho$ and unit vector $\mathbf{e}_z$ along the tube axis for any $\mathrm{Re}$ below the turbulent transition~\cite{Jamali2019,Wicke2018,schaefer21b}, see Fig.~\ref{fig:advection_diffusion_transport_intuition}(b).

\emph{Pulsatile flow:} In arteries, the heartbeat drives a periodic flow described by the Womersley solution \cite{Womersley1955}, whose effect on solute transport \cite{Wang2025} was recently studied for \ac{SMC} \cite{Wille2025pulsatile,Symeonidis2026Pulsatile}.

\emph{Turbulent flow:} In airborne \ac{SMC}, flow is often turbulent, e.g., $\mathrm{Re}\approx10^5$ for $\ell\approx\qty{1}{\meter}$ and $\bar{v}\approx\qty{1}{\meter\per\second}$. The mean concentration then approximately obeys an advection-diffusion equation of the form in \eqref{eq:math-advection-diffusion} below, in which the molecular diffusion coefficient $D$ is replaced by an eddy diffusivity that exceeds $D$ by orders of magnitude, whereas the instantaneous concentration is intermittent, i.e., short whiffs are separated by periods without signal \cite{Pope2000,carde2021navigation,heinlein2026source} (see Section~\ref{sec:examples-airborne}).

\textbf{Advection-diffusion equation:} If flow and diffusion are present, the concentration obeys for a given velocity field \cite{Jamali2019}
\begin{equation}
    \frac{\partial c(\mathbf{r},t)}{\partial t} = D\,\nabla^2 c(\mathbf{r},t) - \nabla\cdot\big(\mathbf{v}(\mathbf{r},t)\,c(\mathbf{r},t)\big),
    \label{eq:math-advection-diffusion}
\end{equation}
where the last term equals $\mathbf{v}(\mathbf{r},t)\cdot\nabla c(\mathbf{r},t)$ for incompressible flow, since $\nabla\cdot(\mathbf{v}c)=\mathbf{v}\cdot\nabla c+c\,\nabla\cdot\mathbf{v}$ and $\nabla\cdot\mathbf{v}=0$ by \eqref{eq:math-ns-continuity}. The microscopic description \eqref{eq:math-sde-diffusion} then takes the form
\begin{equation}
    \mathrm{d}\mathbf{r}_i(t) = \mathbf{v}\big(\mathbf{r}_i(t),t\big)\,\mathrm{d}t + \sqrt{2D}\,\mathrm{d}\mathbf{W}_i(t).
    \label{eq:math-sde-advection}
\end{equation}
Both \eqref{eq:math-navier-stokes} and \eqref{eq:math-sde-advection} assume \emph{one-way coupling}, i.e., the \acp{SM} do not alter the flow, such that \eqref{eq:math-navier-stokes} can be solved first. This holds for dilute solutions of small tracers, but not, e.g., if concentrated \ac{SM} solutions change the density or viscosity of the fluid, if droplets do not follow the streamlines, or if injection or spraying perturbs the flow locally \cite{Wicke2021,Gulec21pathogen}. The \emph{P\'eclet number} $\mathrm{Pe}=\bar{v}\ell/D$ with characteristic length $\ell$ compares advection with diffusion \cite{Jamali2019}: For $\mathrm{Pe}\ll1$, \eqref{eq:math-advection-diffusion} reduces to \eqref{eq:math-diffusion}. Fluidic testbeds typically operate at $\mathrm{Pe}\gg1$, e.g., $\mathrm{Pe}\approx10^5$ for $\ell=\qty{1}{\milli\meter}$, $\bar{v}=\qty{1}{\centi\meter\per\second}$, and $D=\qty{e-10}{\square\meter\per\second}$. Even then, diffusion matters in laminar flow, as radial diffusion across the nonuniform velocity profile causes strong axial spreading, known as Taylor-Aris dispersion \cite{Taylor1953,Aris56}, see Fig.~\ref{fig:advection_diffusion_transport_intuition}(b).

\subsection{Chemical Reactions}
\label{ssec:mathematical-reaction}
Chemical reactions may affect the \acp{SM} throughout the end-to-end channel: at the \ac{Tx}, e.g., when \acp{SM} are activated or transported across membranes \cite{Brand2022MediaModulation,tomdieck2026nanodevices}; during propagation, e.g., by (enzymatic) degradation \cite{Noel2014a,scherer2025closed}; and at the \ac{Rx}, e.g., by ligand-receptor binding. We consider $M$ species $S_1,\dots,S_M$ participating in $J$ reactions
\begin{equation}
    \sum_{m=1}^{M}\nu^{-}_{m,j}\,S_m \;\xrightarrow{\;k_j\;}\; \sum_{m=1}^{M}\nu^{+}_{m,j}\,S_m,\quad j=1,\dots,J,
    \label{eq:math-reaction}
\end{equation}
with stoichiometric coefficients $\nu^{-}_{m,j}$ and $\nu^{+}_{m,j}$, rate constants $k_j$, and change vectors $\boldsymbol{\nu}_j$ with elements $\nu_{m,j}=\nu^{+}_{m,j}-\nu^{-}_{m,j}$.

\textbf{Microscopic description:} In a well-mixed volume $V$, the \ac{SM} numbers $\mathbf{x}=[x_1,\dots,x_M]^\mathsf{T}$ form a Markov jump process, in which reaction $j$ occurs within time step $\mathrm{d}t$ with probability $a_j(\mathbf{x})\,\mathrm{d}t$, where the propensity $a_j(\mathbf{x})$ follows from the law of mass action, e.g., $a_j(\mathbf{x})=k_jx_1$ for reaction $S_1\rightarrow\varnothing$ and $a_j(\mathbf{x})=k_jx_1x_2/V$ for $S_1+S_2\rightarrow S_3$. The probability $P(\mathbf{x},t)$ that $x_m$ \acp{SM} of species $S_m$, $m=1,\dots,M$, are present at time $t$ obeys the \ac{CME} \cite{Gillespie1992}
\begin{equation}
    \frac{\partial P(\mathbf{x},t)}{\partial t} = \sum_{j=1}^{J}\Big[a_j(\mathbf{x}-\boldsymbol{\nu}_j)\,P(\mathbf{x}-\boldsymbol{\nu}_j,t) - a_j(\mathbf{x})\,P(\mathbf{x},t)\Big],
    \label{eq:math-cme}
\end{equation}
i.e., there is one linear \ac{ODE} for each possible state $\mathbf{x}$. Since the number of possible states is typically very large, \eqref{eq:math-cme} can be solved in closed form only for simple networks and is otherwise sampled or approximated.

\textbf{Macroscopic description:} For large \ac{SM} numbers and volumes with uniform concentration, the concentrations $c_m=x_m/V$ obey the \acp{RRE} \cite{vanKampen2007}
\begin{equation}
    \frac{\mathrm{d}c_m(t)}{\mathrm{d}t} = \sum_{j=1}^{J}\nu_{m,j}\,r_j\big(\mathbf{c}(t)\big),\quad r_j\big(\mathbf{c}(t)\big) = k_j\prod_{l=1}^{M}c_l^{\,\nu^{-}_{l,j}}(t),
    \label{eq:math-rre}
\end{equation}
where $\mathbf{c}(t)=[c_1(t),\dots,c_M(t)]^\mathsf{T}$ and $r_j(\mathbf{c}(t))$ is the rate of reaction $j$ according to the law of mass action. The \acp{RRE} describe the mean of the \ac{CME} exactly only for zero- and first-order reactions, which lead to linear equations; for higher-order reactions, they are non-linear and deviate from the \ac{CME} for small numbers of \acp{SM}. An important example in \ac{SMC} is first-order degradation $A\rightarrow\varnothing$ with rate constant $k$, for which the concentration $c_A(t)$ of molecule species $A$ decays as $c_A(t)=c_A(0)\,e^{-kt}, t \geq 0$. Degradation matters if \acp{SM} degrade before arriving at an \ac{Rx} at distance $d$, i.e., if $kd^2/D\gtrsim1$ or, for dominant flow: $kd/\bar{v}\gtrsim1$. First-order kinetics with rates identified from measurements can model losses revealed by experiments, e.g., during long-term operation or at the walls of the propagation environment~\cite{scherer2025closed,Jakumeit2026sorption}.

\textbf{Reaction-advection-diffusion equation:} If reactions occur during propagation, the concentration $c_m$ of molecule species $S_m$ obeys
\begin{align}
    \frac{\partial c_m(\mathbf{r},t)}{\partial t} = \nabla\cdot&\big(D_m(\mathbf{r})\nabla c_m(\mathbf{r},t)\big) - \nabla\cdot\big(\mathbf{v}(\mathbf{r},t)\,c_m(\mathbf{r},t)\big)\nonumber\\
    &+ \sum_{j=1}^{J}\nu_{m,j}\,r_j\big(\mathbf{c}(\mathbf{r},t)\big),
    \label{eq:math-rad}
\end{align}
where $D_m(\mathbf{r})$ is its diffusion coefficient. Equation~\eqref{eq:math-rad} is the most general macroscopic propagation model considered in this tutorial and contains \eqref{eq:math-diffusion}, \eqref{eq:math-diffusion-hetero}, \eqref{eq:math-advection-diffusion}, and \eqref{eq:math-rre} as special cases.

\subsection{Boundary Conditions and Model Properties}
\label{ssec:mathematical-further}
To obtain the concentration from the governing equations, the release and boundary conditions of the domain $\Omega$, have to be taken into account (see Fig.~\ref{fig:mathematical-domain}).
The release is accounted for as an initial condition, e.g., $c(\mathbf{r},0)=N\delta(\mathbf{r}-\boldsymbol{\xi})$ for an impulsive release of $N$ \acp{SM} by a point \ac{Tx} at position $\boldsymbol{\xi}$, where $\delta(\cdot)$ denotes the Dirac delta function. For a release over time, the release density $f(\mathbf{r},t)$, i.e., the number of \acp{SM} released per unit volume and time, is added to the right-hand side of \eqref{eq:math-rad}. Background molecules from other sources manifest as a nonzero initial concentration or a constant source and add an offset to the received signal. Using the outward normal vector $\mathbf{n}(\mathbf{r})$ of $\Omega$ (see Fig.~\ref{fig:mathematical-domain}) and the normal derivative $\partial c(\mathbf{r},t)/\partial n=\nabla c(\mathbf{r},t)\cdot\mathbf{n}(\mathbf{r})$, the most relevant linear boundary conditions are introduced below \cite{Jamali2019,carslaw86}. Each boundary condition holds for $t>0$ and all $\mathbf{r}$ on the part of $\partial\Omega$ it applies to, e.g., the \ac{Rx} surface $\partial\Omega_\mathrm{Rx}$. They are stated for walls with no perpendicular flow component, i.e., $\mathbf{v}(\mathbf{r},t)\cdot\mathbf{n}(\mathbf{r})=0$, at which the normal flux of \acp{SM} reduces to $-D\,\partial c(\mathbf{r},t)/\partial n$.

\emph{Reflection (Neumann):} The normal flux vanishes, i.e., $\partial c(\mathbf{r},t)/\partial n=0$, e.g., at inert tube and vessel walls.

\emph{Absorption (Dirichlet):} $c(\mathbf{r},t)=0$, e.g., for fully absorbing \acp{Rx} and perfect sinks.

\emph{Partial absorption (Robin):} \Acp{SM} are absorbed at a rate proportional to the local concentration \cite{CollinsKimball1949},
    \begin{equation}
        D\,\frac{\partial c(\mathbf{r},t)}{\partial n} + k_\mathrm{a}\,c(\mathbf{r},t) = 0,
        \label{eq:math-robin}
    \end{equation}
    with absorption rate constant $k_\mathrm{a}$ in \unit{\meter\per\second}, which yields the reflecting ($k_\mathrm{a}\to0$) and absorbing ($k_\mathrm{a}\to\infty$) cases as limits and can model, e.g., partially absorbing \ac{Rx} surfaces and vessel walls \cite{Lo19laminar}.

    \emph{Reversible adsorption:} With surface concentration $\sigma(\mathbf{r},t)$ in \unit{\per\square\meter} and desorption rate constant $k_\mathrm{d}$ in \unit{\per\second},
    \begin{equation}
        -D\,\frac{\partial c(\mathbf{r},t)}{\partial n} = \frac{\partial\sigma(\mathbf{r},t)}{\partial t} = k_\mathrm{a}\,c(\mathbf{r},t) - k_\mathrm{d}\,\sigma(\mathbf{r},t),
        \label{eq:math-sorption}
    \end{equation}
    e.g., for receptor binding at \ac{Rx} surfaces with abundant receptors and for sorption at vessel walls \cite{Jiang2022}, which was identified as relevant in fluidic testbeds \cite{Jakumeit2026sorption}.

Membranes, e.g., of cell or vesicle \acp{Tx}, are described by the interface condition $-D\,\partial c_\mathrm{in}(\mathbf{r},t)/\partial n=P_\mathrm{m}\big(c_\mathrm{in}(\mathbf{r},t)-c_\mathrm{out}(\mathbf{r},t)\big)$ for $\mathbf{r}$ on the membrane $\partial\Omega_\mathrm{Tx}$, where $c_\mathrm{in}$ and $c_\mathrm{out}$ denote the concentrations inside and outside the \ac{Tx} at the membrane, $\mathbf{n}(\mathbf{r})$ points out of the \ac{Tx}, and $P_\mathrm{m}$ is the membrane permeability in \unit{\meter\per\second} \cite{crank:1975,schaefer20sphere}.

We note that all mathematical descriptions in this section are linear in the concentration if they comprise only diffusion, advection by a given velocity field, zero- and first-order reactions, and linear boundary conditions, and they are time-invariant if all parameters and the domain $\Omega$ are constant. They become non-linear, e.g., for higher-order reactions such as the saturation of a limited number of receptors, for \acp{SM} that alter the velocity field, or for boundaries whose motion depends on the concentration, such as the surface of dissolving drug carriers \cite{Higuchi1961,Lee80release,schafer2022channel}. They become time-variant, e.g., for pulsatile flow \cite{Symeonidis2026Pulsatile,Wille2025pulsatile}, mobile \acp{Tx} and \acp{Rx} \cite{ahmadzadeh2018stochastic}, or pulsating vessel walls. In these cases, closed-form solutions exist only if strong simplifying assumptions are made, and numerical simulations or experiments are usually required to obtain the output signal.

In the following section, we discuss analytical solution methods for the governing equations presented in this section to obtain analytical channel models for \ac{SMC} systems.

\section{Analytical Models for SMC Systems}
\label{sec:theoretical-models}
Analytical channel models express the \ac{CIR} of an \ac{SMC} system in closed form as a function of the physical parameters and directly link the physical environment to a communication-theoretical description. They provide insight, e.g., how the \ac{CIR} scales with the distance between \ac{Tx} and \ac{Rx}, the diffusion coefficient, or the flow velocity, and allow the identification of dominant physical effects. Moreover, they enable fast evaluation of the \ac{CIR} for different system parameters, which is important for system design and performance analysis \cite{Jamali2019,Gohari2016,Farsad2016}. %

\subsection{Green's Function and Channel Impulse Response}
\label{ssec:theoretical-greens}

\begin{figure}[!t]
\centering
\includegraphics{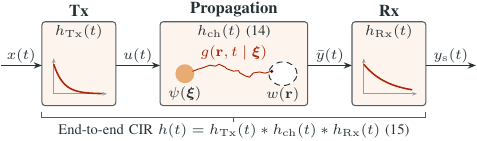}
\caption{\textbf{End-to-end \ac{CIR}} as a cascade of the release kinetics $h_\mathrm{Tx}(t)$, the propagation channel $h_\mathrm{ch}(t)$, which combines the spatial profile $\psi(\boldsymbol{\xi})$ of the \ac{Tx}, the Green's function $g(\mathbf{r}, t\,\vert\,\boldsymbol{\xi})$, and the spatial sensitivity $w(\mathbf{r})$ of the \ac{Rx}, as well as the sensor dynamics $h_\mathrm{Rx}(t)$.}
\label{fig:theoretical-chain}
\vspace*{-0.3cm}
\end{figure}

The central assumption for obtaining \emph{analytical channel models} is linearity. For chemical reactions, linearity restricts the models to zero- and first-order kinetics, e.g., degradation and conversion between molecule species; in the following, we consider first-order degradation with rate constant $k$, whereas higher-order reactions are treated in Section~\ref{ssec:theoretical-special}. If the governing equations are linear and their parameters do not change over time, the channel is \ac{LTI} and fully characterized by its \emph{Green's function} $g(\mathbf{r},t\,\vert\,\boldsymbol{\xi})$, the spatiotemporal analog of the impulse response: \emph{$g(\mathbf{r},t\,\vert\,\boldsymbol{\xi})$ is the concentration at position $\mathbf{r}$ and time $t$ per \ac{SM} released instantaneously at $\mathbf{r}=\boldsymbol{\xi}$ at $t=0$, subject to the boundary conditions of the environment} \cite{Rabenstein23}. Since $c(\mathbf{r},t)=Np(\mathbf{r},t)$, $g(\mathbf{r},t\,\vert\,\boldsymbol{\xi})$ is also the \ac{PDF} of the position of this \ac{SM}.
By superposition, the concentration resulting from an arbitrary release density $f(\boldsymbol{\xi},\tau)$ at the \ac{Tx} follows as (see Fig.~\ref{fig:mathematical-domain})
\begin{equation}
    c(\mathbf{r},t) \! = \!\!\int_{\!0}^{t}\!\!\!\int_{\Omega_\mathrm{Tx}}\!\!\!\!\!\! g(\mathbf{r},t-\tau\,\vert\,\boldsymbol{\xi})\,f(\boldsymbol{\xi},\tau)\,\mathrm{d}\boldsymbol{\xi}\,\mathrm{d}\tau, \,\,\, t\geq 0,\, \mathbf{r}\in\Omega.
    \label{eq:gf1}
\end{equation}
Equation~\eqref{eq:gf1} yields the concentration everywhere in $\Omega$, whereas the \emph{end-to-end channel} maps a channel input to a channel output, e.g., an external stimulus $x(t)$ that triggers the release is mapped to the sensor output signal $y_\mathrm{s}(t)$ of the \ac{Rx} (see Fig.~\ref{fig:theoretical-chain}). Both are connected by the \ac{Tx} and \ac{Rx} models: The \ac{Tx} converts $x(t)$ into the release density $f(\boldsymbol{\xi},t)=\psi(\boldsymbol{\xi})\,u(t)$ with spatial profile $\psi(\boldsymbol{\xi})$ and release rate $u(t)=h_\mathrm{Tx}(t)\ast x(t)$, where $h_\mathrm{Tx}(t)$ describes the release kinetics of the \ac{Tx}.

A transparent \ac{Rx}, i.e., one that does not interact with the \acp{SM}, weights the concentration with its spatial sensitivity $w(\mathbf{r})$, e.g., $w(\mathbf{r})=1$ for a \ac{Rx} that counts the \acp{SM} inside its volume $\Omega_\mathrm{Rx}$ or $w(\mathbf{r})=\delta(\mathbf{r}-\mathbf{r}_\mathrm{Rx})$ for a point \ac{Rx},
and observes the expected signal $\bar{y}(t)=\int_{\Omega_\mathrm{Rx}} w(\mathbf{r})\,c(\mathbf{r},t)\,\mathrm{d}\mathbf{r}=h_\mathrm{ch}(t)\ast u(t)$, where, by \eqref{eq:gf1},
\begin{align}
    &h_\mathrm{ch}(t)\!=\!\int_{\!\Omega_\mathrm{Rx}}\!\int_{\!\Omega_\mathrm{Tx}} \!\!\!\!\!w(\mathbf{r})\,g(\mathbf{r},t\,\vert\,\boldsymbol{\xi})\,\psi(\boldsymbol{\xi})\,\mathrm{d}\boldsymbol{\xi}\,\mathrm{d}\mathbf{r}, &t\geq 0,
    \label{eq:rx-cir}
\end{align}
is the \ac{CIR} of the propagation channel, which thereby contains all spatial effects, such that the remaining \ac{Tx} and \ac{Rx} models act only in time.
A linear sensor with impulse response $h_\mathrm{Rx}(t)$ yields the output $y_\mathrm{s}(t)=h_\mathrm{Rx}(t)\ast\bar{y}(t)$. Hence, $y_\mathrm{s}(t)=h(t)\ast x(t)$ with the end-to-end \ac{CIR}
\begin{equation}
    h(t) = h_\mathrm{Tx}(t)\ast h_\mathrm{ch}(t)\ast h_\mathrm{Rx}(t).
    \label{eq:cascade}
\end{equation}
Fig.~\ref{fig:theoretical-chain} shows the interplay of the individual components of \eqref{eq:cascade}, which are discussed in detail in the following.

\vspace*{-0.2cm}

\subsection{Green's Functions of Unbounded and Bounded Domains}
\label{ssec:theoretical-propagation}
The Green's function $g(\mathbf{r},t\,\vert\,\boldsymbol{\xi})$ in \eqref{eq:gf1} is the solution of the governing equations from Section~\ref{sec:mathematical-models}, i.e., \eqref{eq:math-rad} with zero- and first-order reactions, for the initial condition $c(\mathbf{r},0)=\delta(\mathbf{r}-\boldsymbol{\xi})$ and the boundary conditions of $\Omega$. It can be obtained by standard techniques~\cite{crank:1975,carslaw86,Rabenstein23}. Which technique applies depends mainly on whether the boundaries of the spatial domain $\Omega$ affect the \acp{SM} within the observation time.

\subsubsection{Spatially Unbounded Domains}
\label{ssec:theoretical-unbounded}
In a spatially unbounded, homogeneous medium, $g(\mathbf{r},t\,\vert\,\boldsymbol{\xi})$ can be obtained by spatial Fourier and temporal Laplace transforms, which turn the \ac{PDE} into an algebraic equation. For diffusion, uniform flow with velocity $\mathbf{v}$, and first-order degradation with rate $k$, i.e., \eqref{eq:math-rad} for a single species with constant $D$ and $\mathbf{v}$ and the reaction $A\overset{k}{\rightarrow}\varnothing$, it follows as \cite{Jamali2019,Farsad2016,Noel2014b}
\begin{align}
    g(\mathbf{r},t\,\vert\,\boldsymbol{\xi})=\frac{1}{(4\pi Dt)^{n/2}}\,\exp\!\Big(\!\! -\frac{\lVert\mathbf{r}-\boldsymbol{\xi}-\mathbf{v}t\rVert^2}{4Dt}-kt\Big),\, t\geq 0,
    \label{eq:greens-unbounded}
\end{align}
i.e., a Gaussian centered at $\boldsymbol{\xi}+\mathbf{v}t$, whose width $\sqrt{2Dt}$ grows over time and whose integral decays as $e^{-kt}$. For dimension $n\in\{1,2,3\}$ of $\Omega$, $g(\mathbf{r},t\,\vert\,\boldsymbol{\xi})$ has the unit $\mathrm{m}^{-n}$.

Owing to its closed form, \eqref{eq:greens-unbounded} is used extensively in the \ac{SMC} literature for modulation, detection, and information-theoretic analyses \cite{Jamali2019,Pierobon2010e2e,Pierobon2011noise,Noel2014b,Kadloor2012drift,ahmadzadeh2018stochastic}. It is adequate if the distance $\ell_\mathrm{b}$ of the \ac{Tx} and the \ac{Rx} to the nearest boundary is much larger than the spreading distance $\sqrt{2DT_\mathrm{obs}}$ within the observation time $T_\mathrm{obs}$, which is however not true for many \ac{SMC} application environments, e.g., blood vessels, tissue, microfluidic chips, and pipelines.

\subsubsection{Spatially Bounded Domains}
\label{ssec:theoretical-bounded}
In bounded domains, walls, obstacles, and \ac{Tx} or \ac{Rx} surfaces affect $g(\mathbf{r},t\,\vert\,\boldsymbol{\xi})$ via the boundary conditions specified in Section~\ref{ssec:mathematical-further}, e.g., \eqref{eq:math-robin} and \eqref{eq:math-sorption}, which lead to reflection, partial absorption, or removal of \acp{SM}. For domains of simple shape, $g(\mathbf{r},t\,\vert\,\boldsymbol{\xi})$ follows from a modal expansion \cite{carslaw86,Rabenstein23}, whereas irregular geometries generally require numerical methods. We first consider diffusion and first-order degradation without flow, i.e., the diffusion equation \eqref{eq:math-diffusion} extended by degradation, $\partial c(\mathbf{r},t)/\partial t=D\nabla^2c(\mathbf{r},t)-kc(\mathbf{r},t)$, with reflecting, absorbing, or partially absorbing \eqref{eq:math-robin} walls. For the initial condition $c(\mathbf{r},0)=\delta(\mathbf{r}-\boldsymbol{\xi})$, separation of variables yields \cite{carslaw86,Rabenstein23}
\begin{equation}
    g(\mathbf{r},t\,\vert\,\boldsymbol{\xi})=\sum_{\mu}\frac{\phi_\mu(\mathbf{r})\,\phi_\mu(\boldsymbol{\xi})}{\lVert\phi_\mu\rVert^{2}}\,e^{-(D\lambda_\mu+k)t},\, t\geq 0,
    \label{eq:modal-expansion}
\end{equation}
where $\lVert\phi_\mu\rVert^2=\int_\Omega\phi_\mu^2(\mathbf{r})\,\mathrm{d}\mathbf{r}$. The \emph{modes} $\phi_\mu(\mathbf{r})$, $\mu=1,2,\dots$, are spatial patterns that satisfy the boundary conditions, i.e., eigenfunctions of the Laplace operator, and the eigenvalues $\lambda_1\le\lambda_2\le\dots$ determine how fast the modes vary in space.
The modes $\phi_\mu(\mathbf{r})$ are available in closed form for separable domains, e.g., boxes, cylinders, and spheres \cite{carslaw86,Rabenstein23}.

\begin{figure}[!t]
\centering
\includegraphics{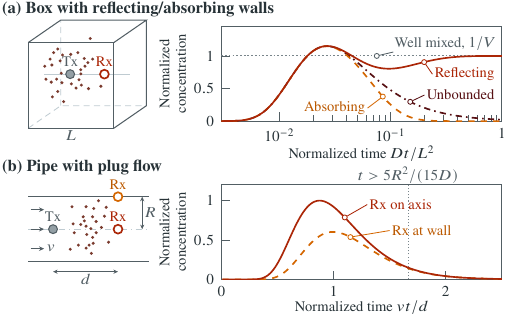}
\caption{\textbf{Green's functions from modal expansions.} Concentration at a point \ac{Rx} per released \ac{SM}: \textbf{(a)} Cube \eqref{eq:greens-box-full} with reflecting and absorbing walls, compared to unbounded space \eqref{eq:greens-unbounded}
normalized to the well-mixed concentration $1/V$; \textbf{(b)} Pipe \eqref{eq:greens-pipe-full} of radius $R=d/2$ with plug flow, $vd/D=20$, and a \ac{Tx} on the axis, for an \ac{Rx} at distance $d$ on the axis and at the wall, respectively, normalized to the peak on the axis.}
\label{fig:theoretical-modal}
\vspace*{-0.3cm}
\end{figure}

\emph{Example 1 (Box with reflecting walls):} For a box $[0,L_x]\times[0,L_y]\times[0,L_z]$ with reflecting walls, the modes $\phi_\mu(\mathbf{r})$ are products of cosines, and \eqref{eq:modal-expansion} factorizes into
\begin{align}
    g(\mathbf{r},t\,\vert\,\boldsymbol{\xi})&=e^{-kt}\prod\nolimits_{i}g_i(r_i,t\,\vert\,\xi_i),\, t\geq 0, \label{eq:greens-box-full}\\
    g_i(r_i,t\,\vert\,\xi_i)&=\frac{1}{L_i}\sum_{\kappa=0}^{\infty}\epsilon_\kappa\cos(q_\kappa r_i)\cos(q_\kappa\xi_i)\,e^{-Dq_\kappa^2t},
    \label{eq:greens-box}
\end{align}
where $r_i$ and $\xi_i$, $i\in\{x,y,z\}$, denote the components of $\mathbf{r}$ and $\boldsymbol{\xi}$, respectively, $q_\kappa=\kappa\pi/L_i$ is the wavenumber of the $\kappa$th mode, i.e., $\lambda_\kappa = q_\kappa^2$, and $\epsilon_0=1$ and $\epsilon_\kappa=2$ for $\kappa\ge1$. The constant mode, i.e., $\kappa=0$ for all coordinates, yields the well-mixed concentration $e^{-kt}/V$ with volume $V=L_xL_yL_z$. For absorbing walls, the modes $\phi_\mu(\mathbf{r})$ in \eqref{eq:greens-box}, i.e., cosines, are replaced by sines and the sum starts at $\kappa=1$. As Fig.~\ref{fig:theoretical-modal}(a) shows, $g(\mathbf{r},t\,\vert\,\boldsymbol{\xi})$ at the \ac{Rx} first follows \eqref{eq:greens-unbounded} until the \acp{SM} reach the walls and then approaches $1/V$ for reflecting (red curve) and zero for absorbing walls (yellow curve).

With flow, i.e., the advection term in \eqref{eq:math-advection-diffusion}, the modes are no longer orthogonal, and instead of \eqref{eq:modal-expansion}, a generalized modal expansion applies \cite{Rabenstein23,schaefer21b}. Nonuniform flow, e.g., laminar flow in tubes, generally requires transfer-function models \cite{schaefer21b}, Taylor--Aris dispersion, or numerical methods, whereas uniform flow along a coordinate in which the domain is unbounded or periodic can be removed by introducing a frame moving with the flow, as the following example shows.

\emph{Example 2 (Pipe with uniform flow):} Consider an infinitely long pipe of radius $R$ with reflecting wall and uniform flow with velocity $v$ along its axis $z$. In the moving frame $z-vt$, the advection term in \eqref{eq:math-advection-diffusion} vanishes. In cylindrical coordinates $(\rho,\theta,z)$ and for a \ac{Tx} on the axis at $\boldsymbol{\xi}=[0,0,\xi_z]^\mathsf{T}$, $g(\mathbf{r},t\,\vert\,\boldsymbol{\xi})$ then factorizes into a radial part, i.e., \eqref{eq:modal-expansion} for a disk, and an axial part as follows \cite{carslaw86,Schaefer19force,Zoofaghari19cylinder}
\begin{align}
    g(\mathbf{r},t\,\vert\,\boldsymbol{\xi})&=g_\mathrm{disk}(\rho,t)\,g_\mathrm{ax}(z,t\,\vert\,\xi_z), \,t\geq 0,\label{eq:greens-pipe-full}\\
    g_\mathrm{disk}(\rho,t)&=\frac{1}{\pi R^2}\sum_{p=0}^{\infty}\frac{J_0(q_p\rho)}{J_0^2(q_pR)}\,e^{-Dq_p^2t},
    \label{eq:greens-pipe}
\end{align}
where $g_\mathrm{ax}(z,t\,\vert\,\xi_z)$ is given by \eqref{eq:greens-unbounded} with $n=1$ along $z$ with velocity $v$ and rate constant $k$, the eigenfunction $J_0(\cdot)$ is the Bessel function of the first kind and order zero, and the wavenumbers $q_p$, i.e., $\lambda_p=q_p^2$, are the solutions of $J_0'(q_pR)=0$.
Fig.~\ref{fig:theoretical-modal}(b) shows the concentration inside the pipe for two different \ac{Rx} locations, where the \ac{Rx} on the axis observes a higher and earlier peak than the \ac{Rx} at the wall, while both signals coincide for $t\gg 1/(Dq_1^2)\approx R^2/(15D)$, see gray dotted vertical line in Fig.~\ref{fig:theoretical-modal}(b).

\subsection{Transmitter and Receiver Models}
\label{ssec:theoretical-txrx}
In the following, we discuss \ac{Tx} and \ac{Rx} models that represent different release proccesses (see Section~\ref{ssec:taxonomy-release}) and connect $g(\mathbf{r},t\,\vert\,\boldsymbol{\xi})$ to the channel input and output (see Fig.~\ref{fig:theoretical-chain}).

\subsubsection{Transmitter Models}
\label{ssec:theoretical-tx}
A \ac{Tx} model specifies the release density $f(\boldsymbol{\xi},\tau)=\psi(\boldsymbol{\xi})\,u(\tau)$ in \eqref{eq:gf1}, i.e., the spatial profile $\psi(\boldsymbol{\xi})$, and the release rate $u(\tau)$. Then, \eqref{eq:gf1} becomes
\begin{equation}
\begin{aligned}
    c(\mathbf{r},t)&=\int_0^t g_\psi(\mathbf{r},t-\tau)\,u(\tau)\,\mathrm{d}\tau, &t\geq 0,\\
    g_\psi(\mathbf{r},t)&=\int_{\Omega_\mathrm{Tx}} g(\mathbf{r},t\,\vert\,\boldsymbol{\xi})\,\psi(\boldsymbol{\xi})\,\mathrm{d}\boldsymbol{\xi}.
\end{aligned}
    \label{eq:tx-conv}
\end{equation}
Common spatial profiles are a point, a sphere~\cite{Noel2016a}, or the cross-section of a tube~\cite{Wicke2021}. The release rate follows from the release mechanism, e.g., impulsive (i.e., $u(\tau)=N\delta(\tau)$), rectangular for an injection of fixed duration~\cite{farsad2017novel}, of first order (i.e., $u(\tau)=Nk_\mathrm{r}e^{-k_\mathrm{r}\tau}$ with rate constant $k_\mathrm{r}$) for membrane-based release or the outward flux from a diffusion-limited carrier~\cite{schafer2022channel}.
If an external stimulus $x(t)$, e.g., light acting on light-driven proton pumps \cite{grebenstein2019biological}, triggers the release linearly, then $u(t)=h_\mathrm{Tx}(t)\ast x(t)$. If the release depends on the concentration outside the \ac{Tx}, \ac{Tx} and channel have to be considered jointly, and if the release mechanism is not known in sufficient detail, $u(t)$ has to be  estimated from measurements \cite{vakilipoor2025cam}.

\begin{figure}[!t]
\centering
\includegraphics{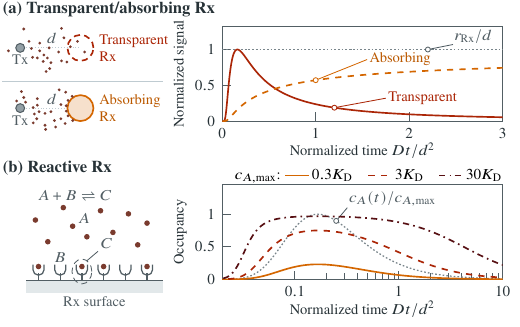}
\vspace*{-0.4cm}
\caption{\textbf{Representative \acp{CIR} obtained from analytical models.} \textbf{(a)}
Fraction of \acp{SM} inside a transparent volume \ac{Rx}, i.e., $w(\mathbf{r})=1$, and absorbed by an absorbing \ac{Rx} in unbounded \ac{3-D} space for \ac{Rx} radius $r_\mathrm{Rx}=0.2d$ at distance $d$, normalized to the maximum value and to $r_\mathrm{Rx}/d$, respectively; \textbf{(b)} Receptor occupancy \eqref{eq:qssa} for the ligand concentration \eqref{eq:greens-unbounded} at a point \ac{Rx}, i.e., $w(\mathbf{r})=\delta(\mathbf{r}-\mathbf{r}_\mathrm{Rx})$, at distance $d$ in unbounded \ac{3-D} space for different peak ligand concentrations $c_{A,\max}\in\{0.3,\,3,\,30\}K_\mathrm{D}$, where $K_\mathrm{D}$ denotes the dissociation constant, compared to the normalized ligand concentration $c_A(t)/c_{A,\max}$ (dotted), on a logarithmic time axis.}
\label{fig:theoretical-cir}
\vspace*{-0.2cm}
\end{figure}

\subsubsection{Receiver Models}
\label{ssec:theoretical-rx}
An \ac{Rx} model maps the concentration to an observable signal.

\textbf{Transparent \acp{Rx}}: A transparent \ac{Rx} observes $\bar{y}(t)=h_\mathrm{ch}(t)\ast u(t)$ with $h_\mathrm{ch}(t)$ given by \eqref{eq:rx-cir};
for $w(\mathbf{r})=1$ inside the \ac{Rx} and zero otherwise, $h_\mathrm{ch}(t)$ is the probability that a released \ac{SM} is inside the \ac{Rx} at time $t$, whereas, e.g., for an optical detection volume, $w(\mathbf{r})$ decays smoothly from its center, since \acp{SM} at the edge contribute less to the measured intensity.

\textbf{Interacting \acp{Rx}}: Absorbing and reactive \acp{Rx} remove or bind \acp{SM} and are therefore boundaries of $\Omega$ with condition \eqref{eq:math-robin} or \eqref{eq:math-sorption}. Their output is the rate at which \acp{SM} are absorbed, $h_\mathrm{ch}(t)=-\oint_{\partial\Omega_\mathrm{Rx}}D\,\partial g_\psi(\mathbf{r},t)/\partial n\,\mathrm{d}A$, i.e., the first-passage time density per released \ac{SM}, where the normal $\mathbf{n}(\mathbf{r})$ points into the \ac{Rx}, or its time integral, i.e., the number of absorbed \acp{SM} per released \ac{SM}. For a point \ac{Tx} at distance $d$ from the center of an absorbing sphere of radius $r_\mathrm{Rx}$ in unbounded space, $h_\mathrm{ch}$(t) follows as \cite{Yilmaz2014absorbing}
\begin{align}
    &h_\mathrm{ch}(t)=\frac{r_\mathrm{Rx}}{d}\,\frac{d-r_\mathrm{Rx}}{\sqrt{4\pi Dt^3}}\,\exp\!\Big(-\frac{(d-r_\mathrm{Rx})^2}{4Dt}\Big), &t\geq 0.
    \label{eq:rx-absorbing}
\end{align}
Hence, the \ac{Rx} model determines the channel output: the \ac{CIR} of a transparent \ac{Rx} rises and decays to zero, whereas for an absorbing \ac{Rx} the fraction of absorbed \acp{SM} accumulates and approaches $r_\mathrm{Rx}/d$ (see Fig.~\ref{fig:theoretical-cir}(a)). Reversible binding to abundant receptors remains linear \cite{deng15}, whereas a finite number of receptors can cause saturation \cite{Ahmadzadeh2016,Lotter21sat}.

\textbf{Signal transduction:} Whereas $w(\mathbf{r})$ determines \emph{which} \acp{SM} the \ac{Rx} observes, the transduction describes \emph{how} the observed quantity $\bar{y}(t)$ is converted into the output signal. Most \acp{Rx} output a transduced signal, e.g., a fluorescence intensity or a sensor resistance, which is often modeled by linear sensor dynamics followed by a static non-linearity,
\begin{equation}
    y_\mathrm{s}(t)=F\big(h_\mathrm{Rx}(t)\ast\bar{y}(t)\big),
    \label{eq:sensor}
\end{equation}
where $h_\mathrm{Rx}(t)$ describes the sensor dynamics, e.g., a first-order low-pass filter with response time $\tau_\mathrm{s}$, and $F(\cdot)$ is the static calibration curve. For example, $F$ saturates for fluorescence \cite{tuccitto2017fluorescent} and follows a power law for \acf{MOS} gas sensors \cite{heinlein:tmbmc2026}. If $F$ is linear, $h_\mathrm{Rx}(t)$ is the last impulse response in \eqref{eq:cascade}.

\subsection{Choosing the Appropriate Level of Detail}
\label{ssec:theoretical-approx}
Analytical models of \ac{SMC} systems can often be simplified by comparing the time scales of the involved processes with each other and with the observation time $T_\mathrm{obs}$. If one process is much faster than another, the corresponding variable or dynamics can be neglected. Table~\ref{tab:approx} groups the most useful approximations, where $d$ denotes the \ac{Tx}-\ac{Rx} distance, $\ell$ is the size of the considered volume, $R$ is the tube radius, $\ell_\mathrm{b}$ is the distance to the nearest wall,
and $K_\mathrm{D}$ and $K_\mathrm{M}$ are respectively the dissociation and Michaelis constants defined below.

\begin{table}[!t]
  \centering
  \footnotesize
  \tablestyleC[O]
  \colorlet{TabGroup}{O3!50!white}
  \setlength{\tabcolsep}{3pt}
  \caption{Approximations of analytical models.}
  \vspace*{-0.2cm}
  \label{tab:approx}
  \begin{tabularx}{\columnwidth}{F>{\raggedright\arraybackslash}X}
  \rowcolor{TabHead}
  \cellcolor{TabHead}\tabhead{Approximation} & \tabhead{Condition} \\
  \rowcolor{TabGroup} \multicolumn{2}{l}{\textcolor{TabHead}{\bfseries\itshape Removes spatial resolution}}\\
  \rowcolor{white}    Well-mixed volume & $\ell^2/D\ll$ all other time scales \\
  \rowcolor{TabZebra} Taylor-Aris & $R^2/D\ll d/\bar{v}$ \\
  \rowcolor{white}    Point \ac{Tx} and \ac{Rx} & \ac{Tx}/\ac{Rx} size $\ll d$ and \ac{Tx}/\ac{Rx} size $\ll\sqrt{2Dt}$ \\
  \rowcolor{TabGroup} \multicolumn{2}{l}{\textcolor{TabHead}{\bfseries\itshape Removes temporal dynamics}}\\
  \rowcolor{white}    Impulsive release & Release duration $\ll$ \ac{CIR} duration \\
  \rowcolor{TabZebra} Dominant impulse response & One impulse response in \eqref{eq:cascade} is much slower \\
  \rowcolor{white}    Quasi-steady state & Fast intermediate chemical reactions \\
  \rowcolor{TabGroup} \multicolumn{2}{l}{\textcolor{TabHead}{\bfseries\itshape Removes boundaries}}\\
  \rowcolor{white}    Unbounded domain & $\ell_\mathrm{b}\gg\sqrt{2DT_\mathrm{obs}}$, no flow towards the wall \\
  \rowcolor{TabGroup} \multicolumn{2}{l}{\textcolor{TabHead}{\bfseries\itshape Removes non-linearity}}\\
  \rowcolor{white}    Pseudo-first order & Abundant reactant or $c(\mathbf{r},t)\ll K_\mathrm{D},K_\mathrm{M}$ \\
  \arrayrulecolor{TabHead}\specialrule{1pt}{0pt}{0pt}
  \end{tabularx}
  \vspace*{-0.4cm}
  \end{table}

\textbf{Well-mixed volume:} If diffusion across a volume $V$ of size $\ell$ is fast compared to all other processes, i.e., if $\ell^2/D$ is much smaller than $1/k$, $T_\mathrm{obs}$, and the exchange time across the surface, then for reflecting walls, the concentrations $c_m(t)$ inside $V$ are spatially uniform and the system of \acp{PDE} becomes a system of \acp{ODE} for $c_m(t)$, which are the \acp{RRE} \eqref{eq:math-rre}. The well-mixed approximation is valid for small volumes, e.g., microscale vesicles \cite{Schaefer2022,tomdieck2026nanodevices}.

\textbf{Taylor-Aris dispersion:} For a tube of radius $R$ with Poiseuille flow (see Section~\ref{ssec:mathematical-flow}), an analytical model requires the solution of the \ac{3-D} advection-diffusion equation \eqref{eq:math-advection-diffusion} for a parabolic velocity profile \cite{schaefer21b}. For $t\gg R^2/D$, radial diffusion causes each \ac{SM} to sample the whole velocity profile, and the cross-sectional mean concentration $\bar{c}(z,t)$ obeys \cite{Aris56}
\begin{equation}
    \frac{\partial\bar{c}(z,t)}{\partial t} = D_\mathrm{eff}\,\frac{\partial^2\bar{c}(z,t)}{\partial z^2} - \bar{v}\,\frac{\partial\bar{c}(z,t)}{\partial z},
    \label{eq:taylor-aris}
\end{equation}
with the effective diffusion coefficient $D_\mathrm{eff}=D+\bar{v}^2R^2/(48D)$ \cite{Taylor1953,Aris56}. The corresponding Green's function is \eqref{eq:greens-unbounded} for $n=1$ with $D$ replaced by $D_\mathrm{eff}$, $v=\bar{v}$, and $k=0$. For a release over the entire cross-section at $z=0$ and a transparent \ac{Rx} of axial length $\ell_\mathrm{Rx}\ll\sqrt{2D_\mathrm{eff}d/\bar{v}}$ that covers the entire cross-section at distance $d$, \eqref{eq:rx-cir} yields \ac{CIR}
\begin{align}
    &h_\mathrm{ch}(t)\approx\frac{\ell_\mathrm{Rx}}{\sqrt{4\pi D_\mathrm{eff}t}}\,\exp\!\Big(-\frac{(d-\bar{v}t)^2}{4D_\mathrm{eff}t}\Big), &t\geq 0.
    \label{eq:taylor-aris-cir}
\end{align}
See Fig.~\ref{fig:advection_diffusion_transport_intuition}(b) for an illustration of Taylor--Aris dispersion.

\textbf{Quasi-steady state:} Reaction networks often contain fast intermediate steps, e.g., the binding of ligands $A$ to receptors $B$ at an \ac{Rx}, $A+B\underset{k_\mathrm{b}}{\overset{k_\mathrm{f}}{\rightleftharpoons}}C$, whose \ac{RRE} \eqref{eq:math-rre} is $\mathrm{d}c_C(t)/\mathrm{d}t=k_\mathrm{f}c_A(t)c_B(t)-k_\mathrm{b}c_C(t)$. If binding equilibrates much faster than the ligand concentration $c_A(t)$ changes and does not noticeably deplete the ligand, the time derivative of $c_C(t)$ can be set to zero in the \ac{RRE} \cite{Segel1989}. With the conservation of receptors, $c_B(t)+c_C(t)=c_{B,\mathrm{tot}}$, where $c_{B,\mathrm{tot}}$ denotes the total receptor concentration, the differential equation simplifies~to
\begin{equation}
    c_C(t)=c_{B,\mathrm{tot}}\,\frac{c_A(t)}{K_\mathrm{D}+c_A(t)},\quad K_\mathrm{D}=\frac{k_\mathrm{b}}{k_\mathrm{f}},
    \label{eq:qssa}
\end{equation}
with dissociation constant $K_\mathrm{D}$.
Hence, the binding dynamics reduce to a static non-linearity: $c_C(t)\approx(c_{B,\mathrm{tot}}/K_\mathrm{D})\,c_A(t)$ is linear for $c_A(t)\ll K_\mathrm{D}$ and saturates to $c_C(t) = c_{B,\mathrm{tot}}$ for $c_A(t)\gg K_\mathrm{D}$.
Fig.~\ref{fig:theoretical-cir}(b) illustrates this for the ligand concentration $c_A(t)$ at a point \ac{Rx} at distance $d$ in unbounded \ac{3-D} space, i.e., \eqref{eq:greens-unbounded} for $n=3$: a 100-fold increase of the peak ligand concentration $c_{A,\max}$ from $0.3K_\mathrm{D}$ to $30K_\mathrm{D}$ raises the peak occupancy only about fourfold, and the response broadens into a plateau.

\subsection{Time-Variant and Non-linear Channel Models}
\label{ssec:theoretical-special}
Obtaining a Green's function requires a linear and time-invariant system. Otherwise, the received signal depends on the release time or on the transmitted signal itself, and a single \ac{CIR} no longer characterizes the channel. In such cases \textit{time-variant} and \textit{non-linear models} have to be employed.

\textbf{Deterministic time variance:} If parameters change over time, $g(\mathbf{r},t\,\vert\,\boldsymbol{\xi})$ depends on both the release time $\tau$ and observation time $t$, i.e., $g(\mathbf{r},t-\tau\,\vert\,\boldsymbol{\xi})$ in \eqref{eq:gf1} becomes $g(\mathbf{r},t\,\vert\,\boldsymbol{\xi},\tau)$. For spatially uniform time-dependent $D(t)$, $\mathbf{v}(t)$, and $k(t)$, a moving frame and a transformed time reduce the governing equation to the time-invariant one, i.e., \eqref{eq:greens-unbounded} holds with $\mathbf{v}t$, $Dt$, and $kt$ replaced by the integrals of $\mathbf{v}(t)$, $D(t)$, and $k(t)$ from $\tau$ to $t$ \cite{crank:1975,Symeonidis2026Pulsatile}.
For pulsatile flow, the \ac{CIR} thus depends on the phase of the release, i.e., the channel is cyclostationary (see Section~\ref{sec:examples-cvs}) \cite{Symeonidis2026Pulsatile}. In general, discrete-time modal or \ac{ODE} models can be extended to time-variant parameters \cite{Rabenstein23,Schaefer2022}.

\textbf{Random time variance:} If the time variance is itself random, the channel remains linear, but $g(\mathbf{r},t\,\vert\,\boldsymbol{\xi},\tau)$ becomes a random process, analogous to doubly selective fading in wireless communications, e.g., for diffusing mobile \acp{Tx} and \acp{Rx} \cite{ahmadzadeh2018stochastic} and in turbulent airborne channels (see Section~\ref{sec:airborne:turbulent}).

\textbf{Non-linear systems:} For non-linear effects, e.g., receptor saturation, a Green's function does not exist. Weak non-linearities can be linearized around an operating point, e.g., by pseudo-first-order kinetics. Block-structured models combine \ac{LTI} systems with static non-linearities \cite{Schoukens2017blockoriented}, e.g., a Wiener model, in which a linear channel is followed by a non-linearity (e.g., saturating receptor binding \eqref{eq:qssa} or the sensor \eqref{eq:sensor}), or a Hammerstein model with the reverse order. Further options are Volterra series \cite{Boyd1985}, step-by-step integration in discrete time \cite{Schaefer2022}, and treating the deviation from a linear model as noise \cite{farsad:channel_noise_models_nonlinearMC}. Strongly non-linear systems and moving boundaries, e.g., of dissolving drug carriers, admit only approximate analytical models \cite{Higuchi1961,Lee80release}, and otherwise require numerical or data-based models (see Sections~\ref{sec:simulation-models} and~\ref{sec:data-based-models}).

\subsection{Stochastic Models of the Received Signal}
\label{ssec:theoretical-statistical}
\label{ssec:theoretical-cme}
The analytical models discussed above only describe the mean concentration.
If only a few \acp{SM} are involved, e.g., at the receptors of a cell or inside a nanoscale \ac{Tx}, the random movements and reactions of individual \acp{SM} cause fluctuations whose relative magnitude scales as $1/\sqrt{\bar{N}}$ for an expected number $\bar{N}$ of \acp{SM} and which can dominate the noise component of the received signal \cite{pierobon11a,lotter22}. Moreover, for bimolecular reactions, the mean of the \ac{CME} \eqref{eq:math-cme} generally deviates from the \acp{RRE} \eqref{eq:math-rre} \cite{schnoerr17}. The most important analytical stochastic models in the literature are summarized~below.

\textbf{Counting statistics:} If $N$ \acp{SM} are released at once from a point source and move independently, each of them is observed at time $t$ with probability $p_\mathrm{obs}(t)$, i.e., $p_\mathrm{obs}(t)=h_\mathrm{ch}(t)$ for a transparent and $p_\mathrm{obs}(t)=\int_0^t h_\mathrm{ch}(\tau)\,\mathrm{d}\tau$ for an absorbing \ac{Rx}. Hence, the number of observed \acp{SM} is binomially distributed, $\mathcal{B}(N,p_\mathrm{obs})$, where $\mathcal{B}(N,p)$ denotes a binomial distribution with $N$ trials and success probability $p$. The binomial distribution can be accurately approximated by a Poisson distribution for large $N$ and small $p_\mathrm{obs}$ and by a Gaussian distribution with mean $Np_\mathrm{obs}$ and variance $Np_\mathrm{obs}(1-p_\mathrm{obs})$ if the mean is large \cite{Jamali2019}.
Likewise, the number of bound receptors among $N_\mathrm{R}$ independent receptors exposed to a deterministic ligand concentration is binomially distributed, with the occupancy from \eqref{eq:qssa} as success probability in steady state.
\textit{Binomial} and \textit{Poisson models} are exact solutions of the \ac{CME} for monomolecular reactions \cite{Jahnke2007CME}, whereas the \textit{Gaussian model} corresponds to the linear noise approximation \cite{vanKampen2007}, which holds for large numbers of \acp{SM} also in non-linear reaction networks; further approximations and numerical methods, e.g., the \ac{SSA}, are reviewed in \cite{schnoerr17} and Section~\ref{sec:simulation-models}, and reaction networks inside cellular \acp{Rx} can also be exploited for detection \cite{Heinlein25}.

For \ac{OOK}, the \ac{Tx} sends bit $a_j\in\{0,1\}$ in symbol interval $j$ of duration $T_\mathrm{sym}$ by releasing $N$ or zero \acp{SM}, and let the \ac{Rx} take one sample per interval at time $jT_\mathrm{sym}+t_0$. Then, \acp{SM} of the previous $I$ symbols cause \ac{ISI}, and background molecules add noise with mean $\lambda_\mathrm{bg}$ \cite{Jamali2019}. For the Poisson approximation, the received sample $y[j]$, i.e., the number of \acp{SM} observed at time $jT_\mathrm{sym}+t_0$, is
\begin{equation}
y[j]\sim\mathcal{P}\Big(\lambda_\mathrm{bg}+N\sum_{l=0}^{I}a_{j-l}\,p_l\Big),
    \label{eq:discrete-channel}
\end{equation}
where $\mathcal{P}(\mu)$ denotes a Poisson distribution with mean $\mu$, $\lambda_\mathrm{bg}+N\sum_{l=0}^{I}a_{j-l}\,p_l\triangleq \bar{y}(jT_\mathrm{sym}+t_0)$ is the mean of $y[j]$, and $p_l=p_\mathrm{obs}(lT_\mathrm{sym}+t_0)$ for a transparent \ac{Rx}.

\textbf{Random arrival times:} If information is encoded in the release time of \acp{SM} instead of their number, the arrival time of each \ac{SM} at an absorbing \ac{Rx} is random with first-passage time density $h_\mathrm{ch}(t)$. In a \ac{1-D} system with flow $v>0$ toward an \ac{Rx} at distance $d$, this is the inverse Gaussian density with mean $d/v$ and variance $2Dd/v^3$, which yields the additive inverse Gaussian noise channel for timing modulation \cite{Srinivas2012IG,Kadloor2012drift}.

\textbf{Extrinsic noise and turbulence:} The models discussed so far capture the intrinsic noise caused by random \ac{SM} movements and reactions, including \ac{ISI} and background molecules in \eqref{eq:discrete-channel}, but assume that all model parameters are fixed and known. In experiments, nominally identical cells, vesicles, or bacterial cultures additionally differ, e.g., in size or expression level. This extrinsic noise is often larger than the intrinsic noise and can be modeled by random model parameters \cite{tomdieck2026nanodevices}. In turbulent airborne channels, the received concentration is intermittent and is described by the statistics of whiffs and blanks instead \cite{celani:odor_landscapes_turbulent_environments,heinlein2026source}.

\section{Simulation Methods for SMC Systems}
\label{sec:simulation-models}
Simulation methods provide a general computational machinery for solving the governing equations introduced in Section~\ref{sec:mathematical-models}.
They are indispensable whenever a closed-form analytical model for the underlying \ac{SMC} system cannot be obtained, e.g., for complex or irregular geometries, non-linear or many-species reaction networks, small \ac{SM} counts where stochastic fluctuations dominate, or turbulent transport;
but they are also routinely required even when a closed-form analytical model \textit{is} available, e.g., to solve a transcendental eigenvalue equation, invert a transform, or truncate a series (see Section~\ref{sec:theoretical-models}).

\begin{figure*}
    \centering
    \includegraphics[width=\linewidth]{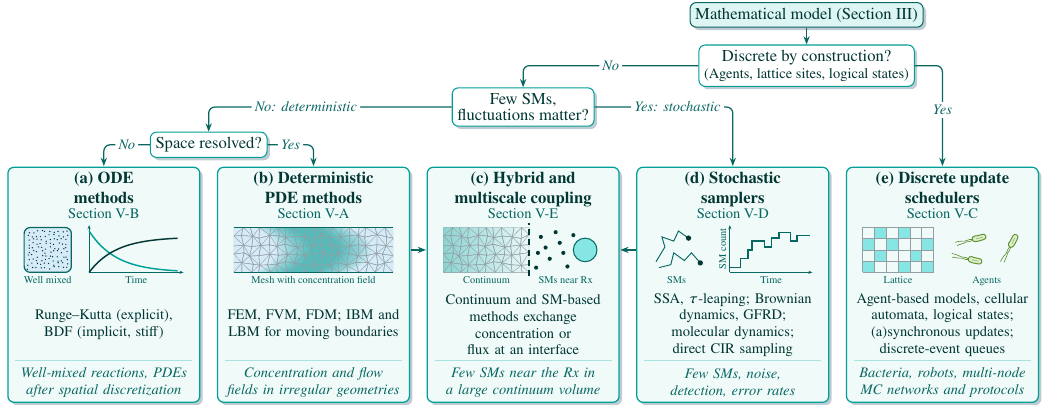}
    \caption{\textbf{Choice of a simulation method.}
    Depending on which properties of the underlying \ac{SMC} system should be represented by a simulation model, an appropriate simulation method has to be chosen.
    Common \ac{SMC} simulation methods can be categorized into five families.}
    \label{fig:simulation-overview}
    \vspace*{-0.3cm}
\end{figure*}

Which simulation method is appropriate depends on which properties of the underlying \ac{SMC} system should be captured by the simulation:
First, are \acp{SM} or agents represented as a continuum or as discrete quantities?
Second, is only the deterministic mean behavior of the \ac{SMC} system relevant or must stochastic dynamics be included?
Third, should space be resolved explicitly or does only the temporal evolution matter?
Based on these considerations, the wide range of simulation methods used in the \ac{MC} literature can be categorized into five families.
In this section, we give an overview of the different families of simulation methods while concurrently outlining how to choose a method.
A flow chart guiding through the method selection process is given in Fig.\ \ref{fig:simulation-overview}, while Table~\ref{tab:solvers} summarizes the properties of the most important methods.

\subsection{Deterministic PDE Methods}\label{sec:deterministic_PDE_solvers}

Simulation methods in this family are used to solve \acp{PDE} that govern continuous fields that depend on space and time, see Fig.\ \ref{fig:simulation-overview}(b).
In \ac{MC}, e.g., the reaction--advection--diffusion equation in~\eqref{eq:math-rad} for a \ac{SM} concentration or the Navier--Stokes equations in~\eqref{eq:math-navier-stokes} for a velocity field can be solved using such methods.
Here, the selection of a simulation method requires two decisions:
how the field is discretized in space (e.g., using a Eularian mesh, grid, lattice, or Lagrangian markers), which is determined from the geometry and which quantity must be conserved exactly;
and how the solution is advanced in time, by explicit or implicit stepping, or by a fixed step tied to the spatial discretization itself.
Below, the most common deterministic methods relevant to \ac{MC} are summarized.

The \textit{\ac{FEM}}, \textit{\ac{FVM}}, and \textit{\ac{FDM}} discretize the domain's interior directly and differ in how that interior is represented and which quantity is conserved exactly~\cite{Quarteroni2017}.
The \ac{FEM} offers flexible meshing and is thus commonly used for fluidic \ac{SMC} systems due to their irregular vessel and microfluidic geometries and mixed boundary conditions, see, e.g., \cite{Jakumeit2026MIGHT}.
The \ac{FVM} is preferred when advective flux must be conserved exactly~\cite{Quarteroni2017}.
The \ac{FDM} is computationally cheaper than the \ac{FEM} since it relies on a less flexible mesh comprised of rectangular grids; however, it suits only simple geometries that can be meshed using rectangular grids with reasonable accuracy~\cite{Quarteroni2017}.

The \textit{\ac{IBM}} and the \textit{\ac{LBM}} are suitable when the boundary itself evolves over time, e.g., in simulations of a compliant vessel wall or deforming \acp{RBC}~\cite{Aidun2010}.
The \ac{IBM} couples deformable, marker-represented boundaries to a fixed background fluid grid~\cite{Peskin_2002}.
In contrast, the \ac{LBM} keeps a fixed lattice and marks each site as fluid or solid, so complex or changing geometries need no re-meshing~\cite{Aidun2010}.

A shared limitation applies to the whole family of deterministic \ac{PDE} simulation methods: fluctuations are absent by construction, so noise, detection, and \ac{BER} studies cannot rely on these simulation methods alone.

\subsection{ODE Methods}\label{sec:ODE_and_algebraic_solvers}

An \ac{ODE} description of an \ac{SMC} system arises once the description of a continuous field (e.g., \ac{SM} concentration) is reduced to a single remaining independent variable, leaving only \textit{ordinary} derivatives, see Fig.\ \ref{fig:simulation-overview}(a).
This occurs, e.g., when space is abstracted away in a well-mixed system (see Section \ref{ssec:theoretical-approx}) or has already been discretized, leaving time as the sole independent variable.
\ac{ODE} formulations can also arise for steady-state spatial problems (see Section \ref{ssec:theoretical-approx}); however, we focus here on time-dependent \acp{ODE}.

For time-dependent \acp{ODE}, the choice of simulation method is largely determined by the \textit{stiffness} of the system.
A system is stiff when it exhibits processes evolving on widely separated time scales, e.g., a fast reaction occurring alongside a much slower one~\cite{Hundsdorfer2003}.
In this case, \textit{implicit} methods are generally preferable.
Implicit methods determine the future state of a system by solving an equation that depends on both the current and the unknown future state, allowing comparatively large time steps even in the presence of fast dynamics.
A prominent example is the \textit{\ac{BDF}} method~\cite{Hundsdorfer2003}.

For non-stiff systems, \textit{explicit} methods are typically more suitable.
They compute the future state directly from known quantities and are generally less computationally expensive than implicit methods.
However, their time step must be sufficiently small to resolve the fastest process in the system.
Consequently, explicit methods such as \textit{Runge--Kutta} can become prohibitively slow for stiff systems~\cite{Hundsdorfer2003}.

\subsection{Discrete Update Schedulers}\label{sec:discrete_update_schedulers}

Discrete update schedulers apply to models that are discrete by construction, rather than derived from a continuous equation, see Fig.\ \ref{fig:simulation-overview}(e). In \ac{SMC}, these models describe systems that are characterized by rules for individual entities rather than by a governing equation, e.g., a microbial population communicating via \acp{SM}~\cite{almpanis2019agent} or the gene regulation inside a biological \ac{Tx} or \ac{Rx}. The objective of simulating them is to predict the system-level behavior that emerges from these rules. One such approach is \ac{ABM}, where each entity, e.g., a bacterium or a robot, is simulated as a separate object with its own position and behavior, revealing how collective behavior emerges from individual actions~\cite{bonabeau2002agent, macal2009agent}. Another approach uses lattice-based models, which divide space into a fixed grid of discrete positions, called sites, each defining a local state, e.g., occupied or empty, and simulate how spatial patterns arise from local rules~\cite{Stephen_1983_Statistical}. Finally, logical-state models assign each entity of a biological regulatory network, e.g., a gene or a protein, one of a few discrete states, e.g., active or inactive, and reveal the stable activity patterns the network settles into~\cite{Stuart_1969_Metabolic}. However, a key challenge in simulating these models is the order in which the states of the entities or sites are updated, since this order can change the outcome of the simulation.

In a \textit{synchronous update}, the system advances in discrete time steps, and the new state of every entity or site is computed based on the system's state at the previous time step. All the states or sites update together at once. This can create artifacts that are not part of the real system's behavior, e.g., a site's state oscillating at every time step instead of settling, producing a checkerboard pattern caused only by the update rule~\cite{Birgitt_1999_Synchronous}.
\textit{Asynchronous} or \textit{random-order updates} avoid this by updating entities or sites one at a time in random order, but lose a well-defined global time step~\cite{Birgitt_1999_Synchronous}. Discrete-event queues instead suit networks of several communicating devices, e.g., \acp{Tx}, relays, and \acp{Rx}. Rather than tracking individual \acp{SM}, each transmission between two devices is treated as a single scheduled event, since analyses of throughput, delay, and protocol behavior across a network are affected by the properties of \acp{SM} transmission and reception, but not directly by the \acp{SM}' trajectories~\cite{Ertan_2010_NanoNS, Yubing_2017_nanoNS3}.

 \begin{table*}[!t]
  \caption{Overview of simulation methods, grouped by family, with their spatial and temporal treatment, the problems they apply to, and representative software tools.}
  \label{tab:solvers}
  \centering\scriptsize
  \tablestyleC[T]
  \colorlet{TabGroup}{T3!50!white}
  \setlength{\tabcolsep}{3pt}
  \renewcommand{\tabularxcolumn}[1]{m{#1}}%
  \begin{tabularx}{\textwidth}{>{\cellcolor{TabFirst}\raggedright\arraybackslash}m{2.6cm}
    >{\centering\arraybackslash}m{2.6cm}
    >{\centering\arraybackslash}m{2.4cm}
    >{\centering\arraybackslash}X
    >{\centering\arraybackslash}m{3.0cm}}
  \rowcolor{TabHead}
  \cellcolor{TabHead}\tabhead{Method} & \tabhead{Spatial treatment} & \tabhead{Temporal treatment} & \tabhead{Applicable to} & \tabhead{Tools} \\
  \rowcolor{TabGroup} \multicolumn{5}{l}{\textcolor{TabHead}{\bfseries\itshape A. Deterministic \ac{PDE} methods (Section~\ref{sec:deterministic_PDE_solvers})}}\\
  \rowcolor{white}     \acs{FEM}~\cite{Quarteroni2017, johnson:1995, Axelsson:2001} & Unstructured mesh & Fixed or adaptive step & Reaction--advection--diffusion, Navier--Stokes; irregular geometries, mixed boundary conditions~\cite{Kuscu2018, Bi2020, Jakumeit2026MIGHT}
  & COMSOL Multiphysics, FEniCSx, deal.II \\
  \rowcolor{TabZebra}  \acs{FDM}~\cite{Quarteroni2017} & Structured grid & Fixed or adaptive step & Reaction--advection--diffusion, simple geometries~\cite{Jamali2018reactive} & Custom code (MATLAB, Python) \\
  \rowcolor{white}     \acs{FVM}~\cite{Quarteroni2017} & Control volumes & Fixed or adaptive step & Advection-dominated transport, Navier--Stokes, e.g., airborne \acf{CFD}~\cite{Gulec2023airborne} & OpenFOAM, Ansys Fluent \\
  \rowcolor{TabZebra}  \acs{IBM}~\cite{Peskin_2002} & Moving markers on a fixed grid & Fixed step & Flow with deformable boundaries, e.g., vessel walls, flagella~\cite{Hou2025IBLBM} & IBAMR, IB2d \\
  \rowcolor{white}     \acs{LBM}~\cite{LBM_Ref} & Regular lattice & Fixed step, set by the lattice & Navier--Stokes and advection--diffusion in complex geometries~\cite{Chopard2009} & Palabos, waLBerla, OpenLB
  \\
  \rowcolor{TabGroup} \multicolumn{5}{l}{\textcolor{TabHead}{\bfseries\itshape B. \Ac{ODE} methods (Section~\ref{sec:ODE_and_algebraic_solvers})}}\\
  \rowcolor{white}     Explicit Runge-Kutta~\cite{Hundsdorfer2003, butcher1996history} / implicit \acs{BDF}~\cite{Hundsdorfer2003} & \na & Adaptive step & \acs{RRE}; reaction--diffusion after spatial discretization~\cite{chou15} & MATLAB
  \texttt{ode45}/\texttt{ode15s}, SciPy \texttt{solve\_ivp}, SUNDIALS CVODE, DifferentialEquations.jl \\
  \rowcolor{TabGroup} \multicolumn{5}{l}{\textcolor{TabHead}{\bfseries\itshape C. Discrete update schedulers (Section~\ref{sec:discrete_update_schedulers})}}\\
  \rowcolor{white}     Synchronous / asynchronous update~\cite{Birgitt_1999_Synchronous, Stephen_1983_Statistical} & Lattice or continuum & Discrete time steps & \acs{CA}, \acs{ABM}~\cite{almpanis2019agent, Nagarajan2022} & NetLogo, Mesa, BSim, PhysiCell; BoolNet, MaBoSS \\
  \rowcolor{TabZebra}  Discrete-event queue~\cite{Yubing_2017_nanoNS3, Ertan_2010_NanoNS} & \na & Event-driven & Multi-node \ac{MC} networks; protocols, throughput, delay~\cite{femminella21} & ns-3 (nanoNS3), OMNeT++ \\
  \rowcolor{TabGroup} \multicolumn{5}{l}{\textcolor{TabHead}{\bfseries\itshape D. Stochastic samplers (Section~\ref{sec:stochastic_samplers})}}\\
  \rowcolor{white}     \acs{SSA}~\cite{Daniel_1977_Exact, Michael_2000_Efficient} & \na & Event-driven or fixed step & \acs{CME}, \acs{RDME}~\cite{noel2018algorithm, noel2017simulating, Egan22} & GillesPy2, COPASI, URDME, MesoRD \\
  \rowcolor{TabZebra}  \acs{BD}~\cite{Steven_2004_Stochastic} & Continuous positions & Fixed step & Particle diffusion and reactions, including drift and active motion~\cite{andrews09, deng15} &
  Smoldyn, MCell, ReaDDy, AcCoRD, MUCIN \\
  \rowcolor{white}     Direct \ac{CIR} sampling~\cite{Jamali2019} & \na & Per symbol & Link-level detection and error-rate studies~\cite{Vakilipoor2022lowcomplexity} & Custom code (MATLAB, Python) \\
  \rowcolor{TabGroup} \multicolumn{5}{l}{\textcolor{TabHead}{\bfseries\itshape E. Hybrid and multiscale models (Section~\ref{sec:hybrid_and_multiscale_coupling})}}\\
  \rowcolor{white}     Continuum--particle hybrid~\cite{Adam_2015_Multi, Mark_2012_Two, Benjamin_2013_Multiscale} & Mesh in the bulk, particles near the \ac{Rx} & Synchronized at the interface & Continuum bulk coupled to a particle-resolved
  \ac{Rx} region~\cite{Flegg2015convergence} & AcCoRD \\
  \arrayrulecolor{TabHead}\specialrule{1pt}{0pt}{0pt}
  \end{tabularx}
  \vspace*{-0.3cm}
  \end{table*}

\subsection{Stochastic Samplers}\label{sec:stochastic_samplers}

Stochastic samplers are needed whenever the randomness of individual \acp{SM} affects the quantity of interest, which in \ac{SMC} is typically the case when only few \acp{SM} are involved, see Fig.\ \ref{fig:simulation-overview}(d). This randomness mainly stems from Brownian motion as well as from reaction and binding events~\cite{Pierobon2011noise, pierobon11a}. Capturing the resulting fluctuations of the received signal is essential for noise, detection, and \ac{BER} studies, for which the deterministic methods in Sections~\ref{sec:deterministic_PDE_solvers} and~\ref{sec:ODE_and_algebraic_solvers} cannot be used alone.
Selecting a stochastic sampler involves two design choices. The first is whether only the number of \acp{SM} of each species is tracked or also the position of each \ac{SM}. Tracking only the numbers suffices for a well-mixed system, whereas positions are needed when the spatial distribution of the \acp{SM} matters, e.g., near the \ac{Rx}. The second design choice is whether the sampling is exact or approximate. Exact samplers generate random outcomes with exactly the probabilities predicted by the underlying stochastic model, whereas approximate samplers advance time in finite steps and thereby introduce an error, which decreases as the time step gets smaller, in exchange for a faster simulation. In the following, we first discuss samplers that track only \ac{SM} counts, then samplers that track \ac{SM} positions, and finally direct sampling of the \ac{Rx} signal from a known \ac{CIR}.

For a well-mixed system, i.e., a homogeneous distribution of the species, the \textit{\acf{SSA}} can be used, and it exists in several variants. The simplest is the \textit{naive \ac{SSA}}, which checks at every small time step whether a reaction has occurred~\cite{Erban2007}. This is very inefficient, since most time steps contain no reaction, and only approximate, since it is exact only in the limit of a vanishing time step. The \textit{Gillespie algorithm} instead samples the time of the next reaction and then decides which reaction occurs at that moment, and each simulation run is a statistically exact sample of the species counts over time~\cite{Daniel_1977_Exact}.
The core idea of \ac{SSA} can be applied to systems that are not well mixed by dividing space into small subvolumes, each assumed to be well mixed. This yields the \ac{RDME}, in which diffusion is modeled as \acp{SM} jumping between neighboring subvolumes, and these jumps are sampled together with the reactions by \ac{SSA}~\cite{Egan22, noel2017simulating}.
However, the \ac{RDME} is confined to a spatial grid and cannot resolve the positions of the \acp{SM} within a subvolume. To track the continuous positions of individual \acp{SM}, \textit{\acf{BD}} can be used. At every time step, this method moves each \ac{SM} by a random displacement and then lets two \acp{SM} react if they are closer than a given distance. In reality, however, \acp{SM} move and react continuously in time. Hence, \ac{BD} approximates this continuous process by discrete time steps, and it becomes more accurate as the time step gets shorter~\cite{Steven_2004_Stochastic}.

\textit{Direct \ac{CIR} sampling}, finally, assumes the propagation physics are already known and interprets the \ac{CIR} as the probability that a single released \ac{SM} is observed at the \ac{Rx} (see Section~\ref{ssec:theoretical-rx}).
Rather than simulating trajectories, it draws the number of observed \acp{SM} per symbol interval from a binomial distribution or its Poisson or Gaussian approximations~\cite{Jamali2019}.
This preserves the randomness of propagation and reception at low cost and is a typical approach for computing error rates over long symbol sequences.

\subsection{Hybrid and Multiscale Models}\label{sec:hybrid_and_multiscale_coupling}

In \ac{SMC}, often only the number of \acp{SM} arriving at the \ac{Rx} is of interest, and simulating the entire environment with a particle-based method would be inefficient. Hybrid and multiscale methods address this by coupling two of the above methods, so that the more detailed method is used only where it is needed, see Fig.\ \ref{fig:simulation-overview}(c).

An example is the \textit{continuum-particle hybrid}, which splits the simulation domain into two coupled regions~\cite{Mark_2012_Two, Benjamin_2013_Multiscale, Adam_2015_Multi}.
A continuum solver for, e.g., a reaction-advection-diffusion equation covers the bulk, where \ac{SM} counts remain large enough for that description to hold.
A particle-resolved solver, such as \ac{BD}, covers a small subregion where \ac{SM} counts are low, e.g., in the immediate vicinity of an \ac{Rx}.
The two solvers are coupled at the interface between these regions, where the concentration or flux computed by one side is passed to the other so that both sides agree on the state there.
Where to place this interface is the central design choice.
It must lie far enough from the \ac{Rx} such that the continuum assumption still holds, but close enough such that the particle-resolved region stays affordable to simulate.

\section{Data-based Models for SMC Systems}
\label{sec:data-based-models}
Data-based channel models can complement analytical, simulation-based, and experimental approaches by addressing several practical challenges. First, analytical models (see Section~\ref{sec:theoretical-models}) can become intractable when systems are non-linear or the topology is complex. Additionally, experiments may be affected by physical processes that are unknown or hard to characterize, prohibiting analytical or simulation-based modeling.
Second, some physical processes such as turbulence can only be simulated with conventional approaches (see Section~\ref{sec:airborne:turbulent}) at high computational cost while generative data-based models can achieve comparable accuracy at a fraction of that cost~\cite{saydemir:unfolding_time_generative_modeling_turbulent_flows_4d}.
Third, data-based models are typically differentiable with respect to their inputs and/or parameters, enabling end-to-end optimization of communication systems~\cite{heinlein:spawc2026}.
Lastly, some frameworks, e.g., \acp{FSC} or \ac{RL}, can model decision making of natural agents~\cite{nicoletti:decoding_beahvior_minimal_interpretable_agent_models,singh:emergent_behavior_neural_dynamics_artificial_agents_tracking_odor_plumes}.

\begin{table*}[!t]
  \tablestyleC[P]
  \setlength{\tabcolsep}{4pt}
  \begin{threeparttable}
      \centering
      \caption{Overview of selected data-based modeling schemes and their properties.}
      \label{tab:databasedmodels}
      \begin{tabular}{>{\cellcolor{TabFirst}}c C{1.5cm} C{3.8cm} C{1.7cm} C{1.1cm} C{1.5cm} C{1.9cm} C{3.2cm}}
          \rowcolor{TabHead}
          \cellcolor{TabHead}\tabhead{Family} & \tabhead{Scheme} & \tabhead{Description} & \tabhead{Training signal} & \tabhead{Data need} & \tabhead{Comput. Complexity} &
  \tabhead{Output} & \tabhead{Exemplary use cases} \\
          \rowcolor{white}
          & \acs*{ANN} (\acs*{MLP}) & Universal approximator for non-linear mappings & Labeled pairs & High & Medium & Point estimate\tnote{a} & Channel correction factors,
  simulator surrogate~\cite{Cheng2024,Ghanbari2025} \\
          \rowcolor{TabZebra}
          & \acs*{RNN} / Transformer & Sequence model & Labeled sequences & High & Medium & Point estimate\tnote{a} & Capturing temporal dependencies~\cite{Schottlender2025,Ozdemir2024} \\
          \rowcolor{white}
          & \acs*{GNN} & Operates on graph structures & Labeled pairs & Medium & Medium & Point estimate\tnote{a} & Modeling of blood-flow channels
  \cite{ahmet:pignn_1d_equations_blood_flow} \\
          \rowcolor{TabZebra}
          & \acs*{PINN} & \acs*{ANN} with \ac{PDE} residual in the loss & \ac{PDE} residual (+ few labels) & Low & High & Point estimate & Channel modeling \cite{Baydas2026} \\
          \rowcolor{white}
          & \acs*{GAN} & Generative \acs*{ANN} architecture learning the data distribution & & High & Medium & Samples & Synthetic channel realizations \cite{Zhu2023} \\
          \arrayrulecolor{TabFirst}\cline{2-3}\cline{5-8}
          \rowcolor{white}
          & Diffusion model & Iteratively refines noise into a sample of the data distribution & \multirow{-3}{1.7cm}{\centering Samples (labeled if conditional)} & High &
  High & Samples & Generation of turbulent flow \cite{saydemir:unfolding_time_generative_modeling_turbulent_flows_4d} \\
          \rowcolor{TabZebra}
          \multirow{-11}{*}{\rotatebox{90}{\textcolor{TabHead}{\bfseries \textit{\Ac{DL}-based}}}}
          & \acs*{RL} & Policy maximizing a reward & Reward signal & High & High & Stochastic policy & Olfactory searcher
  \cite{singh:emergent_behavior_neural_dynamics_artificial_agents_tracking_odor_plumes} \\
          \arrayrulecolor{TabHead}\specialrule{0.5pt}{0pt}{0pt}
          \rowcolor{white}
          & \ac{LTI} system identification & Impulse response of a linear channel & Labeled pairs & Low & Low & Point estimate & Channel characterization
  \cite{jamali:channel_estimation_diffusive_mc} \\
          \rowcolor{TabZebra}
          & \acs*{GP} & Distribution over functions, prior set by kernel & Labeled pairs & Low & Medium & Explicit density & Time-series/spatial modeling \cite{Han2022} \\
          \rowcolor{white}
          \multirow{-4}{*}{\rotatebox{90}{\textcolor{TabHead}{\bfseries \textit{Traditional}}}}
          & \acs*{FSC} & Interpretable agent with finite internal states & Trajectories & Low & Low & Distribution (if stochastic) & Chemotaxis agent
  \cite{nicoletti:decoding_beahvior_minimal_interpretable_agent_models} \\
          \arrayrulecolor{TabHead}\specialrule{1pt}{0pt}{0pt}
      \end{tabular}
      \begin{tablenotes}[flushleft]
          \footnotesize
          \item \!All listed schemes except \ac{RL} are suitable for gradient-based end-to-end optimization.
          \item[a] Can yield a distribution if combined with a probabilistic output layer and a likelihood-based loss.
      \end{tablenotes}
  \end{threeparttable}
  \vspace*{-0.3cm}
  \end{table*}

\subsection{Model Families and Model Choice}\label{sec:db_models}
We distinguish two main families of data-based models for \ac{MC}: \Acf{DL}-based schemes and more traditional schemes, as summarized in Table~\ref{tab:databasedmodels}. We first give an overview of the various schemes and their properties, before discussing how to choose a model.

\textbf{\Ac{DL}-based schemes} are built from one or more \acp{ANN} with many hidden layers and non-linear activation functions. Such models typically make few assumptions about the modeled system and therefore need a substantial amount of training data to perform well~\cite{prince:understanding_deep_learning}.
The most basic \ac{DL}-based scheme is the \ac{MLP}, which is a feed-forward \ac{ANN} that approximates non-linear mappings, e.g., from channel parameters to a correction factor to an otherwise fixed analytical channel model, or to a simulator output~\cite{Cheng2024,Ghanbari2025}. \Acp{RNN} and Transformers model sequences and are therefore suited to model temporal dependencies such as channel memory~\cite{prince:understanding_deep_learning,Schottlender2025}. \Acp{GNN} operate on graph-structured data and are a natural fit for channels with network-like topologies such as branching vessels, as they occur, e.g., in the \ac{CVS}~\cite{ahmet:pignn_1d_equations_blood_flow}. \Acp{PINN} are \acp{ANN} whose loss function includes the residual of the governing \ac{PDE}, e.g., the advection-diffusion equation, in addition to the data-fitting term~\cite{Baydas2026}. This embeds physical knowledge into the training, which lowers the data requirements at the cost of a more complex training procedure.
The remaining \ac{DL}-based schemes are generative or decision-making models. \Acp{GAN} train a generator and a discriminator in an adversarial game such that the generator produces samples that follow the training data distribution~\cite{prince:understanding_deep_learning}. Diffusion models learn to iteratively refine noise into samples that follow the data distribution~\cite{prince:understanding_deep_learning}. Both \acp{GAN} and diffusion models can be conditioned on system parameters, e.g., the distance between \ac{Tx} and \ac{Rx}, or the transmitted symbol to generate channel realizations for a given set-up~\cite{Zhu2023,saydemir:unfolding_time_generative_modeling_turbulent_flows_4d}. Finally, \ac{RL} learns a policy that maximizes a cumulative reward through interaction with an environment. In \ac{MC}, it is used to model, e.g., searchers navigating turbulent odor plumes~\cite{singh:emergent_behavior_neural_dynamics_artificial_agents_tracking_odor_plumes}.

\textbf{Traditional schemes} typically have far fewer parameters than \ac{DL}-based schemes and are often easier to interpret. The simplest traditional scheme is the identification of an \ac{LTI} system, i.e., the estimation of the end-to-end \ac{CIR} $h(t)$ in \eqref{eq:cascade} from measured pairs of $x(t)$ and $y_\mathrm{s}(t)$~\cite{jamali:channel_estimation_diffusive_mc}.
\Acp{GP} are a prominent example of kernel-based methods used to model non-linear systems~\cite{Han2022}: they define a distribution over functions whose prior assumptions (e.g., smoothness) are encoded in a kernel. Conditioning on the data yields a posterior with an explicit mean and uncertainty, which makes \acp{GP} suitable for modeling time series and spatial fields together with their distributions. \Acp{FSC} are agent models with a finite number of internal memory states, whose actions and state transitions depend on the current observation~\cite{nicoletti:decoding_beahvior_minimal_interpretable_agent_models}. When fitted to observed trajectories, they can provide compact and interpretable descriptions of agent behavior.

The columns in Table~\ref{tab:databasedmodels} translate into practical guidelines for the choice of a data-based model. The required output is often the first decision: if only the mean \ac{CIR} is needed, point-estimate models suffice, whereas noise or turbulence call for generative or probabilistic models. The training signal follows from the available data: sequences favor \acp{RNN}/Transformers, graphs favor \acp{GNN}, and agent trajectories favor \acp{FSC}. The data need decreases generally with the prior knowledge embedded in the model: known governing equations favor \acp{PINN} or \acp{GP} with physically motivated kernels, and with few measurements, traditional or physics-informed models are usually preferable. Finally, the computational budget limits the complexity, and interpretability requirements may also favor traditional schemes.

\subsection{Data Collection and Model Training}
\label{sec:db_data}

Data-based models require the \textbf{collection of adequate training data}. Depending on the model, the data must have a specific structure, e.g., labeled pairs, labeled sequences, or unlabeled samples (see Table~\ref{tab:databasedmodels}). The data can originate either from experiments (see Section~\ref{sec:experimental-models}) or from simulations (see Section~\ref{sec:simulation-models}). While simulations enable perfectly controlled data generation, they cannot capture \textit{unknown} real-world effects. On the other hand, experimental data are often closer to the target application but depend heavily on the measurement procedure and may contain testbed-specific effects.
A combination of both types, e.g., pre-training on a large simulated dataset followed by fine-tuning on a smaller experimental one or applying data augmentation to experimental data~\cite{scazzoli:mc_data_augmentation_dl_based_detection}, can help to bridge the theory-to-experiment gap.
The \textbf{coverage of the parameter space} is another important consideration. The data must span the entire range of parameters (e.g., the distances between \ac{Tx} and \ac{Rx} or the flow velocities) that the model may encounter, as extrapolation beyond the training regime is challenging.
Moreover, the dataset size must match the model requirements. It depends on the dimensionality of the parameter space, the output space, and the noise in the training data.
The choice of the loss function used to train data-based models also affects model performance. For instance, the \ac{MSE} corresponds to a Gaussian likelihood with constant variance. It therefore yields the conditional mean and focuses on predicting peak values due to its quadratic penalty. Thus, it may not be suitable for non-Gaussian or signal-dependent noise. If the noise distribution matters, likelihood-based or adversarial losses are more appropriate~\cite{prince:understanding_deep_learning}.
Lastly, the data is usually split into \textbf{training, validation, and test sets} to assess model generalization. To get realistic estimates of the out-of-distribution performance, training, validation, and test sets should contain data that have been generated with different parameters (e.g., with different flow velocities)~\cite{prince:understanding_deep_learning}.

\section{Experimental Models for SMC Systems}
\label{sec:experimental-models}
Analytical, simulation, and data-based models describe an \ac{SMC} system mathematically or numerically, whereas experimental models realize it physically. Testbeds demonstrate the feasibility of \ac{SMC} concepts, reveal physical processes that might occur in the envisioned target application, and provide the parameters and data on which the other models build.

\subsection{Experimental Testbeds as Models}
An experimental \ac{SMC} testbed is not the target application itself but a physical model of it. Similar to a mathematical model, it reproduces selected properties of the target environment and abstracts from others. For instance, tube-based testbeds for in-body \ac{SMC} mimic the advection-dominated transport in blood vessels, whereas tube material, channel topology, and carrier fluid typically differ substantially from the real environment \cite{Lotter23b}. Data recorded directly in the target environment, e.g., clinical measurements, do not form an experimental model, since the environment is not artificially designed, the input can be controlled only partially, and the measurement conditions cannot be reproduced. Yet, they are the ground truth that experimental models aim to emulate and can thus be used to assess whether these capture the relevant physical effects.

Developing a testbed involves several interdependent steps, namely, specifying the target application and the research question, selecting suitable \ac{SM} species, designing the propagation environment, establishing the measurement procedure, and designing communication schemes based on the measured data \cite{lotter2023experimental,Lotter23b}. Several steps require expertise beyond communications engineering, e.g., from chemistry, microbiology, fluid dynamics, material science, and sensor technology \cite{lotter2023experimental,Egan23}.
In the following, we focus primarily on testbeds developed for \ac{SMC} settings and their role as experimental models of the target application, and refer to \cite{lotter2023experimental, Lotter23b} for a comprehensive overview of existing testbeds.

\subsection{Classification of Experimental SMC Models}
\label{sec:experimental-classification}

\begin{figure}[!t]
\centering
\resizebox{\columnwidth}{!}{\includegraphics{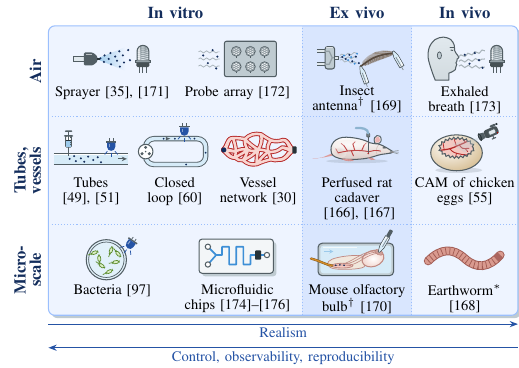}}
\caption{\textbf{Representative experimental \ac{SMC} testbeds.} Testbeds are classified by their proximity to the living organism and by medium and scale. $^*$Electrical rather than molecular signaling; $^\dagger$Developed outside the \ac{SMC} community.}
\label{fig:experimental-testbed-map}
\vspace*{-0.3cm}
\end{figure}

\textbf{Proximity to a living organism:} In biology and medicine, experimental models are commonly classified by their proximity to a living organism. This classification can also be adopted for \ac{SMC} and defines the columns of Fig.~\ref{fig:experimental-testbed-map}. It applies directly to healthcare applications, e.g., in the \ac{CVS}. For applications without a living organism, e.g., bioprocess control or odor-based communication, the same taxonomy can be used to describe how closely a testbed reproduces the real target environment, e.g., a bioreactor or the outdoor atmosphere, where field measurements take the role of \emph{in vivo} experiments (cf.\ Section~\ref{sec:airborne:turbulent}).
\emph{In vivo} models are experiments within a living organism and capture its complexity, e.g., the vascular network and its interaction with tissue, but offer limited control and observability and are subject to ethical and regulatory constraints. \emph{Ex vivo} models employ tissue, organs, or entire bodies that are no longer part of a living organism. For example, rat cadaver models preserve the anatomy of the vascular network, while systemic effects such as active metabolism are absent \cite{zucal2022innovative,Deininger2023}. \emph{In vitro} models, e.g., tubes, microfluidic chips, or air channels, are realized outside living organisms, possibly involving living cells such as bacteria. They allow full control and repeatable measurements but represent the target environment only through the properties deliberately built into them.
\emph{In silico} models, i.e., simulation and data-based models (see Sections~\ref{sec:simulation-models} and~\ref{sec:data-based-models}), are even more strictly limited: they reveal only effects included in their assumptions, whereas in \emph{in vitro} models, physical effects that were not deliberately built in, e.g., wall adsorption, can still occur and be observed.
Hence, from \emph{in silico} to \emph{in vivo}, realism increases, whereas controllability, observability, and reproducibility decrease. Most \ac{SMC} testbeds are \emph{in vitro}. \emph{In vivo} testbeds remain rare, e.g., the \ac{CAM} of chicken eggs \cite{vakilipoor2025cam} and, with electrical rather than molecular signaling, living earthworms \cite{abbasi18}, and \emph{ex vivo} models have not yet been employed by the \ac{SMC} community, though they are used in other fields.
Examples for \emph{ex vivo} models from other fields that could serve as \ac{SMC} testbeds include systems based on insect antennas~\cite{Shigaki2020odor}, perfused rat cadavers~\cite{zucal2022innovative,Deininger2023}, and the olfactory bulb of a mouse~\cite{Meeks2013aob}.

\textbf{Medium and scale:} The rows of Fig.~\ref{fig:experimental-testbed-map} distinguish experimental testbeds by medium and spatial scale, which largely follow from the target application~\cite{lotter2023experimental, Lotter23b}: microscale testbeds in liquids, with distances up to a few millimeters, target lab-on-a-chip and cellular-level in-body applications; macroscale testbeds in liquids, i.e., tubes and vessel networks, target in-body communication in the \ac{CVS} or pipeline monitoring; and air-based testbeds target airborne \ac{SMC} applications. Moreover, \ac{Tx} and \ac{Rx} may be biological or artificial, and the \ac{Rx} may measure invasively or non-invasively \cite{lotter2023experimental}.

\textbf{Evolution of testbeds:} The evolution of \ac{SMC} testbeds reflects a shift from theory-led to application-driven research: early testbeds built from off-the-shelf components demonstrated general feasibility \cite{farsad2013tabletop}, whereas later ones increasingly reproduced selected properties of specific applications, e.g., biocompatible \acp{SM} \cite{bartunik2023development}, biological \acp{Tx} \cite{grebenstein2019biological}, closed-loop propagation \cite{scherer2025closed}, and in vivo environments \cite{vakilipoor2025cam}.

\subsection{Components of Experimental SMC Models}
\label{sec:experimental-methods}
Testbeds physically realize the building blocks of an end-to-end \ac{SMC} system~\eqref{eq:cascade}, i.e., \ac{Tx}, channel, and \ac{Rx}. Complementing Section~\ref{sec:taxonomy}, we summarize their experimental realizations and the implications for analytical and simulation models.

\textbf{Signaling molecules:} The choice of \ac{SM} species largely determines the feasible release and detection mechanisms. In liquids, fluorescent dyes, quantum dots, and fluorescent proteins \cite{tuccitto2017fluorescent, scherer2025closed}, biocompatible \acp{SPION} \cite{bartunik2023development}, acids and bases \cite{farsad2017novel}, and salts \cite{angerbauer2023salinity} are common; in air, volatile compounds such as alcohols \cite{farsad2013tabletop}.

\textbf{Transmitters:} \acp{SM} are typically injected by pumps in liquids and by sprayers in air, which may perturb the flow locally (see Section~\ref{ssec:mathematical-flow}). On microfluidic chips, short pulses can be generated by hydrodynamic gating, i.e., by briefly switching the flow rates at a channel junction such that a plug of \acp{SM} enters the main channel \cite{Bolhassan2024, Akyol2024hydrodynamic}. Alternatively, light can serve as the input, either triggering the release of \acp{SM}, e.g., protons by light-driven proton pumps \cite{grebenstein2019biological, Harder2024,tomdieck2026nanodevices}, or switching the state of photoswitchable proteins already present in the channel  \cite{Brand2022MediaModulation, scherer2025closed}. In both cases, the \ac{Tx} does not perturb the flow. The release kinetics determine $h_\mathrm{Tx}(t)$ in \eqref{eq:cascade}.

\textbf{Channel:} The propagation environments of existing testbeds are shown in Fig.~\ref{fig:experimental-testbed-map}; in liquids, the flow of water, blood substitutes, or blood \cite{debus2025blood} is typically pump-driven. Geometry, flow rate, and wall material affect $h_\mathrm{ch}(t)$ and the relevant dimensionless numbers, e.g., the Péclet and Reynolds numbers, determine which approximations potentially apply (see Table~\ref{tab:approx}) \cite{Bruus2008}.

\textbf{Receivers:} Fluorescent \acp{SM} are detected optically, i.e., non-invasively, via fluorescence or absorbance by cameras, photodiodes, or spectrometers, where spectral sensing allows for detecting multiple \ac{SM} types \cite{wietfeld2024evaluation}. \acp{SPION} are detected by inductive \cite{Wicke2021} or capacitive \cite{Bartunik2023capacitive} sensors, which can also distinguish fluids of different permittivity \cite{huang2024non}. The pH value and conductivity can be measured by electrochemical sensors, e.g., pH microelectrodes, inside the channel, and volatile compounds in air can be detected by \ac{MOS} gas sensors.

\textbf{Measurement chain:} Testbeds usually do not measure the quantity in which analytical and simulation models are formulated, e.g., the number of \acp{SM} at the \ac{Rx}, but a transduced proxy (see Section \ref{sec:taxonomy}), such as a fluorescence intensity or an inductance change, whose possibly non-linear relation to the former, i.e., $F(\cdot)$ in \eqref{eq:sensor}, has to be established by calibration. The sensor response time acts as a low-pass filter, i.e., $h_\mathrm{Rx}(t)$ in \eqref{eq:sensor}, and limits the meaningful sampling rates, and sensor noise has to be characterized separately, e.g., by measurements without \acp{SM} \cite{grebenstein19}.
Some quantities are observable only at limited resolution: vesicles are handled as populations on the order of $10^{12}$ per milliliter, such that mostly ensemble quantities are measured, since single-vesicle assays exist but are hardly feasible at this scale \cite{lehman2023liposome,tomdieck2026nanodevices}, and in the \ac{CAM}, a camera records only relative intensities of a two-dimensional projection \cite{vakilipoor2025cam}.

All these limitations affect the channel output of a testbed and have to be included in a corresponding analytical or simulation model.

\begin{table}[t]
  \centering
  \caption{Publicly available experimental \ac{MC} data sets (DP: IEEE DataPort, Z: Zenodo); in vitro unless stated otherwise.}
  \vspace*{-0.2cm}
  \label{tab:datasets}
  \scriptsize
  \tablestyleC[B]
  \setlength{\tabcolsep}{3pt}
  \renewcommand{\tabularxcolumn}[1]{m{#1}}%
  \begin{tabularx}{\columnwidth}{F
    >{\centering\arraybackslash}X
    >{\centering\arraybackslash}m{2.3cm}
    >{\centering\arraybackslash}m{0.75cm}}
  \rowcolor{TabHead}
  \cellcolor{TabHead}\tabhead{Ref.} & \tabhead{\acsp{SM}} & \tabhead{Medium} & \tabhead{Platf.} \\
  \cite{grebenstein19} & Protons (bacterial \acs{Tx}) & Liquid & DP \\
  \cite{abbaszadeh2020tracking} & Fluorescent dye, tracer particles & Liquid & DP \\
  \cite{cali2023isk} & Carbon nanoparticles & Liquid & DP, Z \\
  \cite{hofmann2023dataset,hofmann2023anc} & Ethanol & Air & DP \\
  \cite{bartunik2023channel,Jakumeit2026sorption} & \leavevmode\acsp{SPION} & Liquid & DP, Z \\
  \cite{Cali24exp} & Graphene quantum dots & Liquid & DP, Z \\
  \cite{walter2023real} & Sodium hydroxide (pH) & Liquid & Z \\
  \cite{scherer2025closed} & \leavevmode\acs{GFPD} & Liquid & Z \\
  \cite{debus2025blood} & \leavevmode\acsp{SPION} & Water, blood substitute, blood & Z \\
  \cite{vakilipoor2025cam} & Indocyanine green & Blood (in vivo) & Z \\
  \cite{Cavallaro2025dataset} & Dimethyl methylphosphonate & Air & Z \\
  \cite{Figueiredo2026dataset} & Rhodamine B, fluorescein & Liquid & DP \\
  \cite{Bhattacharjee2026dataset} & Water vapor & Exhaled breath (human) & Z \\
  \arrayrulecolor{TabHead}\specialrule{1pt}{0pt}{0pt}
  \end{tabularx}
  \vspace*{-0.3cm}
  \end{table}

\subsection{Interplay between Experiments and Theory}
\label{sec:experimental-interplay}
Experiments can benefit from and inform the analytical, simulation, and data-based models discussed in this paper (see Fig.~\ref{fig:intro:model-choice}). Validating analytical, simulation, or data-based models for \ac{SMC} systems by experiments typically involves the following steps:
\begin{enumerate}
    \item \emph{Decide on the quantity of interest}, e.g., the peak time, the tail of the \ac{CIR}, or the \ac{BER}, and the accuracy required by the application.
    \item \emph{Characterize the measurement chain}: perform measurements without \acp{SM} for noise and background, a dilution series in the testbed medium to determine $F(\cdot)$, and a step response to unveil $h_\mathrm{Rx}(t)$ in \eqref{eq:sensor}.
    \item \emph{Fix independent parameters and test assumptions}: geometry, flow rate, $D$, and $k$, obtained from separate measurements or the literature, determine applicable approximations.
    \item \emph{Fit and validate}: fit only the parameters that cannot be determined independently, e.g., an effective dispersion coefficient, to a set of measurements, and validate on additional measurements not used for fitting.
    \item \emph{Analyze the residuals}: systematic deviations between the fitted model and the validation measurements indicate missing effects, which have to be checked for artifacts of the apparatus before any refinement.
\end{enumerate}
The following examples illustrate steps 3)-5). In the closed-loop testbed in \cite{scherer2025closed}, the independently known flow rate, tube radius $R$, and diffusion coefficient $D$ yield $\mathrm{Pe}\approx10^6$ and a radial diffusion time $R^2/D$ much longer than the travel time, which indicates advection-dominated transport rather than Taylor-Aris dispersion (Step 3, see \ Table~\ref{tab:approx}). Measurements at two \ac{Tx}-\ac{Rx} distances confirm this, since the larger distance delays the received pulse without changing its shape (Step 4). In the \ac{CAM}, in contrast, the vessel network cannot be characterized independently, and the measured pulses are strongly broadened \cite{vakilipoor2025cam}. Hence, effective \ac{1-D} closed-loop models \cite{brand25closed} are fitted, and their structure, e.g., the number of loops, is selected by comparison with measurements (Step 4, see Section~\ref{ssec:cvs-cam}). Finally, systematic residuals reveal missing physical effects (Step 5), e.g., signal drift (see Section~\ref{ssec:others-bacteria}) and closed-loop \ac{ISI} (see Section~\ref{ssec:others-closed-loop}).

\textbf{Artifacts of the experimental apparatus:} Not every physical process observed in a testbed is a property of the target application, and modeling artifacts of the apparatus is a common pitfall, cf.\ Step 5. In \cite{mcguiness2019experimental}, deviations between measured and predicted signal amplitudes were presumably caused by interactions between the \acp{SM} and the membrane of the mass-spectrometric detector. In \cite{koo2020deep}, a higher data rate even reduced the \ac{BER}, which the authors of \cite{Lotter23b,koo2020deep} attribute to the more frequent injections washing excess \acp{SM} off the reactive \ac{Rx} surface. Moreover, physical processes present in both testbed and application, such as wall adsorption \cite{Jakumeit2026sorption}, may have very different parameter values, e.g., for plastic tubing and blood vessel walls. Hence, before a model is refined, it has to be clarified whether observed phenomena are relevant for the target application and matter for the task at hand.

\textbf{Data requirements:} Since building an experimental model requires considerable effort \cite{lotter2023experimental}, an increasing number of experimental data sets are made publicly available, see Table~\ref{tab:datasets} and \cite{torresgomez2025nn}. Most stem from \emph{in vitro} testbeds involving liquids, whereas air-based and \emph{in vivo} datasets are more scarce. However, most \ac{SMC} experiments are proof-of-concept studies producing limited amounts of data \cite{lotter2023experimental}. Statistical channel models require many independent realizations rather than a single measurement, since parameters may vary considerably, e.g., between bacterial cultures \cite{grebenstein2019biological}, and data-based models require data covering a wide range of conditions \cite{torresgomez2025nn}.

Sections~\ref{sec:mathematical-models}-\ref{sec:experimental-models} introduced the tools for modeling \ac{SMC} systems individually: the mathematical description of the relevant effects, analytical models derived from it under suitable approximations, simulation and data-based models for cases where closed-form solutions are not available, and experimental models that emulate the target application environment.
Which of these tools are applicable, and how they interact, depends on the application environment. In the following sections, we therefore apply them to concrete \ac{SMC} environments: the \ac{CVS} in Section~\Ref{sec:examples-cvs}, airborne \ac{SMC} in Section~\ref{sec:examples-airborne}, and three further experimental testbeds in Section~\ref{sec:examples-others}. For each, we discuss which analytical, simulation, data-based, and experimental models have been developed, on which approximations and works they build, and whether their development was driven by theory or by experiment.

\section{Modeling SMC Systems in the Cardiovascular System}
\label{sec:examples-cvs}
The \acf{CVS} is the main target environment for the healthcare applications of \ac{SMC}, e.g., targeted drug delivery \cite{Chude-Okonkwo_et_al:IEEECommSurvTutorials:2017,chahibi2017tdd}, continuous in-body health monitoring \cite{felicetti2016applications}, and the coordination of distributed nanodevices in the \ac{IoBNT} \cite{akyildiz15,kuscu21a}.
We consider the transport of \acp{SM} through the network of blood vessels, from the large arteries via the arterioles and capillaries to the venules and veins.
Adopting the taxonomy of Section~\ref{ssec:taxonomy-propagation}, \ac{SM} transport in large parts of the \ac{CVS} (excluding small arterioles, venules, and capillaries) is advection-dominated, within a bounded, branched \ac{VN}, and reception is typically envisioned to occur at a sensor attached to, or embedded in, vessel walls.
The modeling literature covers this range unevenly: comparatively many models exist for individual medium-sized arterial vessels, whereas compliant arteries, the capillary bed, the venous vasculature, and the transitions between vessel scales remain largely unaddressed.

Modeling \ac{SMC} in the \ac{CVS} is challenging due to the interplay of multiple physical effects:
(E1) in arteries close to the heart, blood flow is pulsatile \cite{Womersley1955} and vessels expand during ventricular systole and recoil during diastole;
(E2) the vasculature comprises a complex, branched topology rather than a single straight duct;
(E3) blood is a dense suspension of water, cells, and proteins with non-Newtonian rheology at small scales, e.g., within capillaries, where the diameter of \acfp{RBC} exceeds that of the vessels, forcing the \acp{RBC} to squeeze through the vessel;
and (E4) \acp{SM} may interact with the vessel walls through absorption, adsorption, and desorption \cite{Lo19laminar,Jiang2022,Jakumeit2026sorption}.
Which effects an analytical or simulation model must capture depends on the intended application and task, and the vessels of the \ac{CVS} in which an \ac{SMC} system operates.

\begin{figure}[!t]
    \centering
    \includegraphics[width=\linewidth]{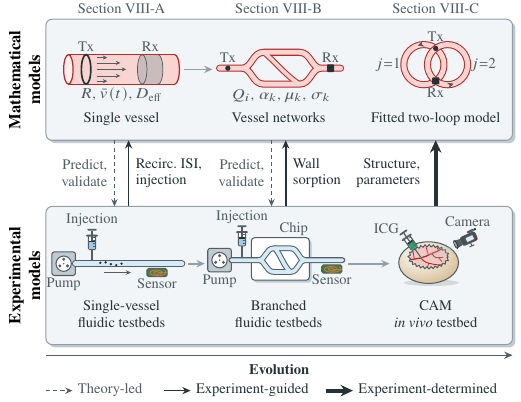}
    \caption{\textbf{Evolution of \ac{SMC} channel models for the \ac{CVS}.} Single-vessel and \ac{VN} models were largely theory-led, with first feedback from fluidic testbeds, whereas for the in vivo \ac{CAM} testbed, the measurements determine the structure and parameters of the model.}
    \label{fig:CVS_model_development}
    \vspace*{-0.3cm}
\end{figure}

To capture the relevant effects, the \ac{SMC} literature has developed a sequence of increasingly realistic \ac{CVS} channel models, see Fig.~\ref{fig:CVS_model_development}. Analytical single-vessel models (see Section~\ref{ssec:cvs-single}) address effects E1 and E4, and analytical \ac{VN} models (see Section~\ref{ssec:cvs-network}) address effect E2. The \emph{in vivo} \ac{CAM} testbed (see Section~\ref{ssec:cvs-cam}), an experimental model of the \ac{CVS}, comprises all effects, but can no longer be described by an analytical model derived from first principles.
For each of these \ac{CVS} channel models, we show representative \acp{CIR} in Fig.\ \ref{fig:cvs-results} and discuss the simulation and experimental models used to validate or inform the models.
In terms of Fig.~\ref{fig:intro:model-choice}, the single-vessel and \ac{VN} models were largely theory-led, i.e., derived first and then validated by simulation and experimental models, whereas for the \ac{CAM}, the experimental model determined the structure and parameters of the analytical model. Hence, analytical \ac{CVS} channel models first became more complex as more effects were included, but as the considered scenario approached the real application environment, experimental models dictated which effects matter, which resulted in a structurally simple analytical model.

\subsection{Single-Vessel SMC Models}\label{ssec:cvs-single}

\begin{figure}[!t]
\centering
\includegraphics{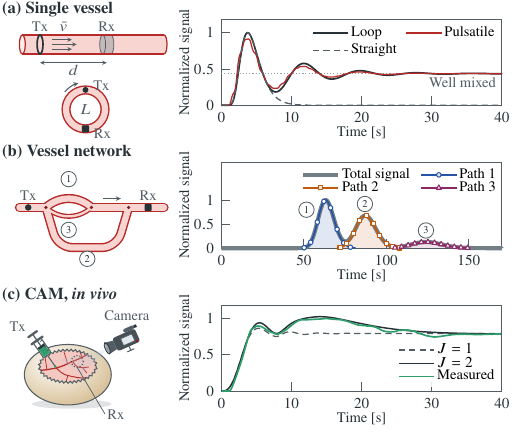}
\caption{\textbf{Normalized received signals of analytical \ac{SMC} models for the \ac{CVS}.} The sketches on the left show the considered scenarios: \textbf{(a)} A single vessel \eqref{eq:cvs-single}, straight and as a closed loop, for $\bar{v}=\qty{2}{\milli\meter\per\second}$, $d=\qty{8}{\milli\meter}$, $L=\qty{16}{\milli\meter}$, and $D_\mathrm{eff}=\qty{e-6}{\square\meter\per\second}$ with constant and pulsatile flow, $\bar{v}(t)=\bar{v}\,(1+0.6\sin(2\pi f t))$ with $f=\qty{1}{\hertz}$, \textbf{(b)} a \ac{VN} \eqref{eq:cvs-network}, recomputed for VN~2 of \cite{Jakumeit2026MIGHT}, with the shaded contributions of the paths, numbered by their arrival, and \textbf{(c)} the \ac{CAM} with fits of \eqref{eq:cam-model} \cite{vakilipoor2025cam}.}
\label{fig:cvs-results}
\vspace*{-0.3cm}
\end{figure}

\textbf{Analytical models:} Early analytical models of \ac{SMC} in the \ac{CVS} abstract a blood vessel as a straight cylindrical duct of radius $R$ with laminar background flow of mean velocity $\bar{v}$. As in Section~\ref{ssec:theoretical-approx}, we consider an impulsive release over the cross-section at $t=0$ and a transparent \ac{Rx} of length $\ell_\mathrm{Rx}$ at distance $d$. Starting from this straight vessel, three analytical models of increasing complexity have been developed, see Fig.~\ref{fig:cvs-results}(a) and are discussed in the following.

\emph{Straight vessel:} For $t\gg R^2/D$, Taylor-Aris dispersion reduces the transport to the \ac{1-D} model \eqref{eq:taylor-aris} \cite{Taylor1953,Aris56}, and the \ac{CIR} is given by \eqref{eq:taylor-aris-cir}. Its transport regimes were characterized as a function of the P\'eclet number and the \ac{Tx}-\ac{Rx} distance in~\cite{Wicke2018}. Complementary \ac{CIR} approximations for \ac{3-D} ducts were obtained in~\cite{Dinc19laminar}, and a transfer-function model for the mixed regime, where both advection and diffusion contribute, was proposed in~\cite{schaefer21b}.

\emph{Closed loop:} Since the \ac{CVS} is a closed system, \acp{SM} passing the \ac{Rx} generally recirculate and arrive repeatedly. Unrolling a loop of length $L$ into an infinite straight vessel, the \ac{Rx} reappears at distances $d+iL$, $i\in\mathbb{Z}$, and the \ac{CIR} becomes the sum of \eqref{eq:taylor-aris-cir} over these distances \cite{vakilipoor2025cam}
\begin{equation}
    h_\mathrm{ch}(t)\approx\sum_{i\in\mathbb{Z}}\frac{\ell_\mathrm{Rx}}{\sqrt{4\pi D_\mathrm{eff}t}}\exp\!\Big(-\frac{\big(d+iL-\bar{v}t\big)^2}{4D_\mathrm{eff}t}\Big),\quad t\geq0,
    \label{eq:cvs-single}
\end{equation}
where the terms with $i\geq1$ correspond to \acp{SM} that have completed $i$ loops, and the terms with $i<0$ are negligible unless diffusion against the flow is significant. Analytical closed-loop models have been formulated for constant flow \cite{vakilipoor2025cam}, with local and global degradation \cite{brand25closed}, and jointly with pulsatile flow \cite{Symeonidis2026Pulsatile}.

\emph{Pulsatile flow:} Since blood flow is pulsatile with angular frequency $\omega$, the Womersley velocity profile~\cite{Womersley1955} can be reduced to an effective \ac{1-D} description with time-variant mean velocity $\bar{v}(t)$ and effective diffusion coefficient $D_\mathrm{eff}(t)$ \cite{Wang2025}.
In this case, the mean distance traveled with the flow, $\bar{v}t$, and half the variance of the axial \ac{SM} position, $D_\mathrm{eff}t$, in \eqref{eq:taylor-aris-cir} and \eqref{eq:cvs-single} are replaced by the time-variant mean and variance
\begin{equation}
    X(t)=\int_0^t\bar{v}(s)\,\mathrm{d}s \quad\text{and}\quad \Sigma(t)=\int_0^t D_\mathrm{eff}(s)\,\mathrm{d}s,
    \label{eq:cvs-pulsatile}
\end{equation}
respectively, which yields the \ac{CIR} for pulsatile flow in a straight vessel and in a closed loop \cite{Symeonidis2026Pulsatile}. %

\textbf{Exemplary CIR:} Fig.~\ref{fig:cvs-results}(a) illustrates the three analytical models. The straight-vessel model \eqref{eq:taylor-aris-cir} yields a single dispersed pulse (dashed curve). In the closed loop \eqref{eq:cvs-single}, the \acp{SM} recirculate and pass the \ac{Rx} repeatedly, such that further, increasingly broadened peaks appear (blue curve). These repetitions can cause inter-loop \ac{ISI} at the \ac{Rx} and converge to the well-mixed level $\ell_\mathrm{Rx}/L$ of the loop (gray dotted curve) \cite{brand25closed}. Pulsatile flow \eqref{eq:cvs-pulsatile} (red curve) superimposes oscillations at the heart rate, which are most visible on the flanks of the peaks, and renders the channel cyclostationary, i.e., its response depends on the phase of the pulsation at the time of release \cite{Symeonidis2026Pulsatile}.

\textbf{Extensions:} Building on these baseline analytical models, individual effects have been added, mostly motivated by theory rather than by measurements: arbitrary release profiles over the cross-section \cite{schaefer21b} and the injection into the background flow \cite{Wicke2021} at the \ac{Tx}; absorption \cite{Lo19laminar}, ligand-receptor binding \cite{Kuscu2018}, and reversible sorption \cite{Jiang2022} at the vessel wall, i.e., effect E4; and drift due to gravity or magnetic forces \cite{Schaefer19force}, which also allows the steering of magnetic \acp{SM} toward the \ac{Rx} \cite{Wicke2019magnetic}.

\textbf{Simulation and experimental models:} Analytical single-vessel models have mainly been validated by particle-based simulation methods (see Section~\ref{sec:stochastic_samplers}), e.g., the transport regimes of \eqref{eq:taylor-aris-cir} in \cite{Wicke2018} and the closed-loop model in \cite{brand25closed}, and the impact of pulsatile flow on nanoparticle propagation has been studied in \cite{Wille2025pulsatile}.
Since simulations reveal only effects included in their assumptions, experimental models are additionally required.
Single-vessel systems have likewise dominated experimental \ac{MC} research, starting from tube-flow testbeds with background flow \cite{farsad2017novel,tuccitto2017fluorescent,Wicke2021}. Already at this stage, experiments influenced the considered analytical models. In the testbed in \cite{Wicke2021}, the received signal was governed by laminar advection and the injection rather than by Taylor-Aris dispersion, which required a dedicated parametric analytical model. Moreover, the closed-loop testbed with the reversibly photoswitchable \ac{GFPD} as \ac{SM} \cite{scherer2025closed} revealed recirculation-induced inter-loop and offset \ac{ISI}, which motivated the analytical closed-loop model in \cite{brand25closed} (see Section~\ref{ssec:others-closed-loop}). However, \acp{SM} in the \ac{CVS} pass many connected vessels rather than a single one, which motivates the extension of analytical single-vessel models to the analytical \ac{VN} models discussed next.

\subsection{Vessel Network SMC Models}\label{ssec:cvs-network}
\textbf{Analytical model:} The \ac{CVS} can be viewed as a branched \ac{VN} of connected vessels spanning several orders of magnitude in length and radius~\cite{aaronson2020cardiovascular}. Since the topology of a \ac{VN} is expected to govern \ac{SM} transport more strongly than the small-scale effects \cite{Jakumeit2026MIGHT}, a parallel line of work has studied \acp{VN} as \ac{SMC} channels, building on vascular graph models of blood flow and scalar transport \cite{Reichold2009}. Most analytical models describe \acp{VN} as directed graphs of cylindrical pipes connected by bifurcations and junctions, and proceed in three steps \cite{Chahibi13,mosayebi2019cancer,Jakumeit2026MIGHT}:
\begin{enumerate}
    \item \emph{Flow:} Each pipe $i$ of length $\ell_i$ and radius $R_i$ exhibits the Hagen-Poiseuille hydraulic resistance $R_{\mathrm{h},i}=8\eta\ell_i/(\pi R_i^4)$, where $\eta$ denotes the fluid viscosity.
    A \ac{VN} is then represented by a circuit with hydraulic resistances, which, in turn, can be interpreted as an electrical circuit, where the pressure drop corresponds to a voltage and the flow rates to currents.
    Kirchhoff's laws then yield the flow rates $Q_i$ and mean velocities $\bar{v}_i=Q_i/(\pi R_i^2)$ for given steady inflows and outlet pressures \cite{Jakumeit2026MIGHT}.
    \item \emph{Paths:} At each bifurcation, the \acp{SM} split proportionally to the outgoing flow rates.
    Hence, a fraction $\alpha_k$ of the released \acp{SM} travels along the $k$-th path from the \ac{Tx} to the \ac{Rx}, given by the product of the flow fractions $Q_j/\sum_{l\in\mathcal{O}_b}Q_l$ at the bifurcations $b$ along the path, where $\mathcal{O}_b$ denotes the set of pipes leaving bifurcation $b$ and $j$ the pipe on the path \cite{mosayebi2019cancer}.
    \item \emph{Transport along a path:} Each vessel acts as a single pipe in the Taylor-Aris regime \eqref{eq:taylor-aris} with effective diffusion coefficient $D_{\mathrm{eff},i}$, which the \acp{SM} pass with mean travel time $\ell_i/\bar{v}_i$ and variance $2D_{\mathrm{eff},i}\ell_i/\bar{v}_i^3$. The mean $\mu_k$ of the travel time along path $k$ is the sum of the mean travel times of its pipes, and since the travel times of consecutive pipes are modeled as independent, its variance $\sigma_k^2$ is the sum of their variances.
\end{enumerate}
The \ac{MIGHT} model approximates the travel time along each path by an inverse Gaussian distribution \cite{Jakumeit2026MIGHT}, which is exact if $\bar{v}_i^2/D_{\mathrm{eff},i}$ is identical for all pipes of a path, e.g., if they share $\bar{v}_i$ and $D_{\mathrm{eff},i}$, and accurate for large path P\'eclet numbers otherwise. The \ac{CIR} then follows as \cite{Jakumeit2026MIGHT}
\begin{equation}
    h_\mathrm{ch}(t)\approx\frac{\ell_\mathrm{Rx}}{\bar{v}_\mathrm{Rx}}\sum_{k}\alpha_k\,\sqrt{\frac{\mu_k^3}{2\pi\sigma_k^2t^3}}\,\exp\!\Big(-\frac{\mu_k\,(t-\mu_k)^2}{2\sigma_k^2\,t}\Big),
    \label{eq:cvs-network}
\end{equation}
for $t\geq0$, where the sum runs over all paths from the \ac{Tx} to the \ac{Rx} and constitutes the density of the arrival time at the \ac{Rx}. Since an \ac{SM} spends time $\ell_\mathrm{Rx}/\bar{v}_\mathrm{Rx}$ inside the \ac{Rx}, where $\bar{v}_\mathrm{Rx}$ denotes the mean velocity in the \ac{Rx} pipe, this factor converts the arrival-time density into the number of observed \acp{SM}, consistent with \eqref{eq:taylor-aris-cir} and \eqref{eq:cvs-single}. For a single pipe, the arrival-time density is the first-passage time density in Section~\ref{ssec:theoretical-rx}, and \eqref{eq:cvs-network} equals \eqref{eq:taylor-aris-cir} multiplied by $d/(\bar{v}t)$. Hence, the analytical \ac{VN} model is a superposition of analytical single-vessel models, where the topology and the hydraulic resistances determine the weights $\alpha_k$, and the pipe lengths and velocities the delays $\mu_k$ and spreads $\sigma_k$.

\textbf{Exemplary CIR:} Fig.~\ref{fig:cvs-results}(b) illustrates \eqref{eq:cvs-network} for a network of six pipes with three paths from \ac{Tx} to \ac{Rx}. The paths have significantly different delays and thus cause separate peaks, each comprising the fraction $\alpha_k$ of the \acp{SM}, i.e., about $0.48$, $0.41$, and $0.12$. In larger networks, e.g., the network of 16 pipes with seven paths in \cite{Jakumeit2026MIGHT}, the contributions of the individual paths may overlap and merge into a single broad signal, from which the paths can no longer be identified. In this manner, different path delays act as an additional dispersion mechanism, i.e., branching dispersion.

\textbf{Related models:} Earlier \ac{VN} models yield time-variant \acp{CIR} under pulsatile flow and wall absorption but only for tree-like network topologies \cite{Chahibi13,Chahibi2015,Chahibi2015a}, assume a quasi-steady \ac{SM} concentration \cite{mosayebi2019cancer,Pal2023}, or yield only stationary \ac{SM} distributions \cite{TORRESGOMEZ2024}; further approaches address the design of microfluidic channels \cite{Bicen2013} or resolve drug uptake by imaging-informed \acf{CFD} \cite{dEsposito2018}.

\textbf{Simulation and experimental models:} Analytical \ac{VN} models are mostly validated against simulation models, e.g., \eqref{eq:cvs-network} against \ac{3-D} \ac{FEM} simulations (see Section~\ref{sec:deterministic_PDE_solvers}) \cite{Jakumeit2026MIGHT}, whereas imaging-informed \ac{CFD} resolves drug uptake in tumor vasculature at the cost of analytical tractability \cite{dEsposito2018}. As for single vessels, however, simulations contain only the modeled effects, and it remains unclear which effects dominate in real \acp{VN}. First experiments in branched fluidic testbeds \cite{Jakumeit2026sorption} revealed the reversible sorption of \acp{SPION} at the pipe walls, which was accounted for in the analytical \ac{VN} model in \cite{Jakumeit2026sorption}; other branched testbeds lack an accompanying analytical \ac{VN} model \cite{Yu2024,wang2020understanding}. As all existing experimental setups simplify the vasculature considerably, the experimental foundation of analytical \ac{VN} models remains thinner than that of single vessels. Experimental models closer to the living organism can close this gap, but there, the topology and flow rates required by a bottom-up analytical \ac{VN} model are no longer known, as the \ac{CAM} testbed discussed subsequently shows.

\subsection{In Vivo SMC Models: The CAM Model}
\label{ssec:cvs-cam}
Unlike in Sections~\ref{ssec:cvs-single} and~\ref{ssec:cvs-network}, in the case of the \ac{CAM} model, the experimental model determines the analytical model, not vice versa.

\textbf{Experimental model:} The fluidic testbeds addressed in the previous subsections lack living tissue, organs, and clearance mechanisms, and the \ac{CAM} testbed narrows this gap \cite{vakilipoor2025cam}. The \ac{CAM} is a highly vascularized membrane in fertilized chicken eggs, connected to the circulation of the embryo, and can thus serve as an experimental model for the human \ac{CVS}. It provides a branched, closed-loop vascular system with organs as natural sinks and complies with the 3R principles (replacement, reduction, refinement) of animal research \cite{vakilipoor2025cam}. However, it is not mammalian, its vasculature changes during embryonic development, and the embryo moves during measurements. In the \ac{MC} interpretation of \cite{vakilipoor2025cam}, a syringe injecting the fluorescent \ac{SM} \ac{ICG} acts as \ac{Tx}, the vasculature as channel, and a near-infrared camera observing a \ac{ROI} as transparent \ac{Rx}, i.e., $w(\mathbf{r})$ in \eqref{eq:rx-cir} is the indicator of the \ac{ROI}; the camera records only relative intensities of a two-dimensional projection of the vessels, i.e., overlapping vessels at different depths cannot be distinguished.

\textbf{Analytical model:} A bottom-up analytical model as in Section~\ref{ssec:cvs-network} cannot be parameterized for the \ac{CAM}, since the vascular topology is highly complex, changes with development, and is not imaged during the experiments, which also applies to the human \ac{CVS}. Instead, the measurements determine the structure of the analytical model \cite{vakilipoor2025cam}. In the \ac{ROI}, the \ac{ICG} intensity reaches a steady state within approximately $25$\,s, partly with intermediate peaks and drops due to recirculation, whereas \ac{ICG} accumulates in the organs over minutes \cite{vakilipoor2025cam}. Exploiting this separation of time scales, the first passage and the recirculation, i.e., the approximately first $25$\,s, are described by the analytical closed-loop model \eqref{eq:cvs-single} with effective loop length $L_\mathrm{eff}$, velocity $v_\mathrm{eff}$, and dispersion coefficient $D_\mathrm{eff}$. With the injection rate $\hat{u}_\mathrm{inj}(t)$ estimated from the measurements and $J$ loops, the modeled normalized intensity in the \ac{ROI} becomes
\begin{equation}
    \hat{I}(t)=\hat{u}_\mathrm{inj}(t)\ast\sum_{j=1}^{J}\alpha_j\,p_\mathrm{wn}(d_j,t;\boldsymbol{\vartheta}_j),
    \label{eq:cam-model}
\end{equation}
where $p_\mathrm{wn}(d_j,t;\boldsymbol{\vartheta}_j)$ denotes \eqref{eq:cvs-single} divided by $\ell_\mathrm{Rx}$, i.e., the wrapped normal density of the \ac{SM} position along the $j$-th loop, with weights $\alpha_j$, $\sum_j\alpha_j=1$, and parameters $\boldsymbol{\vartheta}_j=\{L_{\mathrm{eff},j},v_{\mathrm{eff},j},D_{\mathrm{eff},j}\}$, and $d_j$ is the effective injection-\ac{ROI} distance.

\textbf{Measurement and fit:} In \cite{vakilipoor2025cam}, the analytical model \eqref{eq:cam-model} was fitted to $69$ measurements in $25$ eggs, where $J=2$ yielded a lower root-mean-square error than $J=1$. Fig.~\ref{fig:cvs-results}(c) shows a measurement in which $J=1$ captures the first peak but misses the second, larger maximum around 15\,s, which $J=2$ can represent. Similar to the closed loop in Fig.~\ref{fig:cvs-results}(a), the recirculation peaks are strongly damped, and the signal settles at its well-mixed level within a few loops. Accordingly, the fitted $D_\mathrm{eff}$, which lies within $10^{-6}$--$10^{-5}$\,m$^2$\,s$^{-1}$, exceeds molecular diffusion by orders of magnitude \cite{vakilipoor2025cam}, which is consistent with Taylor-Aris dispersion and with the branching dispersion discussed in Section~\ref{ssec:cvs-network}.

\textbf{Interpretation:} The analytical \ac{CAM} channel model \eqref{eq:cam-model} was obtained by reversing the approach of Sections~\ref{ssec:cvs-single} and~\ref{ssec:cvs-network}: its structure and parameters were deduced from experiments. The price to pay in return is that the parameters describe abstracted whole-body effects rather than individual vessels: they vary strongly between eggs, are partly correlated, and must be refitted for each injection site and \ac{ROI}.
Time-variant blood flow \cite{Symeonidis2026Pulsatile}, plasma-binding of \ac{ICG}, overlapping vessels in the projection, and the physics of the accumulation are not captured \cite{vakilipoor2025cam}. Bottom-up analytical \ac{VN} models and the top-down analytical \ac{CAM} channel model are therefore complementary: both are composed of mixtures of single-vessel responses, whose components are derived from the topology in \eqref{eq:cvs-network} and fitted to data in \eqref{eq:cam-model}, respectively.
Taken together, the single-vessel, \ac{VN}, and \ac{CAM} models exemplify the shift from analytical models validated by experimental models to analytical models determined by experiments, as summarized next.

\subsection{Summary}
\label{ssec:cvs-summary}
In the \ac{CVS}, experimental models have taken on all roles of Fig.~\ref{fig:intro:model-choice}: they revealed effects, e.g., recirculation-induced \ac{ISI} \cite{scherer2025closed} and wall sorption \cite{Jakumeit2026sorption}, and justified approximations \cite{vakilipoor2025cam}. Analytical single-vessel models \eqref{eq:cvs-single}, \eqref{eq:cvs-pulsatile} cover effects E1 and E4 and analytical \ac{VN} models \eqref{eq:cvs-network} effect E2, whereas effect E3, the capillary bed, and the transitions between vessel scales have remained largely unaddressed in \ac{SMC}. Hence, the bottleneck of \ac{CVS} modeling is no longer the availability of analytical solutions, but the experimental evidence for which effects from an \emph{in vivo} model an analytical model must include. Experimental platforms for the \ac{CVS} that could provide this evidence include fluidic chips with vascular-network-inspired topology and functionalizable walls \cite{Yu2024} and \ac{3-D}-printed hydrogel vasculature \cite{Grigoryan2019}, and fitting coarse analytical \ac{VN} models to \ac{CAM} measurements would connect the bottom-up and top-down approaches. Overall, experimental models in the \ac{CVS} have so far mainly refined analytical models, e.g., by unveiling the impact of recirculation and wall sorption, and, for the \ac{CAM}, determined their structure, whereas the advection-dominated transport along the vessels, on which all of the analytical models build, has been confirmed.

Next, we investigate \ac{SMC} in airborne environments, the second application environment considered in this tutorial.
Unlike the \ac{CVS}, airborne environments are macroscale, gaseous, and often turbulent.

\section{Modeling Airborne SMC Systems}
\label{sec:examples-airborne}
Airborne \ac{SMC} is arguably one of the application domains closest to practical deployment due to lower regulatory barriers compared to, e.g., in-body applications, and the availability of affordable hardware for release and reception.
The envisioned target applications range from communication in wave-denied environments~\cite{guo:molecular_physical_layer_wavedenied_environments} to leakage detection~\cite{gulec:localization_passive_source_sensor_network,heinlein2026source} to interfacing natural communication processes, e.g., in the context of pest control or plant communications~\cite{Vakilipoor25,Unluturk17plants}. Many of these envisioned applications require operation outdoors and over distances of several meters.
However, airborne \ac{SMC} comes with its own practical challenges for deployment in general, and channel modeling in particular. First, while affordable hardware exists, the most commonly employed sensors are cross-reactive, non-linear, and have slow response times. Second, especially over longer distances, the molecule propagation is governed by turbulence, a transport phenomenon that is notoriously challenging to model.

\begin{figure}
    \centering
    \includegraphics[width=\linewidth]{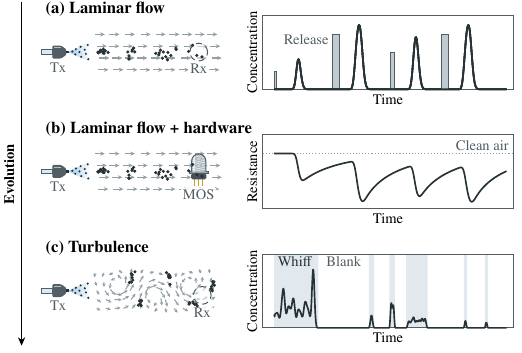}
    \caption{\textbf{Evolution of channel models for airborne \ac{SMC}.} \textbf{(a)} In laminar flow, the received concentration follows from convolving the \ac{CIR}~\eqref{eq:greens-unbounded} with the release pattern, which yields distinct peaks for repeated transmissions.
    \textbf{(b)} \ac{MOS} sensors respond non-linearly and slowly to the concentration, cf.~\eqref{eq:mos_ode}, \eqref{eq:mos_ode_output}, which introduces additional \ac{ISI}.
    \textbf{(c)} In turbulent flow, the received signal is intermittent and consists of whiffs (shaded) and blanks.}
    \label{fig:airborne}
    \vspace*{-0.3cm}
\end{figure}

In contrast to the \ac{CVS}, where experiments have mainly refined analytical models, in airborne \ac{SMC}, experiments repeatedly overturned them, see Fig.~\ref{fig:airborne}. The first analytical models described diffusion in quiescent or uniform flow (see Section~\ref{sec:airborne:analytical_nonturb}, Fig.~\ref{fig:airborne}(a)). Comparisons with measurement results from testbeds then revealed that release and reception hardware cannot be neglected, which led to the development of hardware models (see Section~\ref{sec:airborne:hardware}, Fig.~\ref{fig:airborne}(b)). Finally, real-world measurements revealed that turbulence dominates molecule propagation outdoors and over larger distances, which motivated the derivation of simulation, data-based, and statistical models of the received signal (see Section~\ref{sec:airborne:turbulent}, Fig.~\ref{fig:airborne}(c)).

\subsection{Analytical Modeling of Non-turbulent Flow}
\label{sec:airborne:analytical_nonturb}
The earliest modeling efforts for airborne \ac{SMC} focused on analytical models for molecule propagation under quiescent or uniform flow. Here, the channel is typically modeled as an unbounded medium with diffusion and drift. The \ac{Tx} is usually idealized as an impulsive point source. The resulting \ac{CIR}, given by~\eqref{eq:greens-unbounded}, possibly with first-order degradation, i.e., $k>0$, is adopted in a large part of the analytical airborne \ac{SMC} literature~\cite{farsad:channel_noise_models_nonlinearMC,Bhattacharjee22,mcguiness:experimental_results_openair_transmission}. These closed-form models are mathematically tractable and have enabled foundational results on modulation and detection for airborne \ac{SMC}. However, they rest on strong idealizing assumptions regarding both the release process and the flow field.
Whenever such analytical predictions were compared to measurements from experimental testbeds, a systematic discrepancy between the theoretical and the measured channel response was observed~\cite{farsad:channel_noise_models_nonlinearMC}. This mismatch has typically been absorbed into the model as an additive noise term, mirroring the treatment of channel uncertainty in classical wireless communications, or addressed by fitting the impulse response parameters instead of employing a physical sensor model~\cite{farsad:channel_noise_models_nonlinearMC}.

\subsection{Modeling of Hardware for Airborne \acs{SMC}}\label{sec:airborne:hardware}
Throughout the evolution of the airborne \ac{SMC} literature, mismatches between theory and experiments have motivated the adoption of models that also account for the release and reception processes, rather than for the propagation channel alone. In the following, we discuss important types of release and sensing hardware, the corresponding analytical modeling approaches, and the experimental observations that motivated them.

\subsubsection{Release Hardware}
In practice, molecules do not appear instantaneously in the environment. They first have to be released by the \ac{Tx}, and this release process can take a non-negligible amount of time. It can also be inherently noisy, which is in contrast to the common assumption that only counting noise or sensor noise affects the received signal. In the following, we consider three exemplary release mechanisms, namely microcapsules, spraying, and evaporation, whose release rates enter the channel model as $u(t)$ in \eqref{eq:tx-conv}.

Microcapsules release the encapsulated molecules gradually over time, e.g., for the long-term release of pheromones in pest control~\cite{Vakilipoor25}. Encapsulating molecules in microcapsules allows the release rate to be tuned and prolonged. However, the resulting release dynamics can be temperature-dependent.
This gradual release can be described by the Korsmeyer-Peppas model~\cite{Vakilipoor25}, a semi-empirical model commonly used for the release of molecules from polymeric matrix systems. It models the cumulative number of released molecules $M(t)$ as
\begin{align}
    \frac{M(t)}{M_\infty} = k_\mathrm{KP}\, t^{n_\mathrm{KP}}, \;\;\; 0 \leq t \leq \frac{1}{k_\mathrm{KP}^{1/n_\mathrm{KP}}}\label{eq:peppas},
\end{align}
where $M_\infty$ is the total number of molecules released (equivalent to the initial loading if all molecules are released), $k_\mathrm{KP}$ is the release rate constant, and $n_\mathrm{KP}$ is the release exponent, which indicates the release mechanism. The corresponding release rate is $u(t)=\mathrm{d}M(t)/\mathrm{d}t$.

Spraying instead releases molecules by injecting them into the channel~\cite{gulec:droplet_based_signal_reconstruction_approach,Bhattacharjee22}. This injection influences the velocity field close to the \ac{Tx} (see Fig.~\ref{fig:taxonomy:e2echannel}, left panel). In addition, the amount of molecules released per spray event is not perfectly controllable.

Evaporation from a reservoir, e.g., an open bottle, offers a third mechanism (see Fig.~\ref{fig:taxonomy:e2echannel}, left panel)~\cite{gulec:localization_passive_source_sensor_network,zhu2026experimental}. If the released molecules are dissolved in the liquid, their release rate can be approximated by the first-order release model from Section~\ref{ssec:theoretical-tx}, in which the evaporation rate depends on the amount of molecules remaining in the reservoir, whereas a pure liquid evaporates at approximately constant rate. As molecules evaporate, the reservoir is gradually depleted, so the release rate decreases over time.

Beyond these three mechanisms, other release processes exist as well, such as pheromone release by plants~\cite{Unluturk17plants}.

\subsubsection{Receiver Hardware}
Ideally, an \ac{Rx} would measure the concentration of each molecule species with high temporal resolution. Measurement instruments such as membrane inlet mass spectrometers or systems based on Raman spectroscopy come close to this ideal, but they are prohibitively expensive and bulky for widespread deployment. Instead, low-cost \ac{MOS} sensors~\cite{zhu2026experimental,farsad:channel_noise_models_nonlinearMC,farsad2013tabletop}, and, to a lesser extent, \acp{PID}~\cite{ozmen:high_speed_chemical_vapor_comms_pid} are used in testbeds and are expected to be central in practically deployed \ac{SMC} systems. However, these sensors deviate from the ideal in three aspects.

First, both sensor types respond to several molecule species, i.e., they are \textit{cross-reactive}~\cite{heinlein:tmbmc2026}. As a consequence, \ac{SMC} systems employing such sensors are subject to interference from background molecules not used for signaling~\cite{zhu2026experimental}. A promising way to mitigate this interference is to use arrays comprising several sensors with different affinities to different molecule species.
Second, \ac{MOS} sensors are \textit{non-linear}~\cite{yamazoe:theory_powerlaws_gas_sensors}: their output, the resistance $R(t)$, is not proportional to the concentration $c(t)$ of a target molecule species but depends on $c(t)$ through a power law.
Third, many \ac{MOS} sensors \textit{respond only slowly} to changes of $c(t)$~\cite{DiLello:augmented_switching_linear_dynamical_system_model}.

As a consequence, explicitly modeling the \ac{Rx} hardware is indispensable. Which model is appropriate depends on how fast the concentration changes compared to the sensor response, cf.\ the general sensor model in~\eqref{eq:sensor}.
If the concentration changes slowly, a quasi-steady-state sensor model that captures non-linearity and cross-reactivity is sufficient, i.e., the static non-linearity $F(\cdot)$ in~\eqref{eq:sensor} without the sensor dynamics $h_\mathrm{Rx}(t)$. Several variants of power-law-like functions have been proposed for this purpose~\cite{madrolle:linear_quadratic_model_quantification_mixture_two_diluted_gases_single_MOS}. For example, a commonly employed mathematical model for a sensor exposed to two molecule species $i \in \{1,2\}$ with concentrations $c_i(t)$ is given by~\cite{llobet:steadystate_transient_behavior_of_thickfilm_tin_oxide_sensors_gas_mixtures}
\begin{align}
    \frac{R(t)}{R_0} = \left[ a_1 c_1^{r_1}(t) + a_2 c_2^{r_2}(t) + a_3 c_1^{r_1}(t) c_2^{r_2}(t) +1 \right]^{-1}, \label{eq:mos_steady}
\end{align}
where $R_0$ denotes the sensor's resistance in clean air, i.e., for $c_1(t)=c_2(t)=0$. The exponents $r_i$ capture the non-linearity, the coefficients $a_1$ and $a_2$ specify the sensitivity to the individual species, and $a_3$ characterizes the interaction between both species.

If the concentration changes faster than the sensor can follow, the sensor dynamics cannot be neglected. In this case, the response of a \ac{MOS} sensor can be modeled by an \ac{ODE} (see, e.g.,~\cite{bicelli:model_experimental_characterization_dynamic_behavior_mox_sensor}). A compact mathematical model follows from the kinetics of oxygen adsorption at the sensor surface and its reaction with the target molecules~\cite{yamazoe:theory_powerlaws_gas_sensors}. For a typical \ac{MOS} sensor at constant temperature and oxygen partial pressure, exposed to a single molecule species of concentration $c(t)$, the received signal can be modeled as~\cite{yamazoe:theory_powerlaws_gas_sensors}
\begin{align}
    \tau_\mathrm{MOS} \frac{\mathrm{d} m(t)}{\mathrm{d}t} &= m_0^2  e^{m_0^2 - m^2(t)} - m^2(t) - \gamma \tau_\mathrm{MOS}  m(t) c(t), \label{eq:mos_ode} \\
    \frac{R(t)}{R_0} &= \exp\left(\frac{m^2(t)-m_0^2}{2}\right). \label{eq:mos_ode_output}
\end{align}
Here, $m(t)$ denotes the \textit{reduced depletion depth}, which determines the sensor's resistance, and $m_0$ denotes its steady-state value in clean air, i.e., for $c(t)=0$. Furthermore, $\tau_\mathrm{MOS}$ denotes the intrinsic time scale of the surface kinetics, and $\gamma$ is a rate constant, which depends on the molecule species. All three parameters, i.e., $m_0$, $\tau_\mathrm{MOS}$, and $\gamma$, depend on the sensing material and the operating conditions.

Both discussed sensor models require parameters that typically have to be estimated. Due to manufacturing differences, sensor drift, and temperature changes, the parameters differ not only between individual sensors but can also change over time. Therefore, channel models for airborne \ac{SMC} systems should also capture parameter uncertainty, e.g., via random parameters as in the statistical models of Section~\ref{ssec:theoretical-statistical}, and detection schemes should be designed to be robust to it.

\subsection{Modeling of Turbulent Flow}\label{sec:airborne:turbulent}
Most existing experimental setups for airborne \ac{SMC} systems considered relatively controlled environments with approximately uniform background flow. This choice results in analytical channel models similar to those discussed in Section~\ref{sec:airborne:analytical_nonturb}, which can be represented by a \ac{CIR}.
However, fan-induced, approximately uniform background flow is not always available. Outdoors, especially over longer distances, turbulence dominates molecule propagation~\cite{celani:odor_landscapes_turbulent_environments,Vakilipoor25} (see also Section~\ref{ssec:mathematical-flow}). Since the flow field is itself unsteady and random, the channel is linear but randomly time-variant.

\subsubsection{Full Domain Simulations}
Full numerical simulation (see Section~\ref{sec:simulation-models}) proceeds in two steps, when it is assumed that the released molecules do not perturb the background flow, i.e., one-way coupling. First, the velocity field is obtained by numerically solving the Navier-Stokes equation~\eqref{eq:math-navier-stokes}, e.g., via direct numerical simulation, large-eddy simulation, or Reynolds-averaged simulation, in decreasing order of accuracy and computational cost~\cite{Pope2000}. Second, given this velocity field, molecule transport is simulated either in an Eulerian formulation, by numerically solving an advection-diffusion equation~\eqref{eq:math-advection-diffusion} for the concentration field (see Section~\ref{sec:deterministic_PDE_solvers} and~\cite{Vakilipoor25}), or in a Lagrangian formulation, by advecting individual tracer particles through the velocity field (see Section~\ref{sec:stochastic_samplers} and~\cite{biferale:turb_smoke}).

Because these simulations are computationally expensive, \ac{ML}-based schemes, i.e., data-based models (see Section~\ref{sec:data-based-models}), have been proposed to speed up either step. For the first step, the broad interest in modeling velocity fields across disciplines has produced a wide variety of approaches, ranging from auto-regressive models that predict the velocity field at the next time step~\cite{patil:autoregressive_transformers_datadriven_turbulence} to generative models that directly generate realistic velocity fields for a given geometry~\cite{saydemir:unfolding_time_generative_modeling_turbulent_flows_4d}. Generated velocity fields still require simulating the molecule transport through them. A second line of research therefore skips the velocity field and directly generates molecule trajectories whose statistics resemble those in turbulent flow~\cite{li:synthetic_lagrangian_turbulence_generative_diffusion_models,deWit:datadriven_mori_zwanzig_modeling}. However, while generative models produce realistic flow fields for complex geometries, the generation of Lagrangian particle trajectories is mostly limited to simple geometries or unbounded environments~\cite{li:synthetic_lagrangian_turbulence_generative_diffusion_models}.

\subsubsection{Statistical Modeling of Received Signals}
While the approaches based on simulation or data-based models consider the full particle propagation in the environment, \ac{SMC} research is often only concerned with the concentration $c(t)$ observed at a given \ac{Rx}.
In turbulent flow, $c(t)$ is highly intermittent: it alternates between \textit{whiffs}, i.e., periods in which the concentration exceeds some detection threshold, and \textit{blanks}, i.e., periods when it does not.
Hence, $c(t)$ can be characterized by a few statistical quantities, such as how often and how intensely molecules are observed at the \ac{Rx}, and how long whiffs and blanks last~\cite{celani:odor_landscapes_turbulent_environments} (see Fig.~\ref{fig:airborne}).
In particular, when the \ac{Tx} releases molecules continuously into turbulent flow in the atmospheric boundary layer, i.e., the lowest part of the atmosphere, where the flow is shaped by the ground, the intermittency factor $\chi$, i.e., the fraction of time during which $c(t)>0$, and the expected molecule concentration during these periods, $\mathrm{E}\left[c(t)|c(t)>0\right]$, scale as derived in~\cite{celani:odor_landscapes_turbulent_environments} according to
\vspace*{-0.2cm}
\begin{align}
    \mathrm{E}\left[c(t)|c(t)>0\right] \sim c_0 \left( \frac{U \cdot r_\mathrm{Tx}}{\sigma_v \cdot x} \right)^2\label{eq:airborne:turb:conc},\\
    \chi \sim \cosh^{-2}\left( \frac{U\cdot y}{\sigma_v \cdot x} \right)\label{eq:airborne:turb:chi}.
\end{align}
Here, $x$ and $y$ denote the downwind and crosswind distances between \ac{Tx} and \ac{Rx}, respectively. Moreover, $U$ denotes the mean wind speed and $\sigma_v$ the amplitude of the turbulent velocity fluctuations. Finally, $r_\mathrm{Tx}$ denotes the size of the \ac{Tx} and $c_0$ is the concentration at the \ac{Tx}. Both $\chi$ and $\mathrm{E}\left[c(t)|c(t)>0\right]$ are time averages and thus do not depend on $t$ for statistically stationary turbulence.
From~\eqref{eq:airborne:turb:conc}, it can be seen that the concentration in whiffs depends only on the downwind distance but not on the cross-wind distance. However,~\eqref{eq:airborne:turb:chi} shows that moving crosswind results in observing fewer whiffs.
Due to this dependence on the relative positions of \ac{Tx} and \ac{Rx}, such statistical features of the observed signal can be used for tasks such as source localization or to guide communication system design~\cite{heinlein2026source,rigolli:learning_predict_target_location}.
Importantly, these statistics can be readily compared to experiments, rendering statistical modeling a natural intersection of theoretical and experimental research. For example, it has been shown in~\cite{celani:odor_landscapes_turbulent_environments} that~\eqref{eq:airborne:turb:conc} and~\eqref{eq:airborne:turb:chi} agree well with field measurements.

\subsection{Summary}
In this section, we reviewed how mismatches between experiments and theory drove the evolution of channel modeling for airborne \ac{SMC}.
Mismatches between analytical \ac{CIR} models and testbed measurements revealed that release and reception hardware must be accounted for in models and system design.
Over longer distances and outdoors, however, molecule propagation is turbulent and cannot be described by a single \ac{CIR}.
Besides numerical simulations, statistical models of the received signal are particularly promising due to their low complexity and direct comparability with experiments. Looking ahead, joint models of turbulent flow and sensing hardware, informed by real-world measurements, are needed to understand which information practical low-cost sensors can extract from the received signal. %

\section{Additional Aspects of the Interplay Between Theory and Experiment}
\label{sec:examples-others}
Sections~\ref{sec:examples-cvs} and~\ref{sec:examples-airborne} applied the \ac{SMC} modeling tools of Sections~\ref{sec:theoretical-models}-\ref{sec:experimental-models} to two specific application environments. In this section, to provide a more comprehensive picture, we discuss additional aspects of the interplay between theory and experiment based on the models developed for three experimental \ac{SMC} testbeds, i.e., a bacterial \ac{Tx}, a closed-loop testbed, and electrophysiological signaling in plants.

\subsection{Testbed Based on Genetically Modified Bacteria}
\label{ssec:others-bacteria}
\begin{figure}
    \centering
    \includegraphics[width=\linewidth]{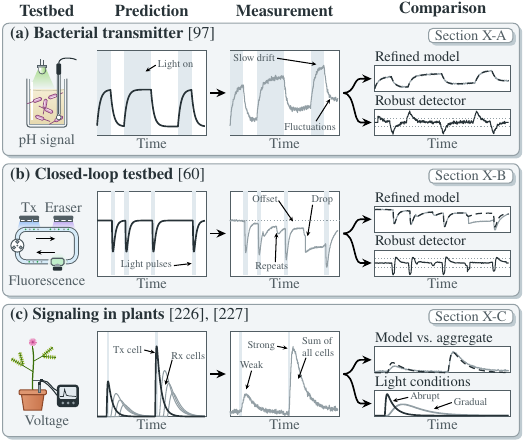}
    \caption{\textbf{Experiments revealing effects in the three exemplary testbeds.} \textbf{(a)} Drift and fluctuations of a bacterial \ac{Tx} lead to a refined model and a robust detector.
    \textbf{(b)} Recirculation in a closed-loop testbed causes repeated dips and an offset, which a refined model captures, whereas an abrupt drop caused by the setup is tolerated by a differential detector.
    \textbf{(c)} In plants, the analytical model describes individual cells, whereas the measured voltage aggregates many cells and agrees better with the analytical model for strong than for weak stimuli, and depends on the external light conditions.}
    \label{fig:others-effects}
    \vspace*{-0.3cm}
\end{figure}

First, we illustrate how experimental observations feed back into the analytical model of a biological \ac{Tx}, see Fig.~\ref{fig:others-effects}(a). Biological \ac{MC} testbeds that encode information via gene expression are limited to data rates of approximately 1\,bit/h \cite{Krishnaswamy2013}. To overcome this bottleneck, the authors of \cite{grebenstein2019biological} realized an optically controlled \ac{Tx} based on \textit{Escherichia coli} bacteria expressing the light-driven proton pump gloeorhodopsin. Upon illumination by a light-emitting diode, the bacteria transport protons into the surrounding medium, and the resulting pH change is measured by a pH microelectrode acting as the \ac{Rx}.

The analytical modeling started from a simplified physical description of the bacterial suspension as a well-mixed volume (see Table~\ref{tab:approx}), in which light-activated proton pumping and passive proton uptake correspond to zero- and first-order reactions, respectively, such that the proton concentration obeys a linear \ac{RRE} \eqref{eq:math-rre}. In line with earlier analytical models \cite{arjmandi2016ion, Zifarelli2008}, this description predicts that the proton concentration converges exponentially toward an equilibrium level after each switching of the light, see \textit{Prediction} in Fig.~\ref{fig:others-effects}(a). The measurements confirmed this behavior but also revealed a slow drift on time scales of approximately 20\,min, i.e., much longer than the symbol duration of 1\,min, as well as fast random fluctuations, see \textit{Measurement} in Fig.~\ref{fig:others-effects}(a). Neither effect was captured by the analytical model. Consequently, the analytical model was heuristically extended by a linear drift term, attributed, e.g., to aging of the bacteria, i.e., a time-variant channel (see Section~\ref{ssec:theoretical-special}), and by a Gaussian noise term \cite{grebenstein2019biological}.
The extended analytical model agreed well with the measurements, although its fitted parameters varied considerably between bacterial cultures, i.e., the \ac{Tx} exhibited extrinsic noise (see  Section~\ref{ssec:theoretical-statistical}).
The identified impairments also shaped the \ac{Rx} design towards a detector model that suppresses the noise by mean filtering and removes the drift by differentiation, see \textit{Comparison} in Fig.~\ref{fig:others-effects}(a). Via this adaptation, correct detection of all symbols of a 600-symbol sequence at 1\,bit/min could be achieved. In contrast, the model-based maximum likelihood detector suffered from parameter estimation inaccuracies, since the time-variant system caused the estimated parameters to become outdated \cite{grebenstein2019biological}. In terms of Fig.~\ref{fig:intro:model-choice}, the experiment thus confirmed the theory-led prediction but also revealed effects missing in the analytical model, i.e., drift and extrinsic noise, which led to a refined analytical model. Since biological \acp{Tx} can be inherently non-stationary and heterogeneous, such effects often become apparent only in experiments.

\subsection{Closed-Loop Testbed with Reversibly Switchable Signaling Molecules}
\label{ssec:others-closed-loop}
The closed-loop testbed in \cite{scherer2025closed} illustrates a complementary aspect of the interplay between theory and experiments: not every effect revealed by an experiment has to be included in the analytical model. It builds on media modulation \cite{Brand2022MediaModulation}, where the \ac{Tx} does not release \acp{SM} but switches the state of \acp{SM} already present in the channel. For the reversibly photoswitchable \ac{GFPD} as \ac{SM}, the authors of \cite{Brand2022MediaModulation} derive a stochastic analytical model of the received signal (see Section~\ref{ssec:theoretical-statistical}) that accounts for, e.g., incomplete and spontaneous state switching, i.e., reactions between the \ac{SM} states (see Section~\ref{ssec:mathematical-reaction}), and \ac{SM} propagation. In the testbed, the \acp{SM} circulate in a closed tube loop, light-emitting diode arrays act as \ac{Tx} and eraser, and the protein state is read out via fluorescence at the \ac{Rx}, see Fig.~\ref{fig:others-effects}(b).

The experiments in \cite{scherer2025closed} revealed effects not captured by this analytical model: since \acp{SM} remained in the loop, switched \acp{SM} passed the \ac{Rx} repeatedly, accumulated as a slowly varying offset, or were degraded by photobleaching, see \textit{Measurement} in Fig.~\ref{fig:others-effects}(b). These forms of \ac{ISI} are inherent to closed-loop systems, see the analytical closed-loop model \eqref{eq:cvs-single}, and are therefore captured by the revised analytical model in \cite{brand25closed}, which solves \eqref{eq:math-rad} with global and localized first-order degradation in a closed loop and shows that localized degradation can mitigate inter-loop and offset \ac{ISI}.

In contrast, an abrupt drop of the fluorescence in a 125-hour experiment, which recovered approximately 14 hours later, was attributed to a temporary agglomeration of photobleached proteins in the \ac{Rx} flow cell \cite{scherer2025closed}, i.e., to an artifact of the measurement setup rather than a property of the channel, see Section~\ref{sec:experimental-interplay}. Whether such an effect has to be modeled depends on the application: if the concentration is of interest, e.g., in targeted drug delivery, abrupt signal changes will be critical, whereas for information transmission, only the distinguishability of the symbols at the \ac{Rx} matters. Since the jump mainly shifts the signal level, a detector operating on the differentiated signal with adaptive thresholds recovers quickly, whereas a detector operating on the original signal produces a burst of errors \cite{scherer2025closed}, see \textit{Comparison} in Fig.~\ref{fig:others-effects}(b). The differential detector also compensates the slow drift, enabling error-free transmission of approximately 90,000 bits over 125 hours without eraser. Hence, instead of refining the analytical model, the communication scheme can be designed to be robust against effects irrelevant to the intended application.

\subsection{Electrophysiological Signaling in Plants}
\label{ssec:others-plants}
Plants constitute natural \ac{MC} systems, where the objective of modeling is to understand rather than to design the communication link. In \cite{awan2019single}, an analytical model for action potentials in plants is developed based on \ac{MC} theory: an external stimulus alters the opening probability of ion channels in a \ac{Tx} cell, and the resulting change of the membrane potential determines the number of released \acp{SM}. These propagate via diffusion or active transport to neighboring \ac{Rx} cells, which are modeled as well-mixed voxels (see Table~\ref{tab:approx}) and detect the \acp{SM} via ligand-receptor reactions, i.e., as reactive \acp{Rx} (see Section~\ref{ssec:theoretical-rx}). Based on this analytical model, the mutual information and the information propagation speed are derived as functions of the number and configuration of the \ac{Rx} cells. In \cite{Awan20exp}, the analytical model is extended to multiple action potentials and to mechanically triggered signals, and compared with the extracellular voltage between stem and soil of \textit{Mimosa pudica} and \textit{Aloe vera} plants \cite[Figs.~5 and~6]{Awan20exp}, see \textit{Prediction} and \textit{Measurement} in Fig.~\ref{fig:others-effects}(c). The measured waveforms agree with the analytical model, particularly for more intense stimuli, and the mutual information computed from the measurements is consistent with the numerically obtained values.

This example illustrates a fundamental challenge of experimental research on natural \ac{MC} systems, which also arises in \ac{SMC} testbeds: the observed quantity differs from the quantity in which the analytical model is formulated, i.e., the measurement chain \eqref{eq:sensor} is largely unknown. While the analytical model describes the number of \acp{SM} in individual cells, the measurement provides a single voltage that aggregates the activity of many cells. Hence, the experiments verify the signal generation at the \ac{Tx} and the mutual information for three \ac{Rx} cells, whereas predictions for larger cell populations, e.g., the saturation of the mutual information beyond 10 to 12 \ac{Rx} cells, rely on numerical results only \cite{Awan20exp}. Moreover, the measurement conditions affect the observed signals: in \cite{bekkari2025modeling}, a Hodgkin-Huxley-type model is fitted to long-term recordings of light-induced action potentials in tobacco and tomato plants, and the differences between the waveforms are attributed to the illumination, i.e., abrupt artificial versus gradual natural light transitions, rather than to the plant species alone. Thus, compared to the design of \ac{SMC} links, the interplay is reversed: rather than validating an analytical model of a designed system, experiments determine which features of the biological system, as opposed to the experimental conditions, an analytical model has to explain.

\section{Challenges and Future Directions}
\label{sec:challenges}
This section presents selected directions for future research in \ac{SMC} channel modeling.

\subsection{Open Modeling Problems}

The maturity of channel modeling varies considerably across the application domains of \ac{SMC}. In this section, we highlight a few important open modeling problems.

\textbf{Heterogeneous and time-varying \ac{SMC} hardware:} Realizations of nominally identical \acp{Tx} and \acp{Rx} vary substantially and change during operation, e.g., in vesicle size and membrane properties \cite{lehman2023liposome} or in the response of living cells \cite{grebenstein2019biological}. Statistically characterizing this extrinsic noise and incorporating it into system modeling, design, and analysis are important directions for future research \cite{tomdieck2026nanodevices}.

\textbf{Statistical models for turbulent channels:} Turbulent transport is relatively underexplored in \ac{SMC}. Numerical fluid-dynamics models can reproduce turbulent behavior \cite{biferale:turb_smoke}; however, tractable statistical models of concentration distributions, temporal correlations, intermittency, and spatial dependencies are still needed for system design and analysis, for which fading, temporal, and spatial/\ac{MIMO} channel models from wireless communications may serve as a blueprint \cite{Tse2005Fundamentals}.

\textbf{Realistic channel models for the \ac{CVS}:} Existing analytical models mostly apply to individual medium-sized vessels and simplified \acp{VN} \cite{schaefer21b,Wille2025pulsatile,Jakumeit2026MIGHT}. More realistic models should capture transport through capillary beds, including blood-cell interactions and non-Newtonian flow, transitions between vessel scales \cite{aaronson2020cardiovascular}, and the lymphatic system. Experimental validation in physiologically representative environments, see, e.g., the \ac{CAM} testbed \cite{vakilipoor2025cam}, is necessary to identify which effects require explicit modeling.

\textbf{Experimental channel models under limited observability:} In many \ac{SMC} systems, the molecular quantities described by analytical models cannot be measured directly; instead, experiments provide indirect or aggregated measurements, such as fluorescence intensities or extracellular voltages, cf.\ \eqref{eq:sensor}. Hence, measurement-aware analytical models are needed that characterize under which conditions the molecular states are observable, i.e., can be reconstructed from the measurements, and that distinguish effects of interest from measurement noise and experimental artifacts.
\vspace*{-0.3cm}

\subsection{Simulation, Emulation, and Digital Twins}

Practical \ac{SMC} systems are often extremely complex environments with interacting physical processes, whose accurate and efficient modeling requires a combination of analytical, simulation, data-based, and experimental models. We discuss three promising directions.

\textbf{Multi-scale, multi-level simulators:} Practical \ac{SMC} environments involve processes at different scales, e.g., microscopic stochastic interactions that can be captured by particle-based simulations and macroscopic transport that is more efficiently described by continuum models (see Section~\ref{sec:hybrid_and_multiscale_coupling}). An important research direction is the development of flexible simulators which adapt the level of fidelity to the application requirements, the computational complexity, and the acceptable modeling error. A complementary direction concerns functional abstraction, i.e., choosing the level of detail according to the performance metrics of interest, e.g., by adapting the physical-layer, link-level, and network-level simulations of wireless communications to \ac{SMC} \cite{Yubing_2017_nanoNS3,Ertan_2010_NanoNS,noel2017simulating}.

\textbf{Channel emulators:} In wireless communications, channel emulators enable reproducible testing under conditions that are difficult to realize experimentally, e.g., satellite links \cite{pugliese2026tutorial}. In \ac{SMC}, emulators that reproduce the input-output behavior of, e.g., the \ac{CVS} or turbulent channels would enable hardware-in-the-loop testing of transceivers and algorithms without access to the physical channel.

\textbf{Digital twins:} Since simulations cannot capture all physical effects and conducting experiments for every scenario is often infeasible, digital twins combine analytical and simulation models with real-time measurements from a specific physical system and continuously update them, where data-based models (see Section~\ref{sec:data-based-models}) can serve as fast surrogates or learn the deviations between model and measurements. In \ac{SMC}, digital twins could support continuous health monitoring, the prediction of system behavior, and the evaluation of what-if scenarios.
\vspace*{-0.3cm}

\subsection{Emerging AI-based Modeling Methodologies}

\Ac{AI}- and \ac{ML}-based methods can advance \ac{SMC} channel modeling particularly for complex systems that are difficult to describe by conventional modeling approaches. We discuss two promising directions.

\textbf{Hybrid physics- and data-driven models:} Data-based models (see Section~\ref{sec:data-based-models}) can efficiently approximate complex physical simulations and capture effects that are difficult to describe analytically or to simulate numerically, but purely data-driven approaches often require substantial training data, can output non-physical results, and may generalize poorly beyond the training conditions. A promising research direction is therefore the development of hybrid models that retain established physical laws and exploit \ac{ML} only for unknown or difficult-to-model effects \cite{Karniadakis2021}, see, e.g., \cite{ahmet:pignn_1d_equations_blood_flow,Baydas2026} for initial examples of physics-informed learning relevant for \ac{SMC}.

\textbf{Generative models:} Most \ac{ML} models predict typical channel outputs but may not capture stochastic fluctuations and temporal correlations, which are important for assessing the performance of \ac{SMC} systems. Generative models, e.g., the \acp{GAN} and diffusion models in Section~\ref{sec:data-based-models}, learn these distributions and generate statistically representative channel realizations conditioned on system parameters, as demonstrated in, e.g., the modeling of turbulent flows \cite{raonic:generative_ai_efficient_statistical_computation_fluids,saydemir:unfolding_time_generative_modeling_turbulent_flows_4d}. An important research direction for \ac{SMC} is the development of generative channel models that capture the statistics relevant for communication, e.g., molecule-counting noise, channel memory, spatiotemporal correlations, and complex flow fields.
\vspace*{-0.3cm}

\subsection{Interdisciplinary, Reproducibility, and Accessibility}

The interdisciplinary nature of \ac{SMC} research makes experimental models difficult to develop and reproduce, which limits access to experimental validation across the research community. A promising development is the increasing availability of open-source simulators and publicly accessible experimental data sets including information about measurement procedures and calibration, see Table~\ref{tab:datasets}. Examples include the ``Methods and Data'' papers in the IEEE Transactions on Molecular, Biological, and Multi-Scale Communications, see, e.g., \cite{grebenstein19,debus2025blood,vakilipoor2025cam}, and the Zenodo community of the research training group SyMoCADS \cite{symocads_zenodo}, which collects experimental and simulation data sets together with evaluation code, e.g., for \cite{scherer2025closed,Wille2025pulsatile}. Extending these practices to a broader range of \ac{SMC} scenarios, e.g., airborne and \emph{in vivo} systems, where experimental data are scarce, would help accelerate progress.
\vspace*{-0.3cm}

\section{Conclusions}
\label{sec:conclusion}
In this tutorial, we have provided a comprehensive overview of \ac{SMC} channel modeling. We first introduced a general end-to-end \ac{SMC} channel model and the mathematical fundamentals required to describe the underlying physical and chemical processes. Subsequently, we introduced analytical, simulation-based, data-based, and experimental models, and discussed their capabilities, limitations, and interrelationships. In particular, we argued that the choice of the model type as well as the level of abstraction should be guided by the intended application and the specific research question. We then applied these modeling approaches to two application environments, the \ac{CVS} and airborne \ac{SMC}, and investigated additional aspects of the interplay between theory-led and experiment-guided modeling based on three experimental testbeds.
Finally, we identified open modeling problems and outlined promising directions for future research.

As a key takeaway of this tutorial, we analyzed the \textit{evolving} relation between theory-led and experiment-guided modeling in the \ac{SMC} literature, where the role of experiments has changed from merely validating analytical models to actively guiding the model development by revealing relevant physical effects and justifying simplifying assumptions. In other words, this continuous interplay determines the relevant modeling questions and, correspondingly, the appropriate choice of analytical, simulation-based, and data-based models. As a relevant future direction, the closer integration of theory and experiment, as well as that of the corresponding modeling approaches are essential for developing practically relevant channel models that support the transition of \ac{SMC} from theoretical concepts to real-world applications.

\section*{Acknowledgment}
This work was funded in part by the German Federal Ministry of Research, Technology and Space (BMFTR) through Project Internet of Bio-Nano-Things (IoBNT) – grant numbers 16KIS1987 and 16KIS1992, in part by the German Research Foundation (Deutsche Forschungsgemeinschaft, DFG) under GRK 2950 – ProjectID 509922606 and under grant numbers SCHA~\mbox{2350/2-1}, SCHA~\mbox{2350/4-1}, SCHA~\mbox{2350/5-1}, SCHO~\mbox{831/24-1}, SCHO~\mbox{831/25-1}, JA~\mbox{3104/6-1}, LO~\mbox{3505/1-1}, and in part by the European Union’s Horizon Europe – HORIZON-EIC-2024-PATHFINDEROPEN-01 under grant agreement Project N. 101185661.

\end{document}